\documentclass[acmsmall,screen]{acmart}
\AtBeginDocument{%
  }

\setcopyright{none}
\renewcommand\footnotetextcopyrightpermission[1]{}
\acmJournal{TOSEM}
\usepackage{graphicx}
\usepackage{multirow}
\usepackage{longtable}
\usepackage{listings}
\usepackage{xcolor}
\usepackage{tcolorbox}
\tcbuselibrary{breakable}
\usepackage{booktabs, tabularx, url}
\usepackage{subcaption}
\usepackage{pdflscape}
\usepackage{enumitem}
\definecolor{diffstart}{named}{gray}
\definecolor{diffadded}{RGB}{230,255,230}
\definecolor{diffremoved}{RGB}{255,235,235}
\definecolor{addedtext}{RGB}{0,100,0}
\definecolor{removedtext}{RGB}{180,0,0}
\definecolor{notecolor}{RGB}{180,90,0}
\definecolor{commentcolor}{RGB}{100,100,100}
\begin{document}

\title{Breaking Changes in Software Ecosystems: A Systematic Literature Review}
\author{Juntao Chen}

\orcid{0009-0006-7815-540X}
\affiliation{%
  \institution{The University of Melbourne}
  \city{Melbourne}
  \country{Australia}}
\email{juntao.chen@student.unimelb.edu.au}

\author{Tingting Bi}
\authornote{Corresponding author.}
\affiliation{%
  \institution{The University of Melbourne}
  \city{Melbourne}
  \country{Australia}}
\email{tingting.bi@unimelb.edu.au}

\author{Yanlin Wang}
\affiliation{%
  \institution{Sun Yat-sen University}
  \city{Guangzhou}
  \country{China}}
\email{wangylin36@mail.sysu.edu.cn}

\author{Patanamon Thongtanunam}
\affiliation{%
  \institution{The University of Melbourne}
  \city{Melbourne}
  \country{Australia}}
\email{patanamon.t@unimelb.edu.au}

\renewcommand{\shortauthors}{Chen et al.}

\begin{abstract}
Modern software systems rely on dependency networks of reusable libraries, where breaking changes propagate and cause downstream consumers to fail. Despite growing research across ecosystems, no comprehensive synthesis exists. We conduct a systematic literature review of 97 primary studies, answering four research questions across five ecosystems: Maven/Java, npm/JavaScript, Python, Web APIs, and Linux distributions. The synthesis yields four results. First, a four-dimensional taxonomy along Nature, Detectability, Scope, and Visibility. Second, five reason categories and five impact dimensions, where maintenance and design improvements account for a larger share of breaking changes than new feature work. Third, 43 detection approaches that reach high accuracy on syntactic breaks but limited coverage on behavioral ones. Fourth, 66 strategies for communicating, preventing, and recovering from breaking changes, organized by the actor's role. Based on these findings, we identify three open challenges and three research opportunities. The challenges are behavioral break detection at scale, the failure of semantic versioning as a trust mechanism, and transitive dependency propagation under information asymmetry. The opportunities are LLM-augmented behavioral contract inference, ecosystem-level dependency graph intelligence, and domain-specific tooling for ML and data science.
\end{abstract}

\ccsdesc[500]{Software and its engineering~Software evolution}
\ccsdesc[500]{Software and its engineering~Software libraries and repositories}
\ccsdesc[300]{Software and its engineering~Software version control}
\ccsdesc[300]{Software and its engineering~Maintaining software}
\ccsdesc[300]{General and reference~Surveys and overviews}

\keywords{Breaking Changes, API Evolution, Backward Incompatibility,
  Software Ecosystems, Systematic Literature Review}


\maketitle

\section{Introduction}
\label{sec:introduction}

Modern software development relies on reusable libraries and packages~\cite{mockus2007reuse, heinemann2011reuse} distributed through software ecosystems such as npm, Maven Central, and PyPI. Applications depend on dozens or hundreds of third-party libraries, both directly and through deep graphs of transitive dependencies~\cite{decan2019ecosystems, wittern2016jsdynamics}. Some direct dependencies are trivial single-purpose packages that are nonetheless widely adopted across projects~\cite{abdalkareem2017trivial}. Empirical studies show that transitive dependencies often include code that is never reached from client functionality~\cite{sotovalero2021bloated}. This reuse model accelerates development, but it also tightly couples library providers and their downstream consumers~\cite{cox2019dependencies}. Consumers commonly keep outdated dependency versions in production rather than upgrade promptly~\cite{kula2018outdated, decan2018lag}. When a library changes its interface or behavior, the effects can propagate through the dependency graph. They reach applications whose authors have no involvement in the change.

Software systems undergo continuous change throughout their lifetime~\cite{lehman1980laws, mens2008evolution}, and libraries are no exception. As libraries evolve, their application programming interfaces (APIs) change continuously~\cite{dig2006apievolution}. Maintainers add new methods, modify existing signatures, deprecate outdated interfaces~\cite{brito2018deprecation, robbes2012deprecation}, and remove obsolete functionality. Downstream consumers, however, do not always respond to deprecation signals in a timely manner, and deprecated APIs frequently remain in use for extended periods~\cite{sawant2019react}. To help consumers reason about the risk of an upgrade, the community has converged on Semantic Versioning (SemVer)~\cite{semverspec}. SemVer is the dominant convention for signaling the compatibility impact of these changes. Under SemVer, a major version increment signals backward-incompatible changes. Minor and patch increments signal backward-compatible additions and fixes, respectively. In practice, however, version numbers alone do not reliably indicate whether an upgrade is safe. SemVer conventions are frequently violated, and downstream consumers are routinely broken by minor or patch releases that should have been backward-compatible.

These compatibility-violating modifications are commonly referred to as \textit{breaking changes}~\cite{desrivieres2007apis}. More precisely, a breaking change is a modification to a library's interface or behavior that causes existing client code to fail. Such failures may surface at build time as compilation or linking errors. They may also surface at runtime as unexpected return values, exceptions, or shifts in side effects~\cite{desrivieres2007apis}. Even unintended observable behaviors are frequently relied upon by downstream consumers, which makes the boundary between intended and incidental behavior difficult to maintain across releases~\cite{winters2020sweatgoogle}. Libraries are released independently from the applications that depend on them. As a result, even a small change introduced by a maintainer can propagate through the dependency graph and break downstream consumers that have not initiated the upgrade~\cite{cox2019dependencies, decan2018vulnerabilities}.

Breaking changes are a central concern for the integrity of the modern software supply chain~\cite{ladisa2023sok, ohm2020backstabber}. Applications rely on increasingly deep stacks of third-party libraries, and only a portion of the embedded code is reachable from typical client usage~\cite{pashchenko2018vulnerable}. A single incompatible change in a widely used package can cascade through transitive dependents and disrupt downstream projects whose authors have not initiated the upgrade~\cite{zimmermann2019npm}. The resulting build failures, runtime errors, and migration costs make breaking changes a recurring source of friction in dependency management. Tools that automate routine update pull requests have seen mixed adoption among consumers who are reluctant to merge upgrades they have not validated~\cite{mirhosseini2017automated}. Supply chain security and reliability now receive sustained attention from both industry and regulators~\cite{whitehouse2021eo14028, souppaya2022ssdf}. Understanding how breaking changes arise, propagate, and can be controlled has become essential to sustaining the open-source ecosystems on which modern software depends. This is especially the case as contributor turnover leaves widely used packages reliant on a small group of maintainers~\cite{constantinou2017ecosystems}.

These concerns have driven a growing body of research on breaking changes. However, to the best of our knowledge, there is no comprehensive survey that synthesizes the topic across ecosystems. Existing primary studies treat the problem in isolation. They focus on detection techniques for specific languages or package managers, on the impact of breaking changes on downstream consumers, or on the reasons library maintainers introduce them. The most closely related secondary study is a systematic review of API evolution literature~\cite{10.1145/3470133}. It surveys 369 publications and groups them into API maintenance, API usability, and other topics. In that review, breaking changes appear as one issue within API usability, alongside integration problems, API misuse, and documentation. The review does not classify breaking changes or analyze them as its central subject. Adjacent surveys cover related but distinct dimensions of API research. Examples include the automated inference of API properties from code and tests~\cite{robillard2013apiinference} and the role of refactorings in shaping how APIs evolve~\cite{kim2011refactorings}. Comparative empirical analyses of dependency issues across packaging ecosystems likewise touch on breaking changes as one symptom among many~\cite{decan2017empirical}. Our study differs by focusing specifically on breaking changes and covering their full lifecycle: classification, reasons, detection, and management. It spans five major software ecosystems (Maven/Java, npm/JavaScript, Python, Web APIs, and Linux distributions).

We conduct a systematic literature review on breaking changes in software ecosystems. We follow the established guidelines of Kitchenham and Charters~\cite{kitchenham2007slr}. We search three major academic databases (IEEE Xplore, ACM Digital Library, and Springer) using structured keyword queries targeting breaking change terminology. We then apply inclusion and exclusion criteria and perform both backward and forward snowballing. The process yields 97 primary studies, from which we extract data to answer our research questions.

We organize our investigation around four research questions. We first ask what types of breaking changes occur in practice. We then ask why such changes are introduced and what consequences they create for downstream consumers. We next ask what techniques exist to detect and analyze breaking changes before they propagate. Finally, we ask how breaking changes are managed and responded to in practice. We formalize these four questions as follows:

\begin{itemize}
    \item \textbf{RQ1:} What are the prevalent types of breaking changes observed in software ecosystems?

    \item \textbf{RQ2:} What are the underlying reasons for introducing breaking changes, and how do these changes impact downstream consumers and the broader ecosystem?

    \item \textbf{RQ3:} What approaches exist to detect and analyze breaking changes, and what are their effectiveness and limitations?

    \item \textbf{RQ4:} What strategies exist to communicate, prevent, and recover from breaking changes in software ecosystems?
\end{itemize}

Our synthesis across the four research questions reveals three recurring findings. First, breaking changes divide into syntactic and behavioral forms. Detection of syntactic breaks is comparatively mature, while behavioral breaks remain difficult to detect automatically. Second, Semantic Versioning is the dominant convention for signaling compatibility, yet it is violated systematically. Non-major releases routinely carry breaking changes that the version number does not advertise, which leads downstream consumers to delay upgrades. Third, transitive dependencies are the largest single source of client-impacting breaks. An information asymmetry between maintainers and consumers makes such breakage hard to diagnose. These findings motivate three directions for future work. Behavioral contract inference, potentially assisted by large language models, could help distinguish intentional improvements from regressions. Ecosystem-level analysis of the dependency graph could predict and trace transitive breakage before it propagates. Domain-specific tooling could address the silent behavioral drift observed in machine learning and data science libraries, which general-purpose tools do not address. We develop these challenges and opportunities further in Section~\ref{sec:challenges}.

The main contributions of this survey are as follows:
\begin{itemize}
    \item We present the first comprehensive survey on breaking changes in software ecosystems, systematically reviewing 97 primary studies.
    \item We identify a four-dimensional taxonomy of breaking changes, covering their nature, detectability, scope, and visibility, drawn from five major software ecosystems: Maven/Java, npm/JavaScript, Python, Web APIs, and Linux distributions.
    \item We examine the reasons behind breaking changes and their downstream impacts, covering five reason categories and five impact dimensions.
    \item We review and compare 43 detection and analysis approaches, covering static, dynamic, learning-based, and hybrid techniques, and provide a comparative evaluation of their trade-offs.
    \item We synthesize 66 strategies for communicating, preventing, and recovering from breaking changes, and organize them by the role of the actor that applies them.
\end{itemize}

The remainder of this paper is organized as follows. Section~\ref{sec:methodology} describes the methodology of our systematic literature review. Section~\ref{sec:rq1} presents the taxonomy of breaking changes (RQ1). Section~\ref{sec:rq2} analyzes the reasons and impacts of breaking changes (RQ2). Section~\ref{sec:rq3} surveys detection and analysis approaches (RQ3). Section~\ref{sec:rq4} examines communication, prevention, and recovery strategies (RQ4). Section~\ref{sec:challenges} discusses open challenges and future research opportunities. Section~\ref{sec:threats} addresses threats to validity. Section~\ref{sec:conclusion} concludes the paper. Appendix~\ref{appendix:primary-studies} lists all 97 selected studies.

\section{Methodology}
\label{sec:methodology}
\begin{figure}[t]
    \centering
    \includegraphics[width=0.8\textwidth]{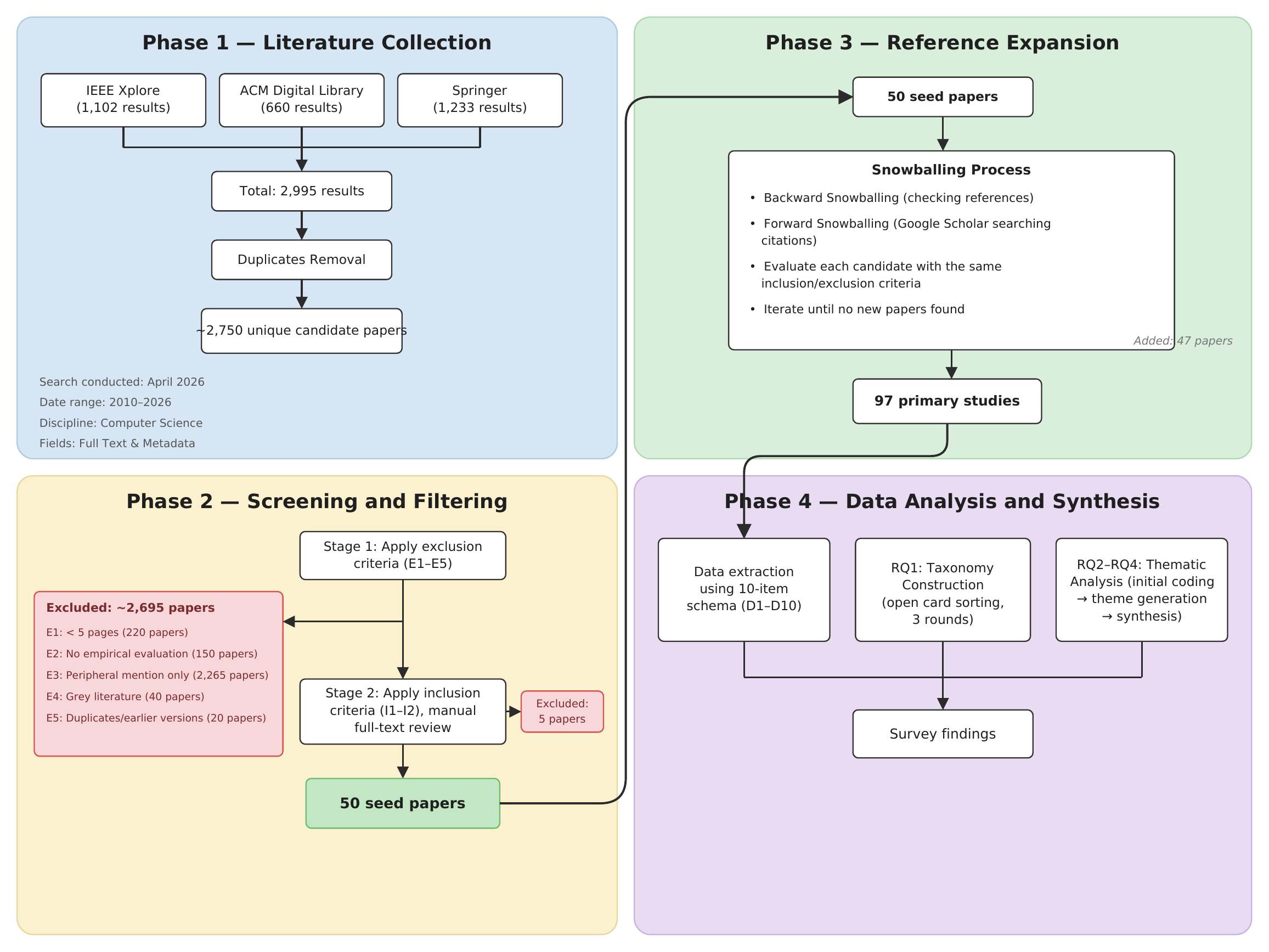}
    \caption{Schematic of the Overall Systematic Literature Review Process}
    \label{fig:methodology1}
\end{figure}

In this section, we describe our methodology for selecting and analyzing papers on breaking changes. We adopt several established systematic literature review practices. These include structured keyword searching, snowballing, and explicit inclusion/exclusion criteria. Our goal is to synthesize the current knowledge on breaking changes across software ecosystems. The methodology consists of four phases: (1) literature collection, (2) screening and filtering, (3) reference expansion, and (4) data analysis and synthesis. Figure~\ref{fig:methodology1} illustrates the overall process.

\subsection{Literature Collection}
We begin by searching three major academic databases: IEEE Xplore, ACM Digital Library, and Springer. These databases cover the primary venues for software engineering research. They include top conferences (e.g., ICSE, FSE, ISSTA, and ASE) and journals (e.g., TSE and TOSEM).

We construct the search string by following the guidelines of Kitchenham and Charters~\cite{kitchenham2007slr}. We first identify the core concept of the search from the research questions of this survey: breaking changes in software libraries, APIs, and ecosystems. We then draw up a list of synonyms and alternative spellings for this concept. The literature refers to breaking changes under several different names across ecosystems and research communities. For each term, we also add plural forms.

We connect all resulting search terms with the Boolean OR operator to form the following search string:

\begin{quote}
\textit{``Breaking Change''} OR \textit{``Breaking Update''} OR \textit{``Breaking Updates''} OR \textit{``Breaking Dependency''} OR \textit{``Breaking API Change''} OR \textit{``Breaking API Changes''} OR \textit{``Library Breaking Change''} OR \textit{``Dependency Breaking Change''} OR \textit{``Backward Incompatible Change''} OR \textit{``Backward Incompatibility''} OR \textit{``Backwards Incompatible''} OR \textit{``Backwards Incompatibility''} OR \textit{``Non-backward Compatible''} OR \textit{``Non-backwards Compatible''} OR \textit{``Incompatible API Change''} OR \textit{``Incompatible Library Change''} OR \textit{``API Incompatibility''} OR \textit{``API Incompatibilities''} OR \textit{``API Breakage''} OR \textit{``API Breakages''} OR \textit{``Library Breakage''} OR \textit{``Dependency Breakage''}
\end{quote}

The search is conducted in April 2026 and restricted to papers published between 2010 and 2026. We choose 2010 as the starting year for two reasons. First, modern package ecosystems emerge around this time (npm launches in 2010, and Maven Central reaches widespread adoption in the same period). Second, breaking change research largely appears after this point. We acknowledge that this cutoff may exclude earlier foundational work. To mitigate this risk, the snowballing phase (Section~\ref{sec:ref-expansion}) can recover influential pre-2010 studies that are heavily cited by papers within our search window. We also limit the search to the Computer Science discipline in each database to exclude results from unrelated fields. Some studies discuss breaking changes in the body of the paper but not in the metadata. To capture these, we apply the search string to the \textit{full text and metadata} of papers in each database. This covers the document title, abstract, author keywords, index terms, and body text. The search returns 1{,}102 results from IEEE Xplore, 660 from the ACM Digital Library, and 1{,}233 from Springer, yielding 2{,}995 results in total. After removing duplicates that appear in more than one database, we obtain 2{,}750 unique candidate papers.

\subsection{Screening and Filtering}
We apply a two-stage filtering process to the 2{,}750 candidate papers. The first stage applies five exclusion criteria (E1--E5) to remove papers outside our scope:

\begin{itemize}
    \item \textbf{E1}: the paper has fewer than five pages, as short-form or preliminary studies often lack the technical depth, novelty, or detailed evaluation required for this survey.
    \item \textbf{E2}: the paper does not include any evaluation or empirical evidence related to breaking changes.
    \item \textbf{E3}: the paper only mentions breaking changes peripherally (e.g., in related work or future directions) without making them a core focus.
    \item \textbf{E4}: the paper is a grey literature source (e.g., blog post, white paper, or non-peer-reviewed technical report).
    \item \textbf{E5}: the paper is a duplicate or an earlier version of a study already included.
\end{itemize}

The second stage involves manual relevance assessment against two inclusion criteria (I1--I2):

\begin{itemize}
    \item \textbf{I1}: the paper presents an empirical study, technique, framework, or tool that directly addresses breaking changes in software libraries, APIs, or ecosystems.
    \item \textbf{I2}: the paper provides actionable evidence on the classification, reasons, impact, detection, communication, prevention, or recovery of breaking changes.
\end{itemize}

In the first stage, the first author screens each candidate paper against the exclusion criteria (E1--E5) by reading its title and abstract. Because the number of candidates is large, the first author uses a large language model (Claude Opus 4.7) to support this stage. The model is provided with the full definitions of the five exclusion criteria. For each paper, it produces an initial assessment against these criteria from the title and abstract. The first author then reviews the title and abstract of every paper to verify each assessment and makes the final exclusion decision. The model serves only as a screening aid and does not make any final decision. This step excludes approximately 2{,}695 papers. About 220 have fewer than five pages (E1), 150 lack empirical evaluation related to breaking changes (E2), 2{,}265 mention breaking changes only peripherally (E3), 40 are grey literature sources (E4), and 20 are duplicates or earlier versions of included studies (E5). The dominant share of E3 reflects the broader recall of the full-text search. Because we search the full text rather than the metadata alone, the search surfaces many papers that use a search term only in passing. We observe three recurring cases among these papers. First, some papers study dependency update practices, such as technical lag or automated update bots, and treat breaking changes only as one risk of updating. Second, some papers analyze semantic versioning compliance in specific ecosystems and mention breaking changes only as the event that a major version increment should signal. Third, some papers examine release notes, API deprecation, or vulnerability patching, where breaking changes appear as one topic among several. In none of these cases are breaking changes the object of study. The search also returns false-positive matches from unrelated fields, such as telecommunications standards, where a term like ``backward incompatibility'' carries a different meaning. In the second stage, the first author reads each of the remaining 55 papers in full to determine whether it meets the inclusion criteria (I1--I2). After this process, 50 papers remain and form the seed set for reference expansion.

\subsection{Reference Expansion}
\label{sec:ref-expansion}
We expand the seed set of 50 papers through both backward and forward snowballing~\cite{Wohlin2014-snowballing}. In the backward pass, we examine the reference lists of all 50 seed papers to find cited works within our scope. In the forward pass, we use Google Scholar to find papers that cite the seed papers. Each candidate identified through snowballing is assessed against the same inclusion and exclusion criteria. The 2010 year restriction is relaxed for foundational works that are heavily cited by papers within our search window. This process continues iteratively until no new relevant papers are found. The snowballing phase adds 47 papers that meet all criteria, bringing the total to 97 primary studies.

When both a conference paper and its journal extension are identified, we include both only in specific cases. The extended version must introduce new research questions, additional ecosystems, or expanded datasets that provide distinct contributions beyond the original study. Snowballing also identifies a paper accepted at a peer-reviewed venue but available only as an arXiv preprint at the time of our search. We include it because it has already passed peer review. All selected studies are listed in Appendix~\ref{appendix:primary-studies}.

\subsection{Data Analysis and Synthesis}

\begin{figure}[htbp]
    \centering
    \begin{minipage}[t]{0.48\textwidth}
        \centering
        \includegraphics[width=\textwidth]{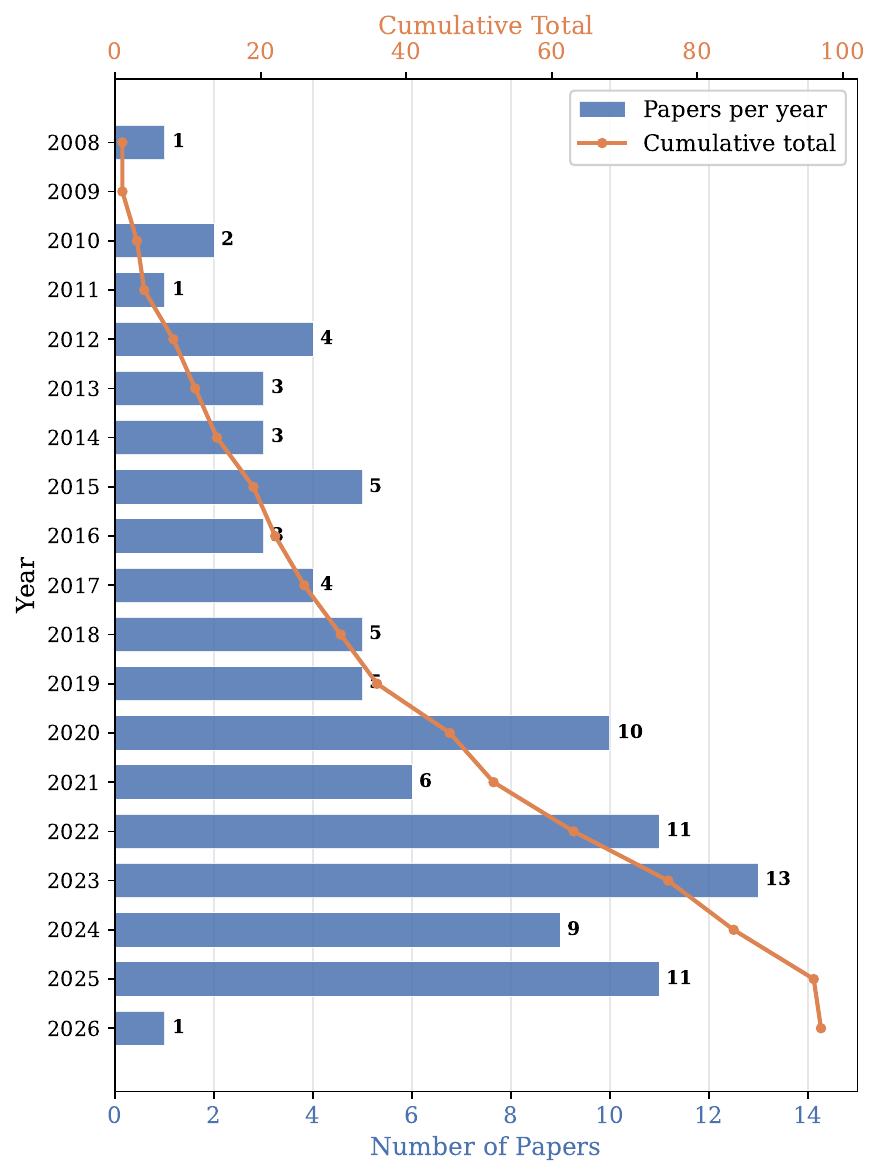}
        \caption{Temporal distribution of the 97 selected studies. Bars indicate the number of papers published each year (left axis); the line shows the cumulative total (right axis).}
        \label{fig:year-distribution}
    \end{minipage}
    \hfill
    \begin{minipage}[t]{0.48\textwidth}
        \centering
        \includegraphics[width=\textwidth]{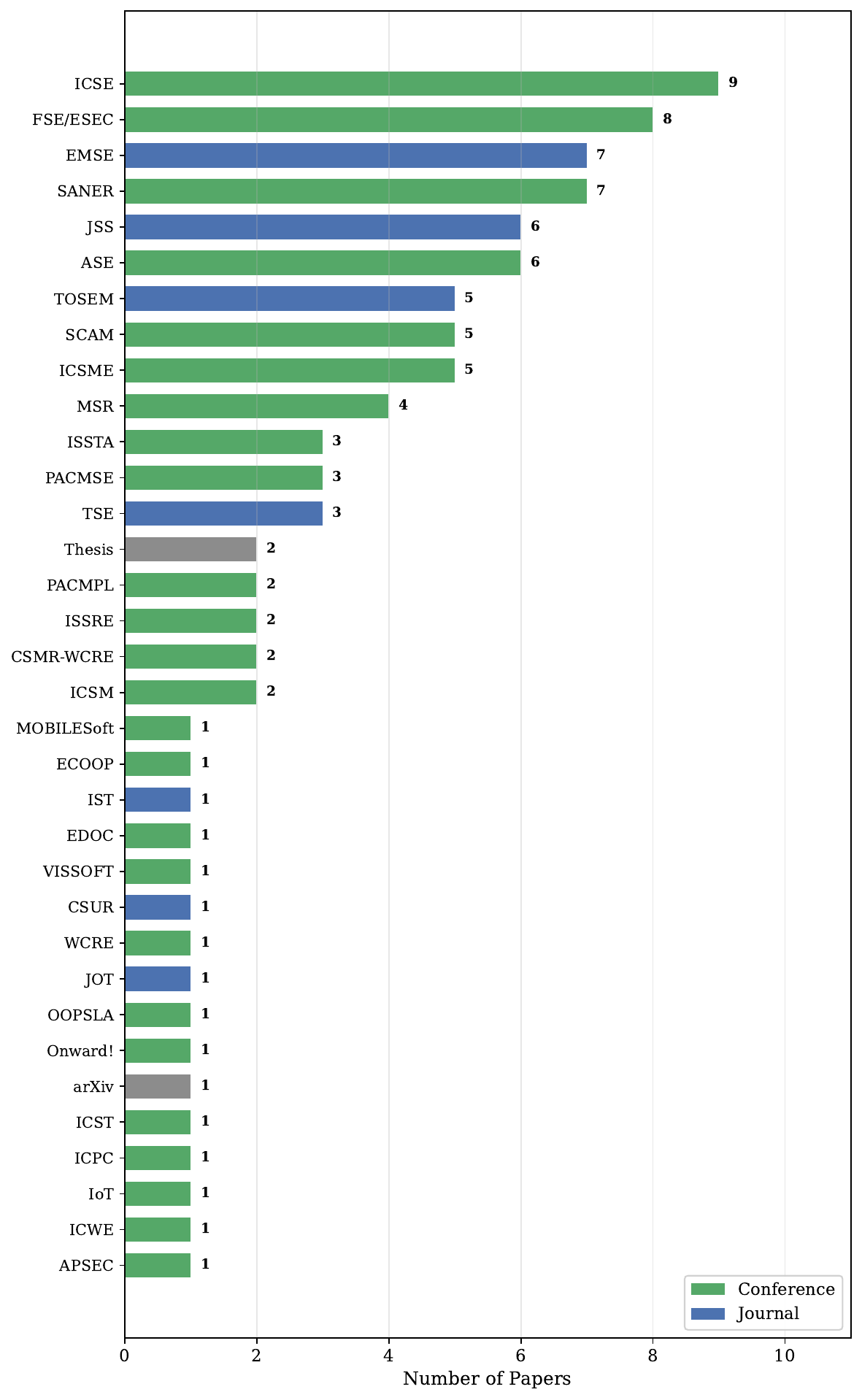}
        \caption{Distribution of the 97 selected studies across publication venues. Green bars indicate conferences, blue bars indicate journals, and gray bars indicate other sources (PhD theses and preprints).}
        \label{fig:venue-distribution}
    \end{minipage}
\end{figure}

\textbf{Statistics of selected papers}: we first analyze the descriptive statistics of the 97 selected studies. Figure~\ref{fig:year-distribution} shows their temporal distribution. Research on breaking changes has grown over the past decade, with an increase since 2017. The period from 2020 to 2026 accounts for 61 of the 97 papers (63\%). This trend shows that breaking change research remains an active area.

Figure~\ref{fig:venue-distribution} shows the distribution of selected studies across publication venues. Most papers (72\%) appear in peer-reviewed conferences. The most frequent conference venues are ICSE (9 papers), FSE/ESEC (8), and SANER (7). Journal publications account for 25\% of the selected studies, led by EMSE (7 papers), JSS (6), TOSEM (5), and TSE (3). The remaining three studies are two PhD theses and one preprint. The selected studies are concentrated in top-tier software engineering venues.

\textbf{Data analysis of RQs}: To answer the RQs listed in Section~\ref{sec:introduction}, we design a data extraction schema. Table~\ref{tab:data_items} presents the ten data items we record from each of the 97 studies. Items D1--D4 capture bibliographic and contextual information. Items D5--D10 map to specific research questions. Breaking change types (D5) inform the taxonomy for RQ1. Reasons and impacts (D6--D7) form the basis for RQ2. Detection techniques (D8) are analyzed in RQ3. Prevention strategies and mitigation approaches (D9--D10) are analyzed in RQ4. We record all papers against this schema before conducting further analysis.

\begin{table}[ht]
  \caption{Data Items Extracted from Selected Studies}
  \label{tab:data_items}
  \scriptsize
  \renewcommand{\arraystretch}{1.3}
  \begin{tabularx}{\textwidth}{l p{2.8cm} X p{2.4cm}}
    \toprule
    \textbf{ID} & \textbf{Data Item} & \textbf{Description} & \textbf{Relevant Part in Survey} \\
    \midrule
    \multicolumn{4}{l}{\textit{Category A: Bibliographic \& Contextual Information}} \\
    \hline
    D1 & Title & Full title of the study. & Overview \\
    D2 & Year & Publication year of the study. & Overview \\
    D3 & Venue & Publication venue (conference or journal). & Overview \\
    D4 & Summary & Main idea and core contribution of the study. & Overview \\
    \midrule
    \multicolumn{4}{l}{\textit{Category B: Research Content}} \\
    \hline
    D5 & Breaking Change Types & Types or categories of breaking changes studied or classified. & Section~\ref{sec:rq1} (RQ1) \\
    D6 & Reasons & Drivers behind the introduction of breaking changes. & Section~\ref{sec:rq2} (RQ2) \\
    D7 & Impacts & Effects of breaking changes on downstream consumers and software ecosystems. & Section~\ref{sec:rq2} (RQ2) \\
    D8 & Detection Techniques & Methods or tools used to identify breaking changes. & Section~\ref{sec:rq3} (RQ3) \\
    D9 & Prevention Strategies & Development practices or policies to avoid introducing breaking changes. & Section~\ref{sec:rq4} (RQ4) \\
    D10 & Mitigation Approaches & Strategies, tools, or frameworks to handle or recover from breaking changes. & Section~\ref{sec:rq4} (RQ4) \\
    \bottomrule
  \end{tabularx}
\end{table}

For the detailed data analysis, we use qualitative methods to analyze the content for each research question. We validate the results in two ways. First, for the data extraction and coding, the first two authors independently extract and code each study and then cross-check their results. Second, for the synthesized taxonomy and themes, the two authors refine them through repeated iterations, as detailed below. In both steps, all disagreements are resolved through discussion until consensus is reached.

\begin{itemize}
    \item Taxonomy Construction (RQ1): We use an open card sorting approach~\cite{Spencer2009-cardsorting} to construct the taxonomy. We conduct this process in three rounds:
    \begin{enumerate}
        \item \textit{Extraction (Round 1):} We scan the selected papers to extract all terms and descriptions related to breaking changes. Each unique term is recorded as an individual card with its original context.
        \item \textit{Clustering (Round 2):} We use a constant comparative method to analyze the cards. We group similar terms together to resolve synonyms and inconsistencies. This produces a set of normalized change types.
        \item \textit{Hierarchical Organization (Round 3):} We aggregate the normalized types into higher-level categories based on shared characteristics. We refine these categories using both top-down and bottom-up approaches until the taxonomy is stable. See Section~\ref{sec:rq1} for the detailed structure.
    \end{enumerate}

    \item Thematic Analysis (RQ2--RQ4): For reasons, impacts, detection methods, and communication, prevention, and recovery strategies, we follow the thematic analysis guidelines of Cruzes and Dyba~\cite{cruzes2011recommended}. This analysis has three phases:
    \begin{enumerate}
        \item \textit{Initial Coding:} We perform line-by-line coding on the extracted text segments (D6--D10). For example, text describing ``security fixes'' or ``refactoring'' as reasons for breaking changes is assigned descriptive labels.
        \item \textit{Theme Generation:} We examine relationships between codes to find recurring patterns. Codes with similar concepts are grouped into candidate themes. For example, ``dependency conflict'' and ``version mismatch'' are grouped under ``Ecosystem Complexity''.
        \item \textit{Synthesis and Structuring:} We refine each theme to ensure the themes are distinct and representative. We then organize these themes into comparison tables and narrative summaries. The summaries support cross-study comparison of detection techniques (RQ3) and communication, prevention, and recovery strategies (RQ4).
    \end{enumerate}
\end{itemize}

\section{RQ1: Taxonomy of Breaking Changes}
\label{sec:rq1}
\begin{figure}[htbp]
    \centering
    \includegraphics[width=1\textwidth]{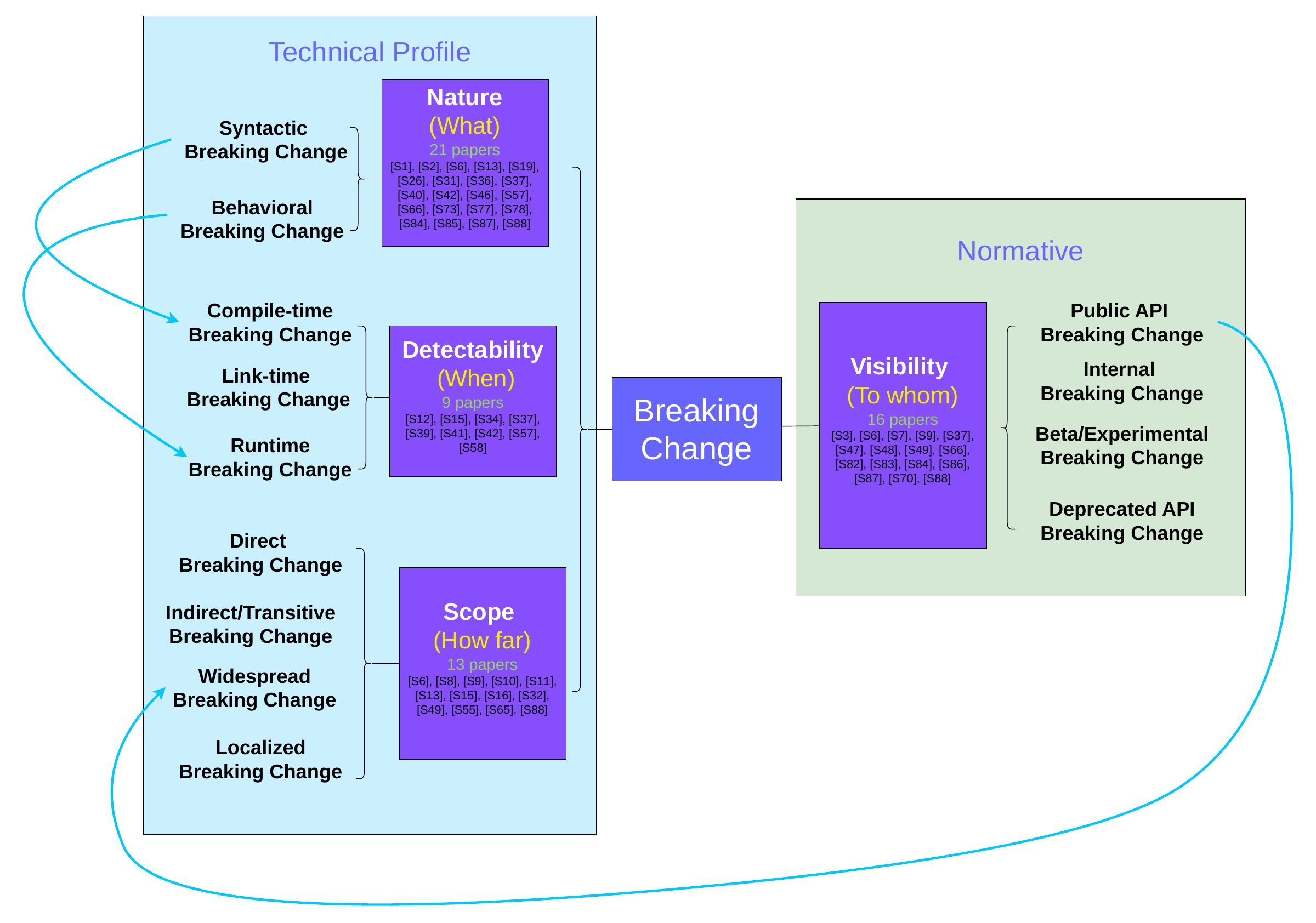}
    \caption{Breaking Change Taxonomy.
    The taxonomy is grouped into a technical profile (Nature, Detectability, Scope) and a normative dimension (Visibility).
    Blue arrows indicate cross-dimension interactions:
    (1) Nature $\rightarrow$ Detectability: syntactic changes in statically-typed ecosystems are typically caught at compile time, while behavioral changes surface only at runtime.
    (2) Visibility $\rightarrow$ Scope: changes to widely-depended-upon public APIs are almost always widespread regardless of their nature.}
    \label{fig:taxonomy1}
\end{figure}
In this section, we present the types of breaking changes derived from the selected literature. Figure~\ref{fig:taxonomy1} shows the four dimensions of the taxonomy: Nature (\textit{what kind} of breaking change is introduced), Detectability (\textit{when} the failure surfaces), Scope (\textit{how far} the impact propagates), and Visibility (\textit{to whom} the affected API element is exposed).

These four dimensions emerge from the three-round open card sorting process described in Section~\ref{sec:methodology}. The first three dimensions (Nature, Detectability, and Scope) are technical, which capture the technical profile of a breaking change. Visibility is normative that classifies the access contract of the affected element. The four dimensions are complementary, and each answers a distinct question about a breaking change that the others do not address. The four dimensions and their subcategories are detailed below:

\begin{enumerate}[label=\textbf{(\arabic*)}, leftmargin=*, itemsep=4pt]
    \item \textbf{Nature (What)}: \textit{What kind of breaking change is introduced?} This dimension is the primary focus of the literature. 21 out of 43 taxonomy-related papers address it. It distinguishes Syntactic breaking changes (violating type signatures, entity visibility, or interface definitions) from Behavioral breaking changes (preserving the syntactic interface but altering runtime semantics). We discuss this further in Section~\ref{sec:rq1-nature}.

    \item \textbf{Detectability (When)}: \textit{When does the failure surface?} This dimension captures the phase of the development lifecycle at which a breaking change becomes observable. The phases are Compile-time (immediate errors), Link-time (binary compatibility failures), and Runtime (behavioral failures during execution). We discuss this further in Section~\ref{sec:rq1-detectability}.

    \item \textbf{Scope (How far)}: \textit{How far does the impact propagate?} This dimension classifies breaking changes by how widely their impact spreads downstream. It distinguishes Direct vs. Transitive propagation, and Localized vs. Widespread effects. We discuss this further in Section~\ref{sec:rq1-scope}.

    \item \textbf{Visibility (To whom)}: \textit{To whom is the affected API exposed?} This normative dimension classifies changes by the declared access level of the affected element. The categories are Public, Internal, Beta/Experimental, and Deprecated APIs. Each carries different compatibility obligations. We discuss this further in Section~\ref{sec:rq1-visibility}.
\end{enumerate}

For each subcategory within each dimension, we report the number of papers in our selected literature that examine it (e.g., ``18 of the 21 Nature-related papers''). This count reflects the level of attention the subcategory has received in the surveyed studies. A single paper may be counted under multiple dimensions when it examines breaking changes from more than one perspective, so the per-dimension counts in Figure~\ref{fig:taxonomy1} sum to more than the 43 taxonomy-related papers.

We also consider candidate dimensions such as API granularity and ecosystem type during taxonomy construction. However, these either reduce to sub-categories within the four dimensions (e.g., method-level vs. class-level granularity falls under Nature) or represent properties of the study setting rather than of the change itself. The dimensions do interact in practice, as shown by the arrows in Figure~\ref{fig:taxonomy1}. Syntactic changes in statically-typed ecosystems are typically caught at compile time. Behavioral changes that preserve the syntactic interface are detectable only at runtime (Nature $\rightarrow$ Detectability). Similarly, a change to a widely-depended-upon public API is almost always widespread, regardless of its nature (Visibility $\rightarrow$ Scope).

\subsection{Nature of Breaking Changes (What)}
\label{sec:rq1-nature}

Section~\ref{sec:rq1} introduces the four dimensions of our taxonomy. In this section, we describe the Nature dimension, which captures what kind of breaking change is introduced. Across the selected papers, two main categories are identified:

\begin{itemize}
    \item \textbf{Syntactic breaking changes} alter the structural contract of an API. They violate type signatures, entity visibility, or interface definitions, and result in failures at compilation, linking, or static analysis.
    \item \textbf{Behavioral breaking changes} preserve the syntactic interface but alter runtime semantics. Examples include changes to exception behavior, return values, or state management. These changes evade static detection and appear only during execution.
\end{itemize}

The distinction between syntactic and behavioral changes is well-established in the literature. However, the surveyed studies are not evenly distributed across the two categories. Of the 21 Nature-related papers, 18 examine syntactic breaking changes and 9 examine behavioral ones. Most papers focus primarily on the syntactic side. The skew in research attention does not reflect the underlying prevalence. In npm, Kong et al.\ categorize 1,519 breaking changes by type. Behavioral changes account for the largest share (1,034 of 1,519, 68.1\%), followed by Remove (211, 13.9\%), Change Signature (134, 8.8\%), Rename (111, 7.3\%), Move (27, 1.8\%), and Inline (2, 0.1\%) [S1]. The skew reflects measurement difficulty. Syntactic changes can be enumerated through bytecode or AST diffing. Behavioral changes require manual inspection or test-based oracles. Table~\ref{tab:syntactic-behavioral-unified} provides a detailed breakdown of both syntactic and behavioral breaking change types across the five ecosystems studied. The following subsections describe how each category manifests across ecosystems.

\subsubsection{Syntactic Breaking Changes}

Syntactic breaking changes are examined in 18 of the 21 Nature-related papers. The specific forms they take vary across ecosystems. The variation reflects differences in language design, package management conventions, and runtime behavior. Common patterns emerge around API element removal, interface modifications, and visibility restrictions. We highlight three key cross-ecosystem observations below.

(1) Across all ecosystems, \textbf{API element removal} is consistently among the most prevalent forms of syntactic breaking change. The studies converge on removal-related changes (deletion of methods, fields, classes, paths, or packages) as a dominant or top-ranking cause of client-impacting incompatibilities. The exact share varies by ecosystem and by what is measured [S1, S2, S31, S74, S95].

(2) Beyond removal, \textbf{modifications to API contracts} form a second recurring family of syntactic breaking changes across code-level ecosystems. The form they take differs across type systems. In statically-typed ecosystems such as Maven/Java, the literature decomposes these changes along type-system boundaries: type-level, field-level, and method-level signature modifications (return type, parameter type, visibility) [S31, S57]. In dynamically-typed ecosystems such as Python and JavaScript, modifications more commonly take the form of parameter-convention changes (default values, keyword arguments, ordering) and asynchronous-pattern evolution. These languages lack the type-signature surface that statically-typed taxonomies key on [S1, S23, S74].

(3) \textbf{Web APIs and Linux distributions} present further variation. Breaking changes often involve structural or metadata-level evolution rather than code-level modifications. Web APIs experience changes at the endpoint, request, and response levels. Linux distributions face shared resource conflicts and package metadata evolution as packages are renamed, merged, or split during distribution maintenance [S25, S40, S46, S95].

\subsubsection{Behavioral Breaking Changes}

Behavioral breaking changes are studied in 9 of the 21 Nature-related papers. They preserve the syntactic interface but alter runtime semantics. As a result, they bypass type systems and static analysis and surface only when client code executes the affected path. The specific patterns vary across ecosystems. Recurring categories emerge around return-value or response-content changes, exception or error-handling changes, and side-effect or state changes. We highlight three key cross-ecosystem observations below.

(1) Across all ecosystems, \textbf{return-value or response-content changes} are the most prevalent behavioral category. Studies converge on changes to what an API returns (or to the content of API responses) as the most frequent behavioral pattern. The syntactic interface remains unchanged [S1, S42, S93].

(2) \textbf{Detection of behavioral breaks} is resistant to the static safety nets that catch syntactic breaks. The ecosystem may be statically- or dynamically-typed. In either case, behavioral breaks emit no error from the compiler, linker, or runtime exception path on their own. They manifest only when client code asserts the affected return value, observes the side-effect, or relies on the prior exception contract. Detection therefore depends on test coverage and cross-version regression testing, which most projects lack [S12, S19, S42].

(3) \textbf{Web APIs and Linux distributions} present behavioral breaks at the system-interaction level rather than within an API's local return semantics. Web APIs experience alterations to authentication, rate-limiting, and the documented pre- and post-conditions of endpoints. Linux distributions face configuration and data-format incompatibilities and runtime interaction failures between packages sharing system-wide state [S40, S46, S81, S93].

{\scriptsize
\begin{landscape}
\begin{longtable}{|p{1.2cm}|p{1.3cm}|p{2.8cm}|p{12.0cm}|}
\caption{Syntactic and Behavioral breaking change categories across ecosystems (npm: Kong et al. [S1]; Maven syntactic: Jayasuriya et al. [S2], Raemaekers et al. [S31], Dietrich et al. [S57], Reyes et al. [S69], and Lam et al. [S73]; Maven behavioral: Mostafa et al. [S42], Sharma and Lam [S19]; Python: Zhang et al. [S74], Du and Ma [S23], Montandon et al. [S26]; Web API: Ko\c{c}i et al. [S40], Schmiedmayer et al. [S25], Espinha et al. [S81], Godefroid et al. [S93], and Serbout and Pautasso [S95]; Linux: Artho et al. [S46]).}
\label{tab:syntactic-behavioral-unified}\\
\hline
\textbf{Eco-system} & \textbf{Type} & \textbf{Category} & \textbf{Description \& Example} \\ \hline
\endfirsthead
\hline
\multicolumn{4}{c}{\tablename\ \thetable{} -- continued from previous page} \\
\textbf{Eco-system} & \textbf{Type} & \textbf{Category} & \textbf{Description \& Example} \\ \hline
\endhead
\hline
\endfoot
\hline
\endlastfoot

\multirow{2}{1.2cm}{\textbf{npm}} &
\textbf{Syntactic} &
\textbf{Change of Entity Availability \& Change of Interface Contract} &
\textit{Explanation:} Removal or relocation of public API elements (removing exports, moving elements between modules, renaming identifiers), and modifications in method signatures or type definitions (signature changes, asynchronous pattern evolution, configuration object property changes). \newline
\textit{Example:} \texttt{rxjs} relocates types from \texttt{rxjs/interfaces} to the main module, breaking sub-module imports; \texttt{mongoose} changes from callback-style \texttt{createIndexes(options, callback)} to Promise-based \texttt{async createIndexes(options)}. [S1] \\ \cline{2-4}

&
\textbf{Behavioral} &
\textbf{Change of Logic for Option Values \& Change of Default State Initialization} &
\textit{Explanation:} Modifications to the internal logic that interprets configuration objects (altering how option values are processed without changing the syntactic signature), and modifications to default values or logic used when function parameters are omitted or left undefined. \newline
\textit{Example:} Removing support for the \texttt{netrc} authentication mechanism from an \texttt{auth} configuration key causes silent logical failures at runtime; in the npm CLI project, the default value of \texttt{depthToPrint} changes from \texttt{undefined} to zero. [S1] \\ \hline

\multirow{8}{1.2cm}{\textbf{Maven}} &
\multirow{6}{1.3cm}{\textbf{Syntactic}} &
\multirow{2}{2.8cm}{\textbf{Type-level Changes}} &
\textit{Explanation:} Removal or modification of classes and interfaces, including visibility changes and annotation modifications. \newline
\textit{Example:} Removing \texttt{org.apache.commons.lang.StringUtils} causes \texttt{ClassNotFoundException} for all clients. [S2, S31, S57, S69, S73] \\ \cline{3-4}

& &
\multirow{2}{2.8cm}{\textbf{Field-level Changes}} &
\textit{Explanation:} Removal or modification of fields, including type changes, visibility restrictions, and constant value changes. \newline
\textit{Example:} Removing \texttt{public static final String VERSION} causes \texttt{NoSuchFieldError} at runtime. [S2, S31, S57, S69, S73] \\ \cline{3-4}

& &
\multirow{2}{2.8cm}{\textbf{Method-level Changes}} &
\textit{Explanation:} Modifications to method signatures, including removal, return type changes, parameter changes, and interface method additions. \newline
\textit{Example:} Changing \texttt{void process(String)} to \texttt{String process(String)} alters the JVM descriptor, causing \texttt{NoSuchMethodError}. [S2, S31, S57, S69, S73] \\ \cline{2-4}

&
\multirow{6}{1.3cm}{\textbf{Behavioral}} &
\multirow{2}{2.8cm}{\textbf{Return Value Change}} &
\textit{Explanation:} A method preserves its signature but returns a different value or structure. This is the most prevalent behavioral category (162 out of 296 BBIs, 54.7\%). \newline
\textit{Example:} In Jsoup, changing the escaping behaviour of \texttt{doc.body().html()} causes client tests asserting specific HTML output to fail. [S42] \\ \cline{3-4}

& &
\multirow{2}{2.8cm}{\textbf{Exception Behavior Change}} &
\textit{Explanation:} A method begins throwing, stops throwing, or changes the type of exception it raises. This is the second most prevalent category (105 out of 296 BBIs, 35.5\%). \newline
\textit{Example:} A library method that previously returned \texttt{null} for missing keys now throws \texttt{NullPointerException}. [S42, S19] \\ \cline{3-4}

& &
\multirow{2}{2.8cm}{\textbf{Memory State / Side-effect Change}} &
\textit{Explanation:} Internal object state is modified differently, or a side effect is added or removed (29 out of 296 BBIs, 9.8\%). When combined with non-primitive return value changes, BBIs requiring non-trivial assertions to detect account for 105 out of 296 (36\%) and cause 54\% (60 out of 112) of real-world BBI-related bugs. \newline
\textit{Example:} In Android 4.4, a padding-update method must be called before displaying date information; omitting it causes UI clipping. [S42] \\ \hline

\multirow{8}{1.2cm}{\textbf{Python}} &
\multirow{6}{1.3cm}{\textbf{Syntactic}} &
\multirow{2}{2.8cm}{\textbf{API Element Removal \& Relocation}} &
\textit{Explanation:} Removal, renaming, or relocation of modules, classes, functions, methods, fields, and aliases. Removals account for 96.4\% of breaking changes [S74]. Method removal is the most frequent breaking change pattern (28.3\% of all API changes), followed by field removal (16.7\%) [S74]. Module and class removal are graded High severity [S23]. \newline
\textit{Example:} \texttt{tf.histogram\_summary} is relocated and renamed to \texttt{tf.summary.histogram} in TensorFlow. \texttt{pd.rolling\_mean} is relocated to \texttt{pd.DataFrame.rolling} in Pandas 0.18.0. [S74] \\ \cline{3-4}

& &
\multirow{2}{2.8cm}{\textbf{Parameter Changes}} &
\textit{Explanation:} Required and optional parameter addition and removal, parameter reordering, default value addition, removal, and change, and parameter type changes [S74]. Du and Ma further classify parameter changes into type, default value, and position sub-categories [S23]. These changes manifest as \texttt{TypeError} (31.4\% of crashing issues) [S74]. \newline
\textit{Example:} Django's removal of the \texttt{context\_instance} optional parameter in \texttt{render\_to\_response} causes a crash when the keyword argument is used. [S74] \\ \cline{3-4}

& &
\multirow{2}{2.8cm}{\textbf{Type \& Inheritance Changes}} &
\textit{Explanation:} Changes to class inheritance hierarchies, attribute types, and function return types [S23]. Due to Python's duck typing, type incompatibility changes are graded as medium-severity, because the errors occur only when the parameter or attribute is accessed or type-checked, rather than at function invocation [S23]. \newline
\textit{Example:} Removing a base class is graded High severity. Changing an attribute type is graded Medium because the error occurs only at access time. [S23] \\ \cline{2-4}

&
\multirow{2}{1.3cm}{\textbf{Behavioral}} &
\multirow{2}{2.8cm}{\textbf{Default Argument Breaking Change (DABC)}} &
\textit{Explanation:} Changes to default parameter values that alter runtime behavior without modifying function signatures or causing crashes. Zhang et al. [S74] document several instances of this pattern (e.g., default value changes in Django). Montandon et al. [S26] introduce the term DABC and identify 93 DABCs across Scikit-Learn (77), NumPy (5), and Pandas (11). Behavior-based DABCs account for 75\% in Scikit-Learn, 80\% in NumPy, and 45\% in Pandas [S26]. \newline
\textit{Example:} In Django 1.6.0, the \texttt{default} parameter in \texttt{BooleanField} changes from \texttt{False} to \texttt{None} [S74]. Scikit-Learn changes \texttt{gamma} in \texttt{SVC} from \texttt{"auto"} to \texttt{"scale"}, causing accuracy to change from 0.05 to 0.82 [S26]. \\ \hline

\multirow{8}{1.2cm}{\textbf{Web API}} &
\multirow{6}{1.3cm}{\textbf{Syntactic}} &
\multirow{2}{2.8cm}{\textbf{Endpoint Level Changes}} &
\textit{Explanation:} Removal, renaming, or replacement of API endpoints, as well as changes to handler identifiers and communication patterns (e.g., from request-response to streaming) [S40, S25]. Path removal without deprecation is the most widespread breaking change, affecting 1,211 out of 3,075 APIs [S95]. \newline
\textit{Example:} In the DHIS2 API, 5 endpoints are removed between versions 2.26 and 2.27 because they are unused or their functionality has been replaced in previous versions. [S40] \\ \cline{3-4}

& &
\multirow{2}{2.8cm}{\textbf{Request Level Changes}} &
\textit{Explanation:} Adding required parameters or request fields, removing or renaming parameters, changing parameter types, tightening constraints (e.g., changing from optional to mandatory), and modifying authority levels for existing endpoints [S40, S25]. \newline
\textit{Example:} In DHIS2 version 2.27, the authority required to POST to the \texttt{predictor/run} endpoint changes from \texttt{F\_PREDICTOR\_ADD} to \texttt{F\_PREDICTOR\_RUN}. [S40] \\ \cline{3-4}

& &
\multirow{2}{2.8cm}{\textbf{Response Level Changes}} &
\textit{Explanation:} Changes to response property types, removal of response properties or enum values, and changes to response field identifiers. The most frequent breaking change across 3,075 Web APIs is response property type change (23,048 occurrences across 714 APIs), followed by response enum value removal (21,210 occurrences across 319 APIs) [S95, S25]. \newline
\textit{Example:} Response required property removal accounts for 12,587 occurrences across 409 APIs. [S95] \\ \cline{2-4}

&
\multirow{2}{1.3cm}{\textbf{Behavioral}} &
\multirow{2}{2.8cm}{\textbf{Pre/Post Condition Change}} &
\textit{Explanation:} Changes to API response content, format, or semantics that preserve endpoint signatures but alter runtime behavior. These include changes to pre-conditions, post-conditions, and API response content [S40], as well as authentication and rate limiting policy changes [S81]. Godefroid et al. identify response regressions where unspecified properties appear in API responses [S93]. \newline
\textit{Example:} In Azure networking APIs, a new \texttt{ipTags} property begins appearing in responses before being documented in the specification, breaking clients that cannot parse the unknown field [S93]. \\ \hline

\multirow{8}{1.2cm}{\textbf{Linux}} &
\multirow{4}{1.3cm}{\textbf{Syntactic}} &
\multirow{2}{2.8cm}{\textbf{Shared Resource Conflict}} &
\textit{Explanation:} Unavailability or inaccessibility of shared system resources, including direct file-level conflicts, package installers modifying files used by other packages, file/directory name conflicts, and runtime library symbol clashes. \newline
\textit{Example:} When attempting to install \texttt{gcc-avr}, the Debian package manager detects that \texttt{/usr/lib64/libiberty.a} is already provided by \texttt{binutils}, causing an error: ``trying to overwrite '/usr/lib64/libiberty.a', which is also in package binutils.'' [S46] \\ \cline{3-4}

& &
\multirow{2}{2.8cm}{\textbf{Package Meta-data Evolution}} &
\textit{Explanation:} Incorrect or outdated meta-data during package architectural modifications (renaming, merging, splitting), including incorrect dependency declarations and spurious conflict declarations. \newline
\textit{Example:} During the transition from \texttt{ttf-telugu-fonts} to \texttt{fonts-telu}, the maintainer uses incorrect package relations, causing a conflict that prevents installation. [S46] \\ \cline{2-4}

&
\multirow{4}{1.3cm}{\textbf{Behavioral}} &
\multirow{2}{2.8cm}{\textbf{Configuration \& Data Conflict}} &
\textit{Explanation:} Incompatible changes to shared data, configuration information, or information flow between programs that bypass package manager validation. \newline
\textit{Example:} A change in the data format between versions of an application breaks other packages expecting the original format. [S46] \\ \cline{3-4}

& &
\multirow{2}{2.8cm}{\textbf{Runtime Interaction Failure}} &
\textit{Explanation:} Unexpected failures triggered by dynamic interaction between installed software packages, including file/resource conflicts and silent runtime resource clashes. \newline
\textit{Example:} Two packages providing different implementations of the same library function name, causing the dynamic linker to load the wrong version. [S46] \\
\hline
\end{longtable}
\end{landscape}
}

\begin{tcolorbox}[colback=yellow!10!white, colframe=yellow!75!black, title=\textbf{Key Findings: Nature of Breaking Changes (What)}]
\begin{enumerate}
        \item \textbf{Breaking changes can be grouped into syntactic and behavioral categories}: Syntactic changes violate the structural contract of an API, while behavioral changes preserve the interface but alter runtime semantics. Across the Nature-related papers, research attention concentrates on the syntactic side, which 18 of the 21 papers examine.
        \item \textbf{Ecosystem conventions outweigh language paradigm in shaping the breaking change profile}: Across the surveyed ecosystems, the relative prevalence of syntactic and behavioral changes tracks ecosystem norms and release practice more closely than the static vs.\ dynamic typing distinction.
\end{enumerate}
\end{tcolorbox}

\subsection{Detectability of Breaking Changes (When)}
\label{sec:rq1-detectability}

In this subsection, we examine the Detectability dimension, which classifies breaking changes by when the failure is first observed.

Not all ecosystems exhibit every detectability phase. Dynamically-typed languages such as JavaScript and Python lack a separate compilation step. Most breaking changes in these languages surface only at runtime (as discussed in Section~\ref{sec:rq1-nature}). The compile-time, link-time, and runtime distinction is most relevant to statically-typed, compiled ecosystems such as Java and C++. Among these, all 9 detectability-related primary studies focus on the Java/Maven ecosystem. The Java Language Specification (JLS) and the Java Virtual Machine Specification (JVMS) formally distinguish three phases, each governed by a distinct set of consistency rules [S57]. Therefore, this section describes the three detectability categories in the context of Java/Maven: compile-time, link-time, and runtime breaks. Table~\ref{tab:detectability-unified} summarizes the representative breaking change categories for each of the three phases. The table includes their description and example, and the supporting empirical evidence.

\paragraph{Compile-time Breaking Change.}
Compile-time breaking changes are examined empirically by 7 of the 9 Detectability-related papers. They are the easiest to catch. They show up immediately when client code is recompiled against a new library version. The compiler flags the problem directly, so developers know something has changed and where to look. As noted in Section~\ref{sec:rq1} (Nature $\rightarrow$ Detectability), compile-time breaks correspond closely to the Syntactic breaking changes described in Section~\ref{sec:rq1-nature}. Changes such as method removal, type renaming, and access restriction alter the structural contract of an API. They are caught by the compiler's type-checking and name-resolution rules. Even within this most-visible category, some source-level changes can silently alter program semantics or remain binary-incompatible despite passing recompilation. Developers' self-assessed knowledge of these rules is markedly weaker for binary compatibility than for source compatibility [S57].

\paragraph{Link-time Breaking Change.}
Link-time breaking changes, also termed binary breaking changes, are covered by the same 7 of the 9 Detectability-related papers that examine compile-time breaks, since Java/Maven detection tools such as \texttt{japicmp} typically check source and binary compatibility in tandem. They pass compilation without any error. The problem only surfaces when the JVM tries to load and link an un-recompiled client binary against the new library. The result is a \texttt{LinkageError} subclass such as \texttt{NoSuchMethodError} or \texttt{NoClassDefFoundError} [S34, S57]. Among the patterns surveyed, the stale inlined constant is the most subtle. The Java compiler inlines \texttt{public static final} values directly into client bytecode at compile time. Changing such a constant in the library leaves un-recompiled clients silently using the old value with no error [S34]. Listing~\ref{lst:link-inlined-const} illustrates this pattern.

\begin{lstlisting}[caption={Stale Inlined Constant (silent wrong value -- no error thrown)}, label=lst:link-inlined-const, escapeinside={(*@}{@*)}]
(*@\textcolor{blue!70!black}{// Library: constant value changed between releases}@*)
(*@\colorbox{diffremoved}{\makebox[\dimexpr\linewidth-2\fboxsep][l]{\color{removedtext}- public static final int MAX\_RETRIES = 3;}}@*)
(*@\colorbox{diffadded}{\makebox[\dimexpr\linewidth-2\fboxsep][l]{\color{addedtext}+ public static final int MAX\_RETRIES = 5;}}@*)

(*@\textcolor{commentcolor}{// Client compiled against OLD library -- NOT recompiled after upgrade:}@*)
  for (int i = 0; i < Config.MAX_RETRIES; i++) { ... }
(*@\textcolor{commentcolor}{// javac inlined the literal 3 directly into client bytecode at compile time}@*)
(*@\textcolor{commentcolor}{// Client bytecode contains: for (int i = 0; i < 3; i++) -- hardcoded!}@*)

(*@\textcolor{notecolor}{// No NoSuchFieldError. No crash. Client silently uses stale value 3, not 5.}@*)
(*@\textcolor{red!80!black}{// Silent wrong behaviour: retry loop runs 3 times instead of intended 5}@*)
\end{lstlisting}

\paragraph{Runtime Breaking Change.}
Runtime breaking changes are reported empirically by only 4 of the 9 Detectability-related papers. They are the last to surface in the software lifecycle. Unlike compile-time and link-time breaks, they produce no error at compilation or class loading. The code runs, but behaves incorrectly. They require test coverage or dynamic analysis to surface, and can manifest as silent logic drift rather than explicit crashes. Runtime breaks correspond directly to the Behavioral category in the Nature dimension. They recur across ecosystems as exception-behavior, return-value, or state and side-effect changes. Most such changes are introduced during non-major releases that semantic versioning would otherwise deem safe [S12, S42].

{\scriptsize
\begin{longtable}{|p{1.2cm}|p{3.2cm}|p{5.9cm}|p{2.5cm}|}
\caption{Breaking change categories by detectability phase in the Java/Maven ecosystem (Dietrich et al. [S57, S34]; Mostafa et al. [S42]; Jayasuriya et al. [S12]).}
\label{tab:detectability-unified}\\
\hline
\textbf{Phase} & \textbf{Category} & \textbf{Description \& Example} & \textbf{Evidence} \\ \hline
\endfirsthead
\hline
\multicolumn{4}{c}{\tablename\ \thetable{} -- continued from previous page} \\
\textbf{Phase} & \textbf{Category} & \textbf{Description \& Example} & \textbf{Evidence} \\ \hline
\endhead
\hline
\endfoot
\hline
\endlastfoot

\textbf{Compile-time} &
\textbf{Method Added to Interface}; \textbf{Generalise Parameter Type}; \textbf{Source/Binary Asymmetry} &
\textbf{Method Added to Interface}: forces all implementing classes to provide an implementation; the compiler rejects the client until the method is added (e.g., adding \texttt{void validate()} to \texttt{interface Processor}). \newline
\textbf{Generalise Parameter Type}: widening a parameter type (e.g., \texttt{int} to \texttt{float}) can silently alter numeric precision (e.g., \texttt{isEven(int i)} to \texttt{isEven(float i)} rounds \texttt{Integer.MAX\_VALUE}). \newline
\textbf{Source/Binary Asymmetry}: certain changes pass recompilation but break the un-recompiled binary, or vice versa (e.g., \texttt{Collection getColl()} to \texttt{List getColl()} compiles but throws \texttt{NoSuchMethodError} on the old binary). &
76\% aggregate developer correctness rate on source compatibility questions; only 30\% predict parameter widening correctly; only 27\% identify return type specialization as binary incompatible [S57] \\ \hline

\textbf{Link-time} &
\textbf{Method Descriptor Change}; \textbf{Class / Package Removal}; \textbf{Stale Inlined Constant} &
\textbf{Method Descriptor Change}: specialising a return type or generalising a parameter type changes the JVM method descriptor; the linker throws \texttt{NoSuchMethodError} (e.g., \texttt{Collection getColl()} to \texttt{List getColl()}). \newline
\textbf{Class / Package Removal}: removing or renaming a class or package invalidates all bytecode references; the JVM throws \texttt{NoClassDefFoundError} at class-loading time (e.g., renaming \texttt{org.example.utils.StringHelper}). \newline
\textbf{Stale Inlined Constant}: the Java compiler inlines \texttt{public static final} constants into client bytecode; changing the constant value leaves un-recompiled clients silently using the old value (e.g., \texttt{MAX\_RETRIES = 3} to \texttt{5}). &
171 method removals (M1) and 145 incompatible method changes (M2) [S34]; 54,700 SO references for \texttt{NoClassDefFoundError} [S57]; 252 stale-constant occurrences after filtering parser-generated constants in the Qualitas Corpus [S34] \\ \hline

\textbf{Runtime} &
\textbf{Exception Behaviour Change}; \textbf{Return Value / Assertion Change}; \textbf{Memory State / Side-effect Change} &
\textbf{Exception Behaviour Change}: an API method throws a new exception, a different exception, or stops throwing one it previously raised (e.g., a method that previously returned \texttt{null} for missing keys now throws \texttt{NullPointerException}). \newline
\textbf{Return Value / Assertion Change}: the return value or its structure changes while the method signature remains identical (e.g., in Jsoup, changing the escaping behaviour of \texttt{doc.body().html()} causes client assertions to fail). \newline
\textbf{Memory State / Side-effect Change}: internal object state is modified differently, or a side effect is added or removed (e.g., in Android 4.4, a padding-update method must be called before displaying date information; omitting it causes UI clipping). &
Exception and crash changes account for 105 of 296 BBIs (35.5\%); return value changes account for 162 of 296 BBIs (54.7\%); the 105 BBIs (36\%) involving memory state or side effects cause 60 of 112 (54\%) real-world BBI-related bugs [S42] \\
\hline
\end{longtable}
}

\begin{tcolorbox}[colback=yellow!10!white, colframe=yellow!75!black, title=\textbf{Key Findings: Detectability of Breaking Changes (When)}]
\begin{enumerate}
    \item \textbf{Failures surface across three phases of the development lifecycle}: The surveyed literature distinguishes compile-time, link-time, and runtime breaks (see Table~\ref{tab:detectability-unified}), each governed by a different consistency check and observed at a different stage.
    \item \textbf{Surviving compile-time recompilation does not imply binary compatibility}: Across the surveyed Java/Maven studies, source-binary asymmetry recurrently produces changes that pass one check but fail the other (e.g., return type specialization, stale inlined constants), and developer surveys report markedly lower self-assessed correctness for binary compatibility than for source compatibility.
    \item \textbf{Runtime breaks are the only category with no automatic safety net}: Across the surveyed studies, runtime breaks emit no compiler, linker, or runtime exception on their own, and they require the test coverage or dynamic analysis that most downstream projects lack.
\end{enumerate}
\end{tcolorbox}

\subsection{Scope of Breaking Changes (How Far)}
\label{sec:rq1-scope}

In this section, we examine the Scope dimension, which classifies breaking changes by the breadth of their impact across the dependency graph. We distinguish four categories: \textbf{direct}, \textbf{transitive}, \textbf{widespread}, and \textbf{localized} breaking changes. Table~\ref{tab:scope-unified} summarizes the characteristics for each category.

\subsubsection{Direct and Transitive Breaking Changes}

Among the four scope categories, direct and transitive breaking changes describe the propagation path from library to the downstream consumers. 8 of the 13 Scope-related papers focus on this propagation-path axis. A direct breaking change affects downstream consumers that explicitly declare the modified library as a dependency. The breaking change is traceable to a single dependency edge in the downstream consumer's build manifest. This makes it the most straightforward category to diagnose and resolve.

A transitive (indirect) breaking change propagates through one or more intermediate libraries before reaching the downstream consumers. This creates a multi-hop impact chain that is harder to diagnose and resolve. Many consumers directly invoke functionality from transitive dependencies without declaring them explicitly. This makes the downstream consumer vulnerable to changes in libraries it does not nominally depend on [S11]. In ecosystems with permissive version ranges such as npm, provider changes can also manifest as breaking changes long after their introduction. The downstream dependency tree evolves through version range resolution, which surfaces the latent break [S9].

\subsubsection{Widespread and Localized Breaking Changes}

Widespread and localized breaking changes are examined by the remaining 5 of the 13 Scope-related papers. A widespread breaking change affects a high-popularity API used across a large fraction of the ecosystem. It requires coordinated migration. A localized breaking change affects a rarely-used API. The impact is confined to a small subset of consumers.

In practice, most breaking changes are localized, since most publicly defined methods in Maven are never invoked by other libraries [S6, S16]. However, this localized impact is not the result of maintainers actively protecting popular APIs. Popular methods break at roughly the same rate as unpopular ones. One likely explanation is that maintainers are not always aware of which of their methods are widely used [S16].

\begin{table}[h]
\centering
\caption{Breaking change scope categories in the Maven ecosystem (Jayasuriya et al. [S11]; Keshani et al. [S16]; Ochoa et al. [S6]).}
\label{tab:scope-unified}
\footnotesize
\begin{tabular}{|p{1.3cm}|p{4.8cm}|p{6.9cm}|}
\hline
\textbf{Category} & \textbf{Characteristics} & \textbf{Key Empirical Evidence} \\
\hline
Direct &
Affects clients that explicitly declare the modified library as a dependency. &
11.58\% of dependency updates cause client-impacting breaking changes (latest version); 4.35\% for adjacent version. The most common causes are method return type change, package deletion, and class deletion [S11]. \\
\hline
Transitive &
Propagates through one or more intermediate libraries; harder to diagnose and resolve. &
Transitive changes account for approximately one-fifth (20.36\%) of all client-impacting source breaking changes, the most common reason [S11]. Over 60\% of clients (61.24\%) use transitive dependency functionality without declaring it [S11]. \\

\hline
Widespread &
Affects high-popularity APIs used by a large fraction of the ecosystem. Requires coordinated migration. &
67\% of Maven artifacts introduce at least one semantic versioning violation (either a breaking change or an illegal API extension) in their history [S16]. Top library versions (\texttt{commons-io} 2.4, \texttt{slf4j-api} 1.7.21) have over 30,000 clients each [S6]. \\
\hline
Localized &
Affects edge-case or rarely-used APIs. Impact is limited to a small subset of clients. &
On average, 87\% of publicly defined Maven methods are never used by other libraries [S16]. Only 7.9\% of all clients are affected by breaking changes [S6]. \\
\hline
\end{tabular}
\end{table}

\begin{tcolorbox}[colback=yellow!10!white, colframe=yellow!75!black, title=\textbf{Key Findings: Scope of Breaking Changes (How Far)}]
\begin{enumerate}
    \item \textbf{Scope is classified along two complementary axes}: The surveyed literature distinguishes direct vs.\ transitive propagation and widespread vs.\ localized impact (see Table~\ref{tab:scope-unified}), with most studies focusing on the propagation-path axis.
    \item \textbf{Most breaking changes are localized rather than widespread}: Across the surveyed studies, the majority of publicly defined library methods are never invoked by other libraries, so the typical breaking change reaches only a small subset of consumers even when it concerns a public-facing API.
    \item \textbf{API popularity does not predict the likelihood of being broken}: Multiple Maven-ecosystem studies report that popular and rarely-used methods break at similar rates; one explanation that recurs across these studies is that maintainers cannot easily tell which of their methods are widely used.
\end{enumerate}
\end{tcolorbox}

\subsection{Visibility of Breaking Changes (To Whom)}
\label{sec:rq1-visibility}

In this section, we examine the Visibility dimension, which classifies breaking changes according to the stability contract of the modified API element. We distinguish four categories: \textbf{public API}, \textbf{internal API}, \textbf{beta/experimental API}, and \textbf{deprecated API} breaking changes. Table~\ref{tab:visibility-unified} summarizes these categories.

\paragraph{Public API Breaking Changes.}
Public API breaking changes are examined by 9 of the 16 Visibility-related papers. They affect elements that are part of the library's stable, documented contract, typically \texttt{public} or \texttt{protected} members in non-\texttt{internal} packages. Such changes violate the API contract and require a major version bump under SemVer. In practice, this obligation is frequently not honored. Non-major Maven releases still routinely introduce public API breaks. Broken clients form one of the most common categories of breaking-change-related questions on developer Q\&A platforms [S6, S8].

\paragraph{Internal API Breaking Changes.}
Internal API breaking changes are examined by only 3 of the 16 Visibility-related papers. They affect elements marked as implementation details (e.g., \texttt{@Internal} annotations, \texttt{internal} package naming). These changes are not contract violations. Nonetheless, clients still depend on internal APIs in practice. Their breakage rate on update is higher than the overall average for Maven dependency updates [S11]. The boundary between internal and public APIs is not always clear-cut. Many methods classified as non-API are stable enough in practice to be treated as part of the de facto contract. Mature codebases identify large numbers of such methods as candidates for promotion [S66].

\paragraph{Beta/Experimental API Breaking Changes.}
Beta/experimental API breaking changes are reported empirically by 3 of the 16 Visibility-related papers. They affect elements marked as unstable (e.g., \texttt{@Beta}, \texttt{@Exper\allowbreak{} imental}). Clients are warned that changes may occur without a major version bump. These annotations are used widely across mature library ecosystems. However, their practical value is limited by two factors. First, there is no standardization of unstable API mechanisms such as \texttt{@Beta} and \texttt{@Internal}. This makes it difficult for clients to determine which declarations should be considered stable [S6]. Second, existing Java compatibility tools have known shortcomings in detecting certain API incompatibilities. Even tool-assisted clients may miss breaking changes in beta APIs [S64].

\paragraph{Deprecated API Breaking Changes.}
Deprecated API breaking changes are examined by 9 of the 16 Visibility-related papers, comparable to the share that examines public API breaks. They affect elements formally marked for removal via \texttt{@Deprecated}. This formal marking provides clients with advance notice and a migration path. In practice, client adaptation is slow. Method replacements (where the old method is unconditionally superseded) tend to provoke faster client adaptation than method improvements (where the old method remains valid). The latter offers no functional pressure to migrate [S48].

{\footnotesize
\begin{longtable}{|p{1.8cm}|p{2.5cm}|p{3cm}|p{5.7cm}|}
\caption{Unified summary of breaking change visibility categories (Brito et al. [S8]; Ochoa et al. [S6]; Jayasuriya et al. [S11]; Businge et al. [S66]; Jezek and Dietrich [S64]; Hora et al. [S48]).}
\label{tab:visibility-unified}\\
\hline
\textbf{Category} & \textbf{SemVer Obligation} & \textbf{Identification Mechanism} & \textbf{Key Empirical Evidence} \\ \hline
\endfirsthead
\hline
\multicolumn{4}{c}{\tablename\ \thetable{} -- continued from previous page} \\
\textbf{Category} & \textbf{SemVer Obligation} & \textbf{Identification Mechanism} & \textbf{Key Empirical Evidence} \\ \hline
\endhead
\hline
\endfoot
\hline
\endlastfoot

\textbf{Public API} &
Major bump required. Unambiguous contract violation. &
\texttt{public} / \texttt{protected} access in non-\texttt{internal} packages. &
Approximately one in five non-major Maven releases introduce public API breaks [S6]; 45\% of SO breaking change posts are clients seeking fixes [S8]. \\ \hline

\textbf{Internal API} &
No obligation. Not a contract violation, but clients face elevated risk. &
\texttt{@Internal}, \texttt{@InternalApi}, \texttt{internal} package naming. &
Approximately 4.8\% of Maven clients use internal APIs; their breakage rate on update is 20.6\% compared to the overall average of 11.58\% [S11]. Eclipse has 327K stable non-API methods as candidates for promotion [S66]. \\ \hline

\textbf{Beta / Experimental} &
No obligation. Changes are expected and explicitly warned. &
\texttt{@Beta}, \texttt{@Experimental}, \texttt{experimental} package naming. &
\texttt{@Beta} appears 1,451 times in the top 1,000 Maven libraries [S6]. All nine evaluated Java compatibility tools have shortcomings in detecting certain API incompatibilities [S64]. \\ \hline

\textbf{Deprecated} &
No obligation for removal after deprecation. Considered a ``graceful'' break. &
\texttt{@Deprecated} annotation; removal in a subsequent release. &
In Pharo, 53\% of API changes (62 out of 118) caused ecosystem reactions; median reaction time 34 days; method replacements elicit faster adaptation (median 20 days) than method improvements (median 47 days) [S48]. \\
\hline
\end{longtable}
}

\begin{tcolorbox}[breakable, colback=yellow!10!white, colframe=yellow!75!black, title=\textbf{Key Findings: Visibility of Breaking Changes (To Whom)}]
\begin{enumerate}
    \item \textbf{Visibility classifies breaking changes by the access contract of the affected element}: The surveyed literature distinguishes public, internal, beta/experimental, and deprecated APIs (see Table~\ref{tab:visibility-unified}), each carrying a different stability obligation.
    \item \textbf{Public APIs are routinely broken in non-major releases despite SemVer}: Across the surveyed Maven studies, non-major releases regularly introduce changes that violate the public-API stability contract; broken clients are also one of the most common categories of breaking-change-related questions on developer Q\&A platforms.
    \item \textbf{Internal APIs are widely depended upon despite explicit ``do not use'' markers}: Multiple Maven-ecosystem studies show that clients couple to internally marked APIs in practice and that these internal APIs break at a higher rate than the overall average, so the ``internal'' label does not deter downstream coupling.
\end{enumerate}
\end{tcolorbox}

\section{RQ2: Reasons and Impact of Breaking Changes}
\label{sec:rq2}
In this section, we examine what drives breaking changes and how they affect downstream consumers (items D6--D7 in Table~\ref{tab:data_items}). Across the 97 primary studies, reasons are addressed in around 20 papers. The section is organized in two parts. First, we examine the \textbf{reasons} behind breaking changes, grouped by who initiates the change. Second, we examine their downstream \textbf{impact}. The impact part covers the scale of breakage, how version signals and transitive dependencies shape it, and what diagnostic information is available when things go wrong.

\subsection{Reasons behind Breaking Changes}
Across Maven, npm, Python, Web API, and Linux distributions, the literature consistently shows that maintenance and design-related reasons are the most frequent drivers of breaking changes. They collectively exceed feature delivery in the ecosystems where detailed categorizations are available [S8, S1].

We organize the five reason categories into three groups by who initiates the change. The five categories are ordered so that each group spans a contiguous range of numbers. \textbf{(1) Internal maintenance and refactoring} represents deliberate choices to reduce technical debt. \textbf{(2) Defensive design and robustness} reflects an explicit intent to improve API contracts. \textbf{(3) Functional evolution} covers adding features and improving performance. \textbf{(4) Ecosystem integration and constraints} represents breaking changes imposed by upstream dependency evolution or shared-resource constraints. \textbf{(5) Reactive and accidental modifications} are changes the maintainer initiates internally but does not anticipate will break clients. Categories (1), (2), and (3) form \textbf{library-internal decisions}. The maintainer deliberately initiates these changes. Category (4) is the only category in \textbf{externally imposed changes}. The change originates outside the library. Category (5) covers \textbf{unintended consequences}. The breaking effect on clients is not anticipated.

Across the ecosystems studied, maintenance and design-related reasons (categories 1 and 2) consistently account for the largest share of breaking changes. Two ecosystem-level studies provide quantitative breakdowns of maintainer-stated reasons. Brito et al.\ classify Java breaking changes into four reason categories [S8]. Kong et al.\ classify npm breaking changes into three reason categories [S1]. After mapping their categories to our taxonomy, categories (1) and (2) jointly dominate in both ecosystems, exceeding category (3) functional evolution. In Java, API simplification (17 of 59, 29\%) and maintainability improvement (14 of 59, 24\%) together account for 53\% of breaking changes, exceeding new feature implementation (19 of 59, 32\%) [S8]. In npm, improving API design alone accounts for 939 of 1,519 cases (61.8\%). Reducing code redundancy (469, 30.9\%) and improving identifier names (111, 7.3\%) cover the rest [S1]. Each of these percentages derives from a single primary study with the sample size shown in parentheses (59 Java changes in [S8] and 1{,}519 npm changes in [S1]), and the surveyed literature does not yet provide independent replications of these proportions for either ecosystem. Category (5) reactive and accidental modifications represents the smallest identified share. Table~\ref{tab:reason-summary} summarizes each reason category. The columns include initiator group, description, ecosystem coverage, per-ecosystem frequency, and intended benefit, together with the mapping between each surveyed study's labels and our taxonomy.

\paragraph{Internal Maintenance and Refactoring}

The specific operations in this category include removing obsolete methods, moving classes to better-fitting packages, and renaming identifiers to reflect actual semantics. Six primary studies examine this category across the Maven, npm, Python, Web API, Eclipse, R/CRAN, and Linux ecosystems [S1, S8, S26, S40, S46, S52]. Both ecosystem-level studies that quantify reason distributions confirm that this category accounts for a substantial share of breaking changes [S1, S8].

\paragraph{Defensive Design and Robustness}

Defensive design and robustness reflects a deliberate choice to prioritize long-term API correctness over short-term backward compatibility. The category differs from internal refactoring in its explicit intent. Defensive design changes the contract rather than merely the structure. Six primary studies examine this category [S1, S8, S26, S40, S42, S52]. Representative patterns reported across these studies include enforcing stricter input validation, relaxing constraints to accept previously rejected inputs, adjusting error handling behavior, and restricting access modifiers.

\paragraph{Functional Evolution}

Functional evolution covers adding new features. Six primary studies examine this category [S1, S8, S26, S40, S52, S73]. In the two ecosystem-level studies that quantify reason distributions [S1, S8], this category accounts for a smaller share than maintenance and defensive design combined.

These figures may understate the role of feature work. Feature additions often force changes to existing API contracts. The breaking change arises not because the new feature itself is incompatible. It arises because accommodating new functionality cleanly requires restructuring the existing interface. Examples include introducing asynchronous patterns and restructuring configuration objects. In each case the breaking change is a structural consequence of the feature rather than the feature itself [S8, S1, S40].

\paragraph{Ecosystem Integration and Constraints}

Ecosystem integration and constraints is the only category under externally imposed changes. It represents breaking changes that are imposed by upstream dependency evolution or shared-resource constraints rather than chosen by the maintainer. Five primary studies report breaking changes that fit this category [S1, S8, S42, S46, S69]. Reported triggers include shared system resources, incompatible format evolution between packages, transitive dependency upgrades, and trademark or licensing conflicts. Unlike the first three categories, the maintainer cannot avoid these changes by designing the API more carefully. The pressure originates outside the library itself.

\paragraph{Reactive and Accidental Modifications}

Reactive and accidental modifications, the only category under unintended consequences, represents breaking changes whose breaking effect on clients the maintainer did not anticipate, even though the change itself is initiated internally. Six primary studies classify this category among the maintainer-stated reasons they observe [S8, S26, S40, S42, S52, S73]. Three additional studies provide indirect evidence of accidental breakage without classifying reasons [S1, S6, S57]. The most direct case is bug fixes. In the two ecosystem-level studies that quantify reasons [S1, S8], bug-fix-driven breaks are the smallest identified share. Yet bug fixes can still break downstream clients who depend on the pre-fix behavior. As articulated by Hyrum's Law, any observable behavior, including bugs, will eventually be depended upon by some client.

A second pattern arises when maintenance-driven changes produce unintended behavioral effects. Studies on Java behavioral breaking changes [S42] and on Python data-science libraries [S26] both report cases in which maintainers change an API method's behavior under normal input to make it more reasonable. The resulting client breakage is an unintended side effect. Together, these findings illustrate a cross-cutting pattern. Changes driven by correctness or maintainability improvement, rather than by feature delivery, can still produce unintended behavioral breakage for downstream clients [S26, S42, S73].

\paragraph{Cross-Category Observations}

The five reason categories described above are derived from studies that explicitly classify maintainer intent. Broader empirical evidence reveals the overall scale of unintentional breakage. Maven Central history shows that a non-trivial fraction of non-major releases still introduce public API breaks despite SemVer obligations [S6]. Developer surveys show that even self-assessed expert developers cannot reliably predict binary compatibility outcomes [S57]. These findings suggest that a portion of breaking changes result from insufficient tooling and developer awareness rather than deliberate decisions. Unintentional breakage is unlikely to be signaled via SemVer, documented in changelogs, or accompanied by migration guidance, since the maintainer is not aware of the incompatibility.

Across all five categories, maintenance-related reasons collectively account for a larger share of breaking changes than feature delivery in both Java [S8] and npm [S1]. Library maintainers and client developers also perceive these reasons differently. Zaitsev et al.\ survey library and client developers and find that the two groups differ in how they perceive the impact of library evolution and the time required to update [S97]. This perception gap between producers and consumers of breaking changes is further explored in the impact analysis below.

{\scriptsize
\begin{landscape}
\begin{longtable}{|p{1.4cm}|p{1.6cm}|p{1.6cm}|p{4.0cm}|p{2.2cm}|p{1.5cm}|p{2.9cm}|p{1.8cm}|}
\caption{Comprehensive summary of reasons behind breaking changes across software ecosystems, grouped by who initiates the change.}
\label{tab:reason-summary}\\
\hline
\textbf{Initiator Group} & \textbf{Dimension} & \textbf{Reason Category} & \textbf{Description} & \textbf{Ecosystem} & \textbf{Frequency} & \textbf{Intended Benefit} & \textbf{References} \\ \hline
\endfirsthead
\hline
\multicolumn{8}{c}{\tablename\ \thetable{} -- continued from previous page} \\
\textbf{Initiator Group} & \textbf{Dimension} & \textbf{Reason Category} & \textbf{Description} & \textbf{Ecosystem} & \textbf{Frequency} & \textbf{Intended Benefit} & \textbf{References} \\ \hline
\endhead
\hline
\endfoot
\hline
\endlastfoot

\multirow{5}{1.4cm}{\textbf{Library-internal decisions}} &
\multirow{3}{1.6cm}{\textbf{Internal Maintenance \& Refactoring}} &
\textbf{Reduce Code Redundancy} &
Removing obsolete or redundant functionality to reduce technical debt, including deleting unused packages, classes, or methods (``API Simplification''), and phasing out superseded packages. &
Maven, npm, Web API, Eclipse, R/CRAN, Linux &
High in Java [S8]; dominant in npm [S1] &
Reduce maintenance burden; simplify API surface; remove obsolete packages. &
[S1]\allowbreak [S8]\allowbreak [S40]\allowbreak [S52] \\ \cline{3-8}

& &
\textbf{Improve Maintainability} &
Refactoring operations such as moving methods to lighter classes, reducing complexity, splitting large packages, and restructuring class hierarchies. &
Maven, npm, Python, Web API, Eclipse, R/CRAN, Linux &
High in Java [S8] &
Better code structure; easier long-term maintenance; finer-grained modularity. &
[S8]\allowbreak [S26]\allowbreak [S40]\allowbreak [S52] \\ \cline{3-8}

& &
\textbf{Improve Identifier Names} &
Renaming artifacts to improve clarity, accuracy, or structural consistency, from fixing typos to renaming entire packages for structural reasons. &
npm, Maven, Web API, Eclipse, R/CRAN, Linux &
Medium in Java [S8]; notable in npm [S1] &
Eliminate ambiguity; accurate naming; align with language idioms; reflect package evolution. &
[S1]\allowbreak [S8]\allowbreak [S40]\allowbreak [S52]\allowbreak [S46] \\ \cline{2-8}

& \textbf{Defensive Design \& Robustness} &
\textbf{Improve API Design} &
Refining interface contracts to improve robustness, usability, or standards compliance. Includes restricting access modifiers, adding \texttt{final} modifiers, changing parameter lists, and enforcing/relaxing input validation rules. &
npm, Maven, Python, Web API, Eclipse, R/CRAN &
Very High in npm [S1] &
Strengthen postconditions; prevent misuse; improve encapsulation and safety; enforce/relax validation rules. &
[S1]\allowbreak [S8]\allowbreak [S26]\allowbreak [S42]\allowbreak [S40]\allowbreak [S52] \\ \cline{2-8}

& \textbf{Functional Evolution} &
\textbf{Implement New Features} &
Changes to support new functionality requested by users or planned for new releases, often involving moving classes or renaming methods to reflect new capabilities. &
Maven, npm, Python, Web API, Eclipse, R/CRAN &
High in Java [S8] &
Support evolution; fulfill user requests; enrich API capabilities. &
[S8]\allowbreak [S26]\allowbreak [S40]\allowbreak [S1]\allowbreak [S52]\allowbreak [S73] \\ \hline

\textbf{Externally imposed changes} &
\textbf{Ecosystem Integration \& Constraints} &
\textbf{Manage Upstream Dependencies} &
Updating internal dependencies to maintain ecosystem health, fixing bugs, resolving trademark conflicts, removing deprecated dependencies, or managing shared system resources. &
Maven, npm, Eclipse, R/CRAN, Linux &
Very High &
Keep dependencies up-to-date and secure; prevent shared resource conflicts (files, ports). &
[S8]\allowbreak [S1]\allowbreak [S46]\allowbreak [S69]\allowbreak [S42] \\ \hline

\multirow{2}{1.4cm}{\textbf{Unintended consequences}} &
\multirow{2}{1.6cm}{\textbf{Reactive \& Accidental Modifications}} &
\textbf{Fix Bugs} &
Removing elements or changing signatures to correct erratic behavior, such as methods causing memory leaks, iterators returning incorrect collections, or making output/effects more reasonable. &
Maven, npm, Python, Web API, Eclipse, R/CRAN &
Low in Java [S8] &
System stability; removal of unpredicted or dangerous behavior. &
[S8]\allowbreak [S26]\allowbreak [S42]\allowbreak [S40]\allowbreak [S52]\allowbreak [S73] \\ \cline{3-8}

& &
\textbf{Unintentional / Accidental} &
Changes occurring because the developer is unaware that a modification (e.g., to a ``soft-private'' internal method or a subtle behavioral shift) would break downstream clients. &
General / All Ecosystems &
Significant [S6] &
N/A (result of lack of tooling / awareness). &
[S6]\allowbreak [S57]\allowbreak [S1] \\
\hline
\end{longtable}
\end{landscape}
}

\begin{tcolorbox}[colback=yellow!10!white, colframe=yellow!75!black, title=\textbf{Key Findings: Reasons behind Breaking Changes}]
\begin{enumerate}
    \item \textbf{Breaking changes split into three initiator-based groups covering five reason categories}: Library-internal decisions cover internal maintenance and refactoring, defensive design and robustness, and functional evolution; externally imposed changes cover ecosystem integration and constraints; and unintended consequences cover reactive and accidental modifications (see Table~\ref{tab:reason-summary}).
    \item \textbf{Maintenance and design improvements outweigh new features as drivers of breaking changes}: Across the surveyed Java and npm ecosystems, the categories maintainers describe as housekeeping account for the largest share of breaking changes, contradicting the intuition that breaks come primarily from feature delivery.
    \item \textbf{Even the smallest reason category still breaks clients}: Across multiple ecosystems, bug-fix-driven changes account for the smallest share of breaking changes yet still break clients that depended on the pre-fix behavior, empirically confirming Hyrum's Law in the surveyed studies.
\end{enumerate}
\end{tcolorbox}

\subsection{Impact on Downstream Consumers}

In the previous subsection, we examined the reasons behind breaking changes. In this subsection, we examine how these changes impact downstream consumers. We organize the results along five dimensions. \textbf{(1) Magnitude of Impact} covers the gap between theoretical exposure and actual breakage. \textbf{(2) SemVer Violations} covers the reliability of versioning signals. \textbf{(3) Transitive Propagation} covers breakage from undeclared dependencies. \textbf{(4) Technical Manifestation} covers the range from compilation errors to silent behavioral degradation. \textbf{(5) Developer Experience} covers the diagnostic information and documentation gaps downstream consumers face when breakage occurs. These five dimensions are not mutually exclusive. A single breaking change can simultaneously violate SemVer, propagate transitively, and manifest as a compilation error that lacks adequate documentation. Table~\ref{tab:impact-summary} summarizes each impact dimension. The columns include the impact types reported in the literature, their description, ecosystem coverage, and supporting per-study evidence.

\subsubsection{Magnitude of Impact}

A large proportion of clients are theoretically exposed to API changes. However, the actual breakage rate is far lower in both Pharo [S48] and Java [S13]. This order-of-magnitude gap indicates that most breaking changes never manifest in practice. Most clients either do not use the changed API elements or do not upgrade to the breaking version.

When clients do upgrade, the accumulated changes between versions compound the breakage risk. Updating directly to the latest stable version produces a substantially higher failure rate than updating to the adjacent version. Clients who use internal library APIs face an even higher risk [S11]. These figures suggest that the low actual breakage rates reported above are partly a consequence of clients avoiding upgrades altogether.

Even when a single API element changes, the resulting impact can be amplified through inheritance hierarchies. Wu et al.\ find that framework APIs are widely used across client classes. A non-trivial fraction is injected into client code via inheritance and interface implementation [S63]. A breaking change in a widely-used framework method therefore propagates to every client class that extends or implements the affected type. This turns a localized API change into a widespread codebase modification.

\subsubsection{SemVer Violations}

Semantic versioning (SemVer) is the primary contract used to signal compatibility across releases. Major version increments signal breaking changes. Minor and patch releases should be backward compatible. If SemVer is reliably followed, clients can safely adopt non-major releases without compatibility concerns. However, empirical evidence shows that this contract is frequently violated.

Five primary studies provide empirical evidence on this question [S6, S9, S31, S47, S48]. In both Maven [S6, S31] and npm [S9], a non-trivial fraction of minor and patch releases contain breaking changes. There is a positive trend in Maven. Ochoa et al.\ find that the ratio of non-major breaking releases has decreased substantially over the past decade [S6]. Even in recent years a non-trivial share of non-major releases still contains breaking changes. This persistent unreliability has a downstream consequence on client behavior. A majority of new releases are simply ignored by client projects [S47]. The median reaction time to an API change is on the order of weeks [S48]. These low upgrade rates help explain the gap between theoretical and actual breakage discussed in the previous subsubsection.

\subsubsection{Transitive Propagation}

Transitive propagation introduces breakage that arrives from libraries the client never explicitly declared.

Two primary studies, one in npm [S9] and one in Maven [S11], both report two findings. First, a majority of manifesting breaking changes originate from indirect (transitive) providers. Second, a majority of clients invoke transitive dependency functionality without declaring it explicitly. These findings indicate that managing direct dependencies alone is insufficient to prevent breakage. The majority of manifesting breaks come from libraries outside the client's explicit dependency declaration.

The SemVer violations discussed in the previous subsubsection compound this problem. Most client-impacting transitive breaking changes arrive through non-major releases [S11]. Clients typically adopt non-major releases without scrutiny. As a result, transitive breakage arrives precisely in the releases that receive the least review.

\subsubsection{Technical Manifestation}

Manifestation types vary in their detectability. They range from immediate compiler errors to silent behavioral changes that produce no visible signal.

The most visible manifestation is compilation errors. ``Cannot find symbol'' is the dominant compiler-error pattern in Maven, with ``package does not exist'' a distant second [S11]. The compiler points directly to the affected call site. As a result, these errors are the easiest to diagnose and resolve among all manifestation types.

At the opposite extreme, silent logic degradation produces no crash, no compilation error, and no test failure. It still alters program semantics. Montandon et al.\ report that a single default parameter change in a Python data-science library can change a model's accuracy substantially across versions without triggering any exception. A non-trivial fraction of Scikit-Learn and Pandas clients are vulnerable to such changes [S26]. Unlike compilation errors, these changes provide no diagnostic signal. The client's code compiles, runs, and produces output, but the output differs from the previous version. Listing~\ref{lst:silent-degradation-example} illustrates this scenario.

\begin{lstlisting}[float=t, caption={Silent behavioral degradation: default parameter change in machine learning library (Montandon et al.{ [S26]}).}, label=lst:silent-degradation-example, language=Python, escapeinside={(*@}{@*)}, xleftmargin=2.5em, framexleftmargin=2.5em, numbersep=6pt]
(*@\textcolor{blue!70!black}{\# Scikit-Learn version 0.21}@*)
from sklearn.svm import SVC

(*@\textcolor{blue!70!black}{\# Client code - relies on default gamma}@*)
model = SVC()
model.fit(X_train, y_train)
accuracy = model.score(X_test, y_test)
(*@\textcolor{orange!80!black}{\# Result: accuracy = 0.05 (gamma='auto')}@*)

(*@\textcolor{blue!70!black}{\# Scikit-Learn version 0.22 - Default gamma changed}@*)
(*@\textcolor{purple!70!black}{\# Default gamma changed from 'auto' to 'scale'}@*)

model = SVC()
model.fit(X_train, y_train)
accuracy = model.score(X_test, y_test)
(*@\textcolor{red!80!black}{\# Result: accuracy = 0.82 (different gamma strategy)}@*)

(*@\textcolor{red!80!black}{\# No exception raised, no compilation error, no test failure}@*)
(*@\textcolor{red!80!black}{\# Accuracy changed by 77 percentage points across versions}@*)
(*@\textcolor{red!80!black}{\# Client unaware that behavior changed}@*)

\end{lstlisting}

Between these two extremes sit test suite failures. They are detectable but require adequate test coverage to surface. Jayasuriya et al.\ classify outcomes of behavioral breaking changes in client test suites into test errors (runtime and checked exceptions) and test failures (assertion violations). Both types appear in similar proportions [S12]. Unlike silent logic degradation, these manifestations provide a diagnostic signal. The signal appears only if the client has tests that exercise the affected behavior. This condition cannot be assumed in practice.

\subsubsection{Developer Experience and Documentation}

This dimension covers the diagnostic information available to client developers when breakage occurs. Inadequate documentation and diagnostic tooling can make a breaking change difficult to identify and resolve even when it is detectable in principle.

The first gap is advance warning. Raemaekers et al.\ report that only 5.4\% of Maven artifacts (1,196 out of 22,205) use deprecation tags. 33.03\% of public methods are deleted without prior deprecation [S31]. Clients who encounter a missing method therefore have no prior warning and no documented migration path.

The second gap is diagnostic quality. Jayasuriya et al.\ report that nearly half of precondition-violating exceptions triggered by behavioral breaking changes lack meaningful error messages. Almost all such exceptions surface deep in the call stack rather than at the API boundary where the change is made [S12]. This makes it difficult for clients to trace the error back to a dependency change.

The third gap is developer knowledge. Dietrich et al.\ report that even self-assessed expert developers cannot reliably answer binary compatibility questions. Correctness is especially low for scenarios such as return type specialization [S57]. The rules governing what constitutes a breaking change are sufficiently complex that expert developers cannot reliably reason about them without tool support.

\subsubsection{Cross-Dimension Observations}

In summary, the five impact dimensions are interconnected. Low observed impact rates are partly a consequence of low upgrade rates [S47, S48]. SemVer violations mean that version signals do not reliably indicate compatibility [S31, S6]. Transitive propagation introduces breakage from undeclared dependencies [S9, S11]. Silent behavioral regressions evade automated testing unless specific test cases cover the affected behavior [S26]. Documentation gaps make diagnosis difficult even when clients attempt to upgrade [S31, S12, S57].

{\footnotesize
\renewcommand{\arraystretch}{0.9}
\begin{landscape}
\begin{longtable}{|p{1.6cm}|p{1.9cm}|p{7.7cm}|p{1cm}|p{4cm}|p{1cm}|}
\caption{Comprehensive summary of the impact of breaking changes on downstream consumers across software ecosystems.}
\label{tab:impact-summary}\\
\hline
\textbf{Dimension} & \textbf{Impact Type} & \textbf{Description} & \textbf{Ecosystem} & \textbf{Key Evidence} & \textbf{References} \\ \hline
\endfirsthead
\hline
\multicolumn{6}{c}{\tablename\ \thetable{} -- continued from previous page} \\
\textbf{Dimension} & \textbf{Impact Type} & \textbf{Description} & \textbf{Ecosystem} & \textbf{Key Evidence} & \textbf{References} \\ \hline
\endhead
\hline
\endfoot
\hline
\endlastfoot

\multirow{3}{1.9cm}{\textbf{Magnitude of Impact}} &
\textbf{Potential Reach} &
Large proportion of client systems are theoretically exposed to API changes, but most do not experience actual failures. &
Pharo, Maven &
61\% of clients potentially affected; median 14.78\% of API changes break compatibility. &
[S48]\allowbreak [S13] \\ \cline{2-6}

&
\textbf{Actual Manifestation} &
Despite high potential, the actual reaction rate is low, suggesting impact is localized to specific API subsets. &
Pharo, Maven &
Only 5\% of systems actually react; median 2.54\% client impact rate. &
[S48]\allowbreak [S13] \\ \cline{2-6}

&
\textbf{Ripple Effect} &
API changes propagate through client codebases via inheritance and API injection, amplifying the impact beyond direct API consumers. &
Apache, Eclipse &
35\% of client classes use framework APIs; 11\% of API usages cause ripple effects. &
[S63] \\ \hline

\multirow{3}{1.9cm}{\textbf{Update\&Failure Rates}} &
\textbf{Direct Update Failures} &
Updating to the latest stable version has a significantly higher failure rate than adjacent-version updates. &
Maven &
11.58\% failure rate (latest); 4.35\% (adjacent). &
[S11] \\ \cline{2-6}

&
\textbf{Internal API Usage} &
Clients using internal/private APIs face substantially higher breakage rates. &
Maven &
20.6\% failure rate for internal API users. &
[S11] \\ \cline{2-6}

&
\textbf{Release Avoidance} &
Developers selectively ignore new releases to avoid breakage risk. &
Maven, Apache &
69\% of releases ignored; clients upgrade only 60\% of available releases. &
[S47] \\ \hline

\multirow{3}{1.9cm}{\textbf{SemVer \& Compliance Violations}} &
\textbf{Minor/Patch Violations} &
Ostensibly safe releases frequently introduce breaking changes, violating SemVer contracts. &
Maven &
35.7\% of minor and 23.8\% of patch releases contain $\geq$1 BC. &
[S31] \\ \cline{2-6}

&
\textbf{Improving Trend} &
The ratio of non-major breaking releases has decreased significantly over time. &
Maven &
Decreased from 67.7\% (2005) to 16.0\% (2018). &
[S6] \\ \cline{2-6}

&
\textbf{Compliance Rate} &
While approximately 83\% of individual upgrades follow SemVer, approximately 20\% of non-major releases are still breaking. &
Maven &
approximately 83\% of upgrades follow SemVer; approximately 20\% of non-major releases are breaking. &
[S6] \\ \hline

\multirow{3}{1.9cm}{\textbf{Dependency \& Ecosystem Propagation}} &
\textbf{Transitive Impact} &
57.8\% of manifesting breaking changes originate from indirect dependencies, not direct ones. &
npm, Maven &
57.8\% from indirect providers; approximately 20\% of failures traced to transitive dependencies. &
[S9]\allowbreak [S11] \\ \cline{2-6}

&
\textbf{SemVer Level Distribution} &
Non-major updates contain most transitive-dependency-related BCs, violating SemVer expectations. &
Maven &
approximately 42\% of BCs in non-major releases; approximately 33\% minor, approximately 8\% patch. &
[S11] \\ \cline{2-6}

&
\textbf{Indirect Provider Dominance} &
57.8\% of manifesting BCs originate from indirect dependencies, not direct ones. &
npm &
57.8\% from indirect providers. &
[S9] \\ \hline

\multirow{3}{1.9cm}{\textbf{Client Reaction \& Maintenance}} &
\textbf{Reaction Lag} &
Developers need significant time to discover and apply API changes; method improvements take longer than method replacements. &
Pharo &
Median reaction time: 34 days (overall), 47 days (improvements), 20 days (replacements). &
[S48] \\ \cline{2-6}

&
\textbf{Upgrade Reluctance} &
Developers are reluctant to upgrade when APIs are removed and rarely react to deprecation warnings. &
Pharo &
Deprecation warnings rarely heeded ($<$20\% usage). &
[S48] \\ \cline{2-6}

&
\textbf{Support Burden} &
Breaking changes generate significant community support demand and require documentation effort. &
npm &
approximately 78\% of manifesting BCs documented in issue reports, PRs, or changelogs. &
[S9] \\ \hline

\multirow{3}{1.9cm}{\textbf{Technical Manifestation}} &
\textbf{Compilation Errors} &
Static compilation errors dominate, with ``Cannot find symbol'' as the most frequent error type. &
Maven &
approximately 57\% ``Cannot find symbol.'' &
[S11] \\ \cline{2-6}

&
\textbf{Silent Logic Degradation} &
Default argument changes cause no crash but drastically alter results, particularly dangerous in data science. &
Python &
Accuracy changes of up to 77 percentage points across versions without any exception. &
[S26] \\ \cline{2-6}

&
\textbf{Test Failures} &
Behavioral breaking changes manifest as exceptions and assertion failures in test suites. &
Maven &
approximately 59\% Test Errors; approximately 41\% Test Failures. &
[S12] \\ \hline

\multirow{3}{1.9cm}{\textbf{Library-Specific Vulnerabilities}} &
\textbf{Unchecked Exceptions} &
Libraries introduce newly added unchecked exceptions during upgrades, creating latent behavioral breaking changes. &
Maven &
approximately 40\% of libraries (120/302) introduced new unchecked exceptions. &
[S19] \\ \cline{2-6}

&
\textbf{Data Science Defaults} &
Default argument changes in data science libraries silently affect client model outcomes. &
Python &
35\% of Scikit-Learn and 21\% of Pandas clients vulnerable. &
[S26] \\ \cline{2-6}

&
\textbf{Framework Exposure} &
Frameworks inject APIs into client code via inheritance, amplifying the surface area of potential breakage. &
Apache, Eclipse &
On average 35\% of client classes use framework APIs; 8--14\% of APIs injected via inheritance. &
[S63] \\ \hline

\multirow{3}{1.9cm}{\textbf{Developer Experience \& Documentation}} &
\textbf{Documentation Gaps} &
Methods commonly deleted without deprecation tags; disconnect between deprecation practices and actual API removal. &
Maven &
Only 5.4\% of Maven artifacts use deprecation tags; 33.03\% of public methods are deleted without prior deprecation; deprecated methods are never deleted in later versions. &
[S31] \\ \cline{2-6}

&
\textbf{Knowledge Gap} &
Even expert developers struggle to predict binary compatibility outcomes. &
General &
$\sim$60\% correctness rate on compatibility questions. &
[S57] \\ \cline{2-6}

&
\textbf{Diagnosability} &
Error messages are often uninformative and exceptions rarely surface at the API level. &
Maven &
nearly half of precondition-violating exceptions lack meaningful messages; almost all not at API surface. &
[S12] \\
\hline
\end{longtable}
\end{landscape}
}

\begin{tcolorbox}[colback=yellow!10!white, colframe=yellow!75!black, title=\textbf{Key Findings: Impact on Downstream Consumers}]
\begin{enumerate}
    \item \textbf{Client impact is organized along five complementary dimensions}: Across the surveyed studies, the impact of breaking changes is captured by the scale of breakage, the reliability of version signals, propagation through transitive dependencies, the form in which failures manifest, and the diagnostic information available to client developers (see Table~\ref{tab:impact-summary}).
    \item \textbf{Theoretical exposure vastly exceeds actual breakage}: Across multiple ecosystem studies, only a small subset of theoretically-affected clients ever break, partly because most clients deliberately avoid upgrades rather than because libraries are well-behaved.
    \item \textbf{Most manifesting breaks reach clients via libraries they never declared}: Across multiple Maven and npm studies, transitive (indirect) dependencies are the dominant source of client-impacting breakage, so managing direct dependencies alone is insufficient to prevent breakage.
\end{enumerate}
\end{tcolorbox}

\section{RQ3: Detection and Analysis of Breaking Changes}
\label{sec:rq3}

In this section, we examine how breaking changes are detected and analyzed (item D8 in Table~\ref{tab:data_items}). Of the 97 primary studies, 68 papers contribute to detection. One is a systematic literature review on API evolution. The remaining 67 propose detection approaches and collectively describe 43 distinct techniques that we synthesize across Tables~\ref{tab:detection-static}, \ref{tab:detection-dynamic}, \ref{tab:detection-combined}, \ref{tab:detection-learning}, and~\ref{tab:detection-hybrid}. We organize them into three categories distinguished by the source of the detection signal. \textbf{Traditional Approaches} (Section~\ref{sec:traditional}, 51 of 67 papers) rely on predefined rules and direct code analysis, using static signatures, runtime behavior, or both. \textbf{Learning-based Approaches} (Section~\ref{sec:learning}, 8 of 67) infer evolution patterns from historical data through pattern mining, similarity matching, and automated reasoning. \textbf{Hybrid Approaches} (Section~\ref{sec:hybrid}, 8 of 67) integrate multiple detection methods with client-specific impact analysis to filter false positives. Table~\ref{tab:detection-comparison} provides a high-level comparison across the three categories. Section~\ref{sec:comparison} synthesizes the trade-offs and shared limitations.

\subsection{Traditional Approaches}\label{sec:traditional}

Traditional approaches are the largest of the three categories, accounting for 51 of the 67 papers that propose detection approaches. They rely on predefined rules and direct code analysis without learning from historical data. They are distinguished by the signal they observe. \textbf{Static Analysis} (28 out of 51) is the dominant signal within this category. It examines code structure without execution, which makes detection fast and scalable. However, it remains blind to runtime behavior. \textbf{Dynamic Analysis} (13 out of 51) is the second most common. It executes code and observes runtime behavior to capture semantic changes that static analysis misses. It is limited by test coverage. \textbf{Combined Static and Dynamic Analysis} (10 out of 51) is the least common. It integrates both signals to balance their respective trade-offs. We discuss each in the following subsubsections.

\subsubsection{Static Analysis Techniques}

Among traditional approaches, static analysis is the most widely used signal for breaking change detection, primary in 28 of the 67 detection-approach papers. It compares API signatures, type hierarchies, and method contracts across versions without executing code. Most static-analysis tools target the Java, JavaScript, and Python ecosystems. Three technique families appear in the literature. \textbf{Bytecode comparison} operates on compiled artifacts in the Java ecosystem. \textbf{Source-level AST differencing} parses source code for finer-grained structural detection. \textbf{Type inference and built-in SemVer checking} targets dynamically-typed and SemVer-aware ecosystems where bytecode and AST tools do not apply. Table~\ref{tab:detection-static} summarizes the surveyed tools and methods for each family. The columns include the underlying technique, target ecosystem, reported performance, and known limitations.

\paragraph{Bytecode Comparison in Java.}
Within static analysis, bytecode comparison tools are the most mature and widely deployed approach in the Java ecosystem. Six primary studies cover this approach [S2, S3, S7, S11, S34, S64]. Tools such as japicmp, revapi, japi-compliance-checker (Sigtest), and Clirr scan compiled JAR files. They compare method signatures, class hierarchies, field declarations, and visibility modifiers across versions. These tools can analyze large libraries in seconds without requiring source code or test suites. A systematic evaluation by Jezek and Dietrich finds that detection rates vary substantially across tools. The lowest-performing tools' gaps are partly attributable to limited support for Java generics and certain inheritance and member changes [S64].

High syntactic recall does not translate directly into useful warnings for clients. These tools report all potential breaking changes regardless of whether clients actually invoke the affected APIs. Jayasuriya et al. find that japicmp reports far more breaking changes than the fraction of dependency updates that actually cause client compilation failures [S11]. Bytecode comparison also cannot detect behavioral breaking changes. These include logic changes, semantic shifts, and silent calculation errors that preserve the API signature but alter runtime semantics [S64,S7].

\paragraph{Source-Level AST Differencing.}
Beyond bytecode comparison, source-level static analysis tools operate on abstract syntax trees (ASTs). They enable finer-grained detection of structural changes. Five primary studies cover this approach [S2, S7, S12, S14, S29]. The surveyed tools share a source-level abstraction but differ in parsing strategy and how they incorporate client impact. APIDiff compares ASTs of types, methods, and fields extracted from git history [S14]. ROSEAU uses ``Diet Parsing'' (ignoring method bodies) to extract reified API models from source or bytecode. It runs two orders of magnitude faster than compilation-based tools [S29]. BREAKBOT leverages the Spoon framework to statically analyze impact on client source code without building [S7]. Compared to bytecode comparison, source-level tools can detect changes earlier in the development cycle. However, they inherit the same blind spots. They cannot detect behavioral changes and they over-approximate actual client impact.

\paragraph{Type Inference and Built-in SemVer Checking.}
The static techniques discussed above target statically-typed languages. Dynamically-typed and SemVer-aware ecosystems present a different problem. Without explicit type declarations or a fixed compatibility contract at the language level, static tools must either infer types from usage patterns or rely on ecosystem-level tooling. The type-inference route is illustrated by TSINFER-NODE in JavaScript/npm. It generates TypeScript declaration files by analyzing heap snapshots and comparing signatures across versions. Its evaluation reports limited recall and substantial false positives. The errors stem from insufficient exploration and inaccurate type inference, in particular the inability to distinguish native types (e.g., \texttt{Array}) from structurally similar objects [S24]. In Python, AexPy (discussed in Section~\ref{sec:hybrid}) combines dynamic reflection with static analysis to address the same difficulty. The ecosystem-tooling route is taken by Elm and Rust. Their built-in SemVer checkers (elm bump, rust-semverver) automatically bump versions based on exported type changes. These tools are widely adopted but limited to shallow signature-level contracts. They miss behavioral and deep contract changes [S73].

{\scriptsize
\begin{landscape}
\begin{longtable}{|p{2.2cm}|p{4.0cm}|p{2.7cm}|p{2.5cm}|p{3.0cm}|p{2.6cm}|}
\caption{Summary of static analysis approaches for breaking change detection.}
\label{tab:detection-static}\\
\hline
\textbf{Tool/Method} & \textbf{Technique} \& \textbf{Mechanism} & \textbf{Ecosystem} & \textbf{Performance} & \textbf{Limitation} & \textbf{References} \\ \hline
\endfirsthead
\hline
\multicolumn{6}{c}{\tablename\ \thetable{} -- continued from previous page} \\
Tool/Method & Technique \& Mechanism & Ecosystem & Performance & Limitation & References \\ \hline
\endhead
\hline
\endfoot
\hline
\endlastfoot

JAPICMP / RevAPI / Sigtest / Clirr &
Bytecode Comparison: Scans two versions of library binary artifacts to output syntactic or potential semantic breaking changes (method removals, visibility changes). &
Maven (Java) &
Variable detection rates: Sigtest (98.70\%), Japitool (97.40\%), Revapi (82.47\%), Japicmp (65.58\%) &
Context-blind; invisible to behavioral changes; compilation bottleneck; generics \& modifiers issues &
[S2,S3,S7,S34,S11,S12] \\ \hline

CORAL &
Reachability-Guided Compatibility Analysis: Combines binary API checkers (revapi, japicmp, japi-compliance-checker) with static call graph reachability (Soot) to filter unreachable breaking changes. &
Maven (Java) &
approximately 99\% compilation rate, approximately 93\% unit test rate during remediation on 301 projects &
Misses reflection, complex overriding, ghost dependencies; heuristic semantic detection; computationally intensive &
[S78] \\ \hline

Built-in SemVer Tools &
Type/Signature Analysis: Automatically bumps versions based on exported type changes (Elm) or checks crate compatibility (Rust). &
Elm / Rust &
Widespread use in respective ecosystems (elm bump, rust-semverver) &
Shallow contracts only; misses behavioral/deep contract changes &
[S12] \\ \hline

Static Analysis w/ Diff Composition &
Signature Hashing \& Diff Composition: Computes minimal API signatures and uses Diff Composition to linearly combine precomputed consecutive diffs. &
Maven/PyPI/Ruby &
Scalable: supports real-time diff queries &
Over-approximation (CHA/VTA); syntax sensitivity; misses reflection &
[S13,S11] \\ \hline

Incremental Static Analysis (Infer) &
Compositional Shape Analysis \& Incremental Analysis: Uses bi-abduction and incremental state caching to detect bugs and contract violations at scale. &
Java / C / Obj-C &
Highly scalable: verified on tens of millions of lines; fast incremental updates &
Approximation trade-offs; engineering complexity for new languages &
[S12] \\ \hline

UPPDATERA &
AST Differencing \& Reachability Analysis: Uses GumTree for AST differencing and WALA (CHA) for call graphs to find reachable functions with control/data flow changes. &
Maven (Java) &
Detects 74\% direct, 64\% transitive faults (approx. 2x effective as tests); faster/comparable to tests in 72\% of cases &
False positives from over-approximation; reflection blindness; class shading &
[S71,S2] \\ \hline

Type Differencing (TSINFER-NODE) &
Snapshot-based Type Generation: Generates TypeScript declaration files by analyzing heap snapshots and comparing signatures. &
npm (JS) &
Detects 14/176 breaking changes (low recall) &
Insufficient exploration; inaccurate types; false positives &
[S24] \\ \hline

APIDIFF &
Source Code Static Analysis: Compares ASTs of types, methods, and fields from git history. Integrates RefDiff. &
Git (Java) &
Detected over 1,400 breaking changes in MPAndroidChart &
Lack of impact analysis; false positives; misses behavioral changes &
[S14,S29,S12,S2] \\ \hline

BREAKBOT &
Source-Level Static Analysis: Leverages Spoon to statically analyze impact on client source code without building. &
Maven (Java) &
Analyzes 13 clients in 2 minutes &
Focuses on source/binary compatibility; relies on pre-configured client list &
[S7] \\
\hline
\end{longtable}
\end{landscape}
}

\subsubsection{Dynamic Analysis Techniques}

Dynamic analysis executes code against both old and new library versions and compares the observed behavior: return values, exceptions thrown, state mutations, or interaction traces. It is the primary signal in 13 of the 67 detection-approach papers. The surveyed studies employ four strategies. \textbf{Regression testing and client test execution} runs a project's existing test suite against the new library version. \textbf{Crowd-sourced regression testing} aggregates test suites from many downstream dependents to broaden coverage. \textbf{API interaction snapshotting} instruments libraries to record complete call traces and compares them across versions. \textbf{Differential testing and automated test generation} synthesizes targeted tests to reveal breaking changes without relying on existing client tests. Table~\ref{tab:detection-dynamic} summarizes the surveyed tools and methods for each strategy. The columns include the underlying technique, target ecosystem, reported performance, and known limitations.

\paragraph{Regression Testing and Client Test Execution.}
The simplest dynamic detection approach is regression testing. It runs a project's existing test suite against the new library version and observes failures. Six primary studies discuss this approach across npm, Java, and Python ecosystems [S1, S2, S7, S11, S12, S71]. Reported effectiveness varies substantially depending on the ground truth used for evaluation. Detection is far higher when measured against documented breaking changes in client commits [S1] than against injected faults in dependency closures [S71]. The gap reflects a structural limit of regression testing. It can only detect breaking changes exercised by existing tests. Existing tests cover a minority of direct and especially transitive dependency calls [S71]. The approach is also vulnerable to ``skipped tests,'' where developers disable failing tests rather than fixing the underlying breaking change [S1].

\paragraph{Crowd-Sourced Regression Testing.}
To address the test coverage limitation of single-client regression testing, crowd-sourced approaches use test suites from multiple downstream dependents. Mujahid et al.'s approach identifies downstream dependent projects, prioritizes them based on coverage, and executes their test suites against the new library version [S33]. The approach faces practical challenges. A substantial fraction of downstream projects fail to build due to dependency resolution and environment issues. The storage and runtime overheads are also high [S33]. DeBBl extends this concept with IR-based prioritization. It treats library changes as search queries and finds relevant client code via Vector Space Model and Latent Dirichlet Allocation. This substantially reduces the time to detect the first behavioral breaking change [S39].

\paragraph{API Interaction Snapshotting.}
API interaction snapshotting instruments library APIs to record complete call traces (protocol, inputs, outputs, exceptions) during client test execution. It then compares snapshots across versions. This captures behavioral deviations that test assertions may miss. Assertions often only verify return values and ignore exceptions or side effects. GILESI is the representative tool in this category [S75]. Its evaluation reports high recall on seeded behavioral breaks and the discovery of regressions missed by standard client tests. The approach remains blind to I/O side effects. It still depends on client tests to exercise the API [S75].

\paragraph{Differential Testing and Automated Test Generation.}
Differential and generated-test approaches synthesize targeted tests to reveal breaking changes without relying on comprehensive client tests. Six primary studies in this family target both REST APIs and library APIs [S17, S24, S42, S58, S59, S93]. They differ primarily in where they source test inputs. The sources include generated requests against REST endpoints [S93], mined knowledge bases of breaking-change patterns [S59], invocation contexts extracted from client source [S58], and API models inferred from existing client test suites [S17]. Regardless of input source, all automated test generators share the oracle problem. Without explicit specifications, they detect crashes but not silent behavioral changes that produce incorrect results without throwing errors [S42].

{\tiny
\begin{longtable}{|p{2.3cm}|p{3.3cm}|p{1.3cm}|p{1.7cm}|p{1.8cm}|p{0.95cm}|}
\caption{Summary of dynamic analysis approaches for breaking change detection.}
\label{tab:detection-dynamic}\\
\hline
\textbf{Tool/Method} & \textbf{Technique \& Mechanism} & \textbf{Ecosystem} & \textbf{Performance} & \textbf{Limitation} & \textbf{References} \\ \hline
\endfirsthead
\hline
\multicolumn{6}{c}{\tablename\ \thetable{} -- continued from previous page} \\
Tool/Method & Technique \& Mechanism & Ecosystem & Performance & Limitation & References \\ \hline
\endhead
\hline
\endfoot
\hline
\endlastfoot

GILESI &
Dynamic API Snapshotting: Instruments library APIs to record interaction traces (protocol, inputs, outputs, exceptions) during client test execution. Compares snapshots across versions to detect behavioral deviations. &
Java &
96\% recall on seeded behavioral breaks; outperforms client tests (89\%); detects 10 regressions missed by tests &
Blind to I/O side effects; serialization complexity; test-dependent &
[S75] \\ \hline

CompSuite / COMPRUNNER &
Dataset \& Automated Library Upgrade: CompSuite curates 123 incompatibility issues; COMPRUNNER automates dependency upgrades via Maven Versions Plugin and re-executes client test suites. &
Maven (Java) &
Dataset of 123 real-world incompatibility issues across 88 clients and 104 libraries &
Test-dependent; requires manual validation; flaky tests &
[S70] \\ \hline

COMPCHECK &
Knowledge-Guided Test Generation: Mines breaking change knowledge base, performs Context Matching (FSMs), uses Object Reusing and Caller Slicing to generate targeted tests. &
Maven (Java) &
73.1\% recall on revealable sites; reveals 72.7\% more call sites than SENSOR and 94.9\% more than CIA+SBST; zero false positives &
Complex argument generation; slicing limitations; Java-specific; knowledge base dependence &
[S59,S2] \\ \hline

Regression Testing &
Dynamic Analysis: Runs project's test suites on code before and after commit (cross-version testing). &
npm (JS/TS); Java &
81\% recall vs. documentation (npm) [S1]; 47\% direct faults / 35\% transitive faults (Java) [S71] &
Test-dependent; vulnerable to skipped tests; under-approximation; low coverage (58\% direct / 21\% transitive) [S71] &
[S1,S2,\newline S7,S11,\newline S12,S71] \\ \hline

Crowd-Sourced Testing &
Crowd-Sourced Dynamic Analysis: Identifies downstream dependents, prioritizes by coverage, executes their test suites. &
npm (JS) &
Detects 60\% of real-world breaking updates; increases coverage by up to 55\% &
High build failure rate (approximately 38\%); environment rigidity; high storage/runtime overhead &
[S33] \\
\hline
\end{longtable}
}

\subsubsection{Combined Static and Dynamic Analysis}

Combined approaches integrate static and dynamic signals within a single tool: static analysis identifies candidate breaking changes, and dynamic analysis filters false positives and verifies behavioral issues. They appear in 10 of the 67 detection-approach papers and fall into two groups. \textbf{Reachability-guided filtering} uses static call graph analysis to discard breaking changes that clients do not actually invoke. \textbf{Blended analysis and specialized detection} targets specific break types such as compilation errors, exception behavior, binary incompatibilities, and dependency upgrade conflicts. Table~\ref{tab:detection-combined} summarizes the surveyed tools and methods for each group. The columns include the underlying technique, target ecosystem, reported performance, and known limitations.

\vspace{-0.4\baselineskip}
\paragraph{Reachability-Guided Filtering.}
The most common combined approach uses static reachability analysis to filter breaking changes that clients do not actually invoke. CORAL combines off-the-shelf binary API compatibility checkers (revapi, japicmp, japi-compliance-checker) with static call graph reachability using Soot [S78]. UPPDATERA uses GumTree for AST differencing of dependencies and WALA (Class Hierarchy Analysis) for client-dependency call graphs, then finds reachable functions with control/data flow changes [S71]. Both report substantially higher effectiveness than tests alone. However, static reachability analysis has limitations: it misses reflection, complex overriding, and ``ghost dependencies'' (undefined in POM), and over-approximates due to Class Hierarchy Analysis imprecision [S71, S78].

\vspace{-0.4\baselineskip}
\paragraph{Blended Analysis and Specialized Detection.}
Beyond reachability-guided filtering, a second group of combined approaches targets specific break types rather than all breaks within a scope. Each tool pairs static structural detection with a different form of dynamic or contextual analysis. Breaking-Good correlates build log error localization with dependency tree differencing for compilation-related breaks [S18]. UnCheckGuard combines static taint analysis with call graph construction, tracking data flow from client inputs to library exception sites for newly-added unchecked exceptions [S19]. Maracas relies on japicmp and Rascal M3 models to identify binary incompatibilities and their impact on client code [S6]. UPCY applies min-(s,t)-cut on a unified dependency graph to identify simultaneous compatible updates [S38]. These tools achieve higher precision within their target break type at the cost of narrower scope.

{\footnotesize
\begin{landscape}
\begin{longtable}{|p{2.2cm}|p{4.0cm}|p{2.7cm}|p{2.5cm}|p{3.0cm}|p{2.6cm}|}
\caption{Summary of combined static and dynamic analysis approaches for breaking change detection.}
\label{tab:detection-combined}\\
\hline
\textbf{Tool}/\textbf{Method} & \textbf{Technique} \& \textbf{Mechanism} & \textbf{Ecosystem} & \textbf{Performance} & \textbf{Limitation} & \textbf{References} \\ \hline
\endfirsthead
\hline
\multicolumn{6}{c}{\tablename\ \thetable{} -- continued from previous page} \\
Tool/Method & Technique \& Mechanism & Ecosystem & Performance & Limitation & References \\ \hline
\endhead
\hline
\endfoot
\hline
\endlastfoot

Maracas &
Bytecode Comparison \& Impact Analysis: Relies on japicmp and Rascal M3 models to identify binary incompatible breaking changes and impact. &
Maven (Java) &
96.3\% precision, 98.5\% recall in validation benchmarks &
Pessimistic over-approximation; modifier gaps &
[S6] \\ \hline

NoRegrets+ &
Model-Based Test Synthesis: Uses client test suites to dynamically infer reusable API model and synthesizes targeted tests (Type Regression Testing). &
npm (Node.js) &
25x faster than client-based approaches &
Proxy limitations; dependency on client test coverage &
[S17,S28,S12] \\ \hline

Contract-Based Random Testing (QuickCheck/Randoop) &
Dynamic Contract Checking: Generates random test inputs to verify compliance with user-defined or built-in contracts. &
Multi-language (Haskell, Java) &
QuickCheck found hundreds of bugs in industrial C code; Randoop widely used for Java &
Requires specifications; contract-oblivious without explicit properties (oracle problem) &
[S12] \\ \hline

Breaking-Good &
Blended Analysis (Log + Dependency Tree): Correlates build log error localization with dependency tree differencing. &
Maven (Java) &
Automatically identified root causes for 70\% of breaking updates &
Log imprecision; tool dependencies; focused on compilation errors &
[S18] \\ \hline

UnCheckGuard &
Static Taint Analysis: Combines SootUp (call graphs) with FlowDroid (taint analysis) to track data flow from client inputs to library exception sites. &
Maven (Java) &
Reduces candidate callsites by >89\%; identifies 120 libraries with newly-added unchecked exceptions &
Over-approximation; guard blindness; manual verification required &
[S19] \\ \hline

TAPIR &
Pattern-Matching Static Analysis: Uses domain-specific patterns and lightweight field-based alias analysis/access path inference. &
npm (JS) &
100\% recall on 115 affected clients; 86\% precision &
Manual effort (writing patterns); unsoundness; limited scope &
[S20] \\ \hline

AexPy &
Hybrid Analysis (Dynamic Reflection + Static): Combines dynamic reflection (to resolve aliases/assignments) with static analysis (Mypy). &
Python (PyPI) &
86.9\% recall on known breaking changes; 93.5\% precision &
Extraction gaps; model precision; subjectivity &
[S23] \\ \hline

Diagnose &
Dynamic Object Relation Graph w/ Forced Execution: Iteratively constructs graph of runtime structure using forced execution to infer types. &
npm (JS) &
approximately 60\% recall; low false positives; efficient &
Behavioral blindness; refactoring sensitivity &
[S24] \\
\hline
\end{longtable}
\end{landscape}
}

\begin{tcolorbox}[colback=yellow!10!white, colframe=yellow!75!black, title=\textbf{Key Findings: Traditional Approaches (Static, Dynamic, and Combined)}]
\begin{enumerate}
    \item \textbf{Traditional detection forms three families distinguished by the signal source}: Across the surveyed studies, traditional approaches divide into static analysis (compares structural artifacts across versions), dynamic analysis (executes code and observes runtime behavior), and combined techniques (filter static candidates by runtime evidence) (see Tables~\ref{tab:detection-static}, \ref{tab:detection-dynamic}, and~\ref{tab:detection-combined}).
    \item \textbf{High syntactic recall does not translate into useful warnings for clients}: Across multiple Java tooling studies, bytecode comparators report many more breaking changes than the fraction that actually cause client failures, because these tools ignore client reachability and report all potential changes regardless of invocation.
    \item \textbf{Detection maturity is concentrated in a single ecosystem}: Across the surveyed studies, mature bytecode and AST-based tools are heavily concentrated in the JVM, while equivalent tools for dynamically-typed ecosystems remain immature, producing far more false positives and lower recall.
\end{enumerate}
\end{tcolorbox}

\subsection{Learning-based Approaches}\label{sec:learning}

The previous section discusses traditional detection approaches based on predefined rules. Learning-based approaches instead use machine learning, association rule mining, and automated reasoning. They infer breaking changes from historical data and version histories. They are the primary signal in 8 of the 67 detection-approach papers. The surveyed studies adopt three strategies. \textbf{Association rule mining for API evolution} extracts deletion-replacement patterns from commit history. \textbf{Similarity-based API matching} combines signature, text, and call-dependency similarity to recommend replacements for removed APIs. \textbf{Information retrieval-based prioritization and context-driven discovery} formulates detection as a ranking problem over clients, or mines client context to seed targeted tests. Table~\ref{tab:detection-learning} summarizes the surveyed tools and methods for each strategy. The columns include the underlying technique, target ecosystem, reported performance, and known limitations.

\paragraph{Association Rule Mining for API Evolution.}
One category of learning-based approaches uses association rule mining on version control history. APIEvolutionMiner, developed by Hora et al., extracts sets of added and deleted method invocations from commits. It then applies frequent itemset mining to discover evolution rules such as ``deleted call A $\rightarrow$ added call B'' [S68]. LIBSYNC extends this with a graph-based approach. It combines Origin Analysis Tool (OAT) for detecting API declaration changes with GroumDiff for mining adaptation patterns from already-migrated client programs [S65]. Both approaches share a cold-start limitation. They require sufficient historical migration data and cannot handle API usages without analogous adaptations in existing codebases.

\paragraph{Similarity-Based API Matching.}
A second family uses similarity metrics to match APIs across versions and recommend replacements. AURA combines call dependency analysis with text similarity (signature, Levenshtein, LCS) in a multi-iteration algorithm. The algorithm discovers one-to-many and many-to-one replacement rules from anchor matches [S67]. However, AURA is sensitive to major internal implementation changes. It cannot recover replacements when target methods are simply deleted. This limits its applicability in a substantial fraction of cases [S67].

\paragraph{IR-Based Prioritization and Context-Driven Discovery.}
A third family formulates detection as an information retrieval or context-driven mining problem. DeBBI treats library-side API changes as queries and client-side API usages as documents. It uses Vector Space Model and Latent Dirichlet Allocation to rank clients by relevance, and Maximal Marginal Relevance to diversify selection [S39]. SENSOR (also discussed in Section~\ref{sec:traditional}) mines invocation contexts from client source code to seed semantically valid test inputs [S58].

{\scriptsize
\begin{longtable}{|p{2.3cm}|p{3.3cm}|p{1.3cm}|p{1.7cm}|p{1.8cm}|p{0.95cm}|}
\caption{Summary of learning-based breaking change detection approaches.}
\label{tab:detection-learning}\\
\hline
\textbf{Tool}/\textbf{Method} & \textbf{Technique} \& \textbf{Mechanism} & \textbf{Ecosystem} & \textbf{Performance} & \textbf{Limitation} & \textbf{References} \\ \hline
\endfirsthead
\hline
\multicolumn{6}{c}{\tablename\ \thetable{} -- continued from previous page} \\
Tool/Method & Technique \& Mechanism & Ecosystem & Performance & Limitation & References \\ \hline
\endhead
\hline
\endfoot
\hline
\endlastfoot

APIEvolutionMiner &
Revision-Level Association Rule Mining: Extracts deltas (added/deleted method invocations) from commit history. Uses frequent itemset mining to discover evolution rules. &
Smalltalk (Pharo) &
Identifies 169 relevant evolution rules across three systems &
Sensitivity to noisy large commits; relies on co-occurrence in single commits; threshold dependent &
[S68] \\ \hline

DeBBl &
IR-Based Prioritization \& Static RTS: Treats library changes as search queries (VSM/LDA) to find client code. Uses Maximal Marginal Relevance (MMR) for diversity. &
Maven (Java) &
Reduces testing time by 99.1\%; detects 97 BBI bugs &
False positives (POM constraints); requires large pool of accessible clients; redundancy &
[S39,S2] \\ \hline

SENSOR &
Differential Testing w/ Context Seeding: Identifies isomerous conflicting APIs and synthesizes tests using divergence arguments extracted from client source code. &
Maven (Java) &
Precision: 0.80; Recall: 0.76; detected 150 SC issues &
Benign inconsistencies; extraction failures; high computational cost &
[S58,S59] \\ \hline

AURA &
Hybrid Multi-Iteration Analysis: Combines Call Dependency and Text Similarity (Signature, Levenshtein, LCS) to detect one-to-many and many-to-one change rules. &
Java &
over 53\% recall vs. state-of-the-art; ~93\% precision on Eclipse &
Sensitivity to radical internal changes; limited to method-level; dependency on anchors &
[S67] \\
\hline
\end{longtable}
}

\subsection{Hybrid Approaches}\label{sec:hybrid}

The previous sections discuss traditional and learning-based approaches separately. Hybrid approaches instead integrate multiple detection methods with client-specific impact analysis. They filter breaking changes by actual client usage rather than reporting all potential changes. They are the primary signal in 8 of the 67 detection-approach papers. The surveyed studies fall into two groups. \textbf{Ecosystem-specific hybrid tools} combine dynamic reflection with static analysis to handle dynamically-typed ecosystems such as Python and JavaScript. \textbf{Pattern-based and domain-specific detection} uses developer-written patterns or targets specific API types such as REST APIs. Table~\ref{tab:detection-hybrid} summarizes the surveyed tools and methods for each group. The columns include the underlying technique, target ecosystem, reported performance, and known limitations.

\paragraph{Ecosystem-Specific Hybrid Tools.}
Several hybrid tools target dynamically-typed ecosystems where neither bytecode comparison nor pure static type inference is sufficient. The common strategy is to combine two signals. Dynamic reflection or forced execution resolves runtime-only information such as aliases and inferred types. Static analysis extends coverage beyond what a single execution observes. AexPy targets Python by pairing dynamic reflection with Mypy-based static type inference [S23]. Diagnose targets npm/JavaScript by constructing a Dynamic Object Relation Graph through forced execution. The graph infers types in the absence of explicit type declarations [S24]. Both tools substantially outperform purely static or purely dynamic baselines for their respective ecosystems. They share a common limitation. Purely behavioral changes that preserve type structures, such as changes in key ordering or numerical precision, remain undetectable. Beyond these two tools, Breaking-Good and UnCheckGuard (discussed in Section~\ref{sec:traditional}) also employ hybrid techniques. They target compilation error diagnosis and exception-related behavioral breaks, respectively.

\paragraph{Pattern-Based and Domain-Specific Detection.}
A second group of hybrid tools uses developer-written patterns or targets specific API types. TAPIR is a semi-automatic tool. Library developers write patterns describing breaking changes in a dedicated language. A lightweight static analysis then matches these patterns against client code in npm [S20]. However, TAPIR does not discover breaking changes itself. Its effectiveness depends on the completeness of the developer-written patterns [S20]. AutoGuard targets REST API evolution in Java Spring Boot. It statically generates OpenAPI descriptions from source code annotations and diffs them across versions [S30]. It is framework-specific and blind to semantic changes such as altered JSON response structures under the same status code [S30]. Beyond these detection tools, benchmarking infrastructure also supports breaking change research. CompSuite curates a large set of real-world Java client-library incompatibility issues. Each issue has a reproducible test case executed by the companion tool COMPRUNNER [S70].

{\scriptsize
\begin{longtable}{|p{1.5cm}|p{1.6cm}|p{2.9cm}|p{1.4cm}|p{1.6cm}|p{2.0cm}|p{1.0cm}|}
\caption{Summary of hybrid and specialized breaking change detection approaches.}
\label{tab:detection-hybrid}\\
\hline
\textbf{Category} & \textbf{Tool/Method }& \textbf{Technique} \& \textbf{Mechanism} & \textbf{Ecosystem} & \textbf{Performance} & \textbf{Limitation} & \textbf{References} \\ \hline
\endfirsthead
\hline
\multicolumn{7}{c}{\tablename\ \thetable{} -- continued from previous page} \\
Category & Tool/Method & Technique \& Mechanism & Ecosystem & Performance & Limitation & References \\ \hline
\endhead
\hline
\endfoot
\hline
\endlastfoot

\multirow{6}{1.5cm}{Hybrid \& Specialized} &
ROSEAU &
Technology-Agnostic API Model Extraction: Extracts reified API models from source/bytecode using Diet Parsing (ignoring bodies). &
Maven (Java) &
F1=0.99; 2 orders of magnitude faster than compilation-based tools &
Limited to syntactic/binary breaking changes; dangling references &
[S29] \\ \cline{2-7}

&
AutoGuard &
Static OpenAPI Generation \& Diffing: Statically generates OpenAPI descriptions from Spring Boot source code annotations and differences them. &
Java (Spring Boot) &
100\% accuracy for REST methods; runtime ~11s &
Framework-specific; semantic blindness; profile issues &
[S30] \\ \cline{2-7}

&
UPCY &
Graph-Based Min-Cut: Constructs unified dependency graph and applies min-(s,t)-cut to identify simultaneous updates. &
Maven (Java) &
Generates updates with fewer incompatibilities in 41.1\% of cases where naïve updates fail; 70.1\% of generated updates have zero incompatibilities &
Database completeness; search space; timeouts &
[S38] \\ \cline{2-7}

&
Client Compilation Analysis (RDCT) &
Real-world Build Simulation: Retrieves client source code, injects new library version, and builds/tests the client. &
Maven (Java) &
Identifies that approximately 12\% of updates cause actual client build failures &
Resource heavy; fragility; environment sensitivity &
[S11,S2,\newline S7,S38] \\ \cline{2-7}

&
LIBSYNC (OAT) &
Tree-Based Origin Analysis: Uses OAT to detect API declaration changes by aligning project trees via name/structure similarity. &
Java &
OAT: 89-100\% precision; LIBSYNC adaptation recall: 91\% &
Cold start (requires migrated programs); threshold sensitivity &
[S65] \\ \cline{2-7}

\hline

\multirow{6}{1.5cm}{Other Approaches} &
Documentation-Driven Detection &
Automated Text Analysis: Bots analyze issue reports and changelogs to proactively flag risky updates for clients. &
General &
Only 82 out of 296 (27.7\%) BBIs are documented; documentation coverage is low &
Reliance on informal/imprecise documentation; requires NLP; most BBIs undocumented &
[S42] \\ \cline{2-7}

&
Oracle-Free BBC Detection &
Automated Test Generation (Exceptions): Uses automated test generators to identify Test Errors (exceptions) without manual assertion oracles. &
General &
Effective for approximately 59\% of behavioral breaks that manifest as test errors (exceptions) [S12] &
Misses Test Failures (assertion mismatches, approximately 41\%); limited to crashing bugs &
[S12] \\ \cline{2-7}

&
Context-Aware Filtering &
Usage-Based Detection: Cross-references library changes with client's actual functional calls (e.g., using ASM bytecode analysis). &
Java (implied) &
Eliminates irrelevant alerts (high precision) &
Requires access to client source/bytecode; complexity in analyzing transitive calls &
[S11] \\ \cline{2-7}

&
Test-Evolution Analysis &
Historical Test Analysis: Tracks modifications in library's internal test suite to detect undocumented breaking changes and migration patterns. &
General &
267 out of 296 (90.2\%) BBIs are accompanied by internal test code changes that could serve as detection signals &
Requires access to library version control history; inference can be noisy &
[S42] \\ \cline{2-7}

&
Virtualized Installation Testing &
Sandboxed Execution: Installing combinations of packages in virtualized environments to detect script side-effects and resource clashes. &
Linux Distributions &
Potential to detect ~30\% of conflict defects &
Combinatorial complexity; high computational cost; requires virtualized infrastructure &
[S46] \\ \cline{2-7}

&
Automated Meta-data Generation &
Resource Modeling: Using static or runtime analysis to automatically generate fine-grained meta-data for shared resources (files, ports). &
Linux Distributions &
Potential to prevent ~30\% of conflict defects &
Storage overhead for detailed meta-data; requires standardization &
[S46] \\ \cline{2-7}

&
Delta Debugging for Configurations &
Heuristic Search: Installing large subsets of packages and using delta debugging (binary search) to isolate minimal set of conflicting packages. &
Linux Distributions &
Efficiently isolates conflicts in rare package combinations &
Heuristic-based (may not find all); time-consuming process &
[S46] \\ \hline

\multirow{4}{1.5cm}{Additional Specialized} &
APICTURE (via oasdiff) &
Differential Analysis of OAS History: Analyzes git history of OpenAPI specifications using oasdiff to compare consecutive revisions. &
Web APIs (OAS) &
Scalable: successfully applied to 3,271 API histories &
Unclassified changes; atemporal aggregation &
[S51] \\ \cline{2-7}

&
Cross-Project Clone Detection (NiCad) &
Cross-Project Clone Detection: Uses NiCad to detect Type-1 and Type-2 clones of internal methods across releases to track and predict stability. &
Java (Eclipse) &
Precision $\geq$56\%, Recall $\geq$96\%, AUC $\geq$92\% for predicting stability &
Construct validity (complex API usage); Type-2 clones may contain breaks; framework-specific &
[S66] \\ \cline{2-7}

&
Maven Dependency Conflict Detection &
Dependency Conflict Detection: Identifies library APIs with inconsistent semantics between versions due to classpath shading or duplicate classes. &
Maven (Java) &
N/A (method discussed as category) &
Requires dataset modification; false positives if shaded classes identical &
[S11] \\ \cline{2-7}

&
Semantic Version Calculator (Proposed) &
Declarative Datalog Reasoning: Aggregates facts from various static/dynamic analyses and applies community-specific taxonomy rules to compute version numbers. &
General / Language Agnostic &
N/A (proposed conceptual framework) &
Social challenge (requires community agreement); implementation complexity &
[S12] \\
\hline
\end{longtable}
}

\begin{tcolorbox}[colback=yellow!10!white, colframe=yellow!75!black, title=\textbf{Key Findings: Learning-based and Hybrid Approaches}]
\begin{enumerate}
    \item \textbf{Learning-based and hybrid approaches add new signals on top of static and dynamic analysis}: Across the surveyed studies, learning-based methods mine evolution patterns from historical commits and client code (association rule mining, similarity-based matching, IR-based prioritization), while hybrid methods combine multiple detection signals with client-specific impact analysis (see Tables~\ref{tab:detection-learning} and \ref{tab:detection-hybrid}).
    \item \textbf{Learning-based approaches share a cold-start dependency that limits their reach}: Across the surveyed studies, every mining-based and similarity-based technique requires existing migration data or already-adapted client code, so they fail precisely in the early stages of API evolution where no historical patterns yet exist.
    \item \textbf{No surveyed approach reliably detects behavioral changes that preserve type structures}: Across all learning-based and hybrid pipelines studied, semantic shifts that leave signatures intact remain undetectable, which is exactly the dominant breaking change category in the dynamically-typed ecosystems where these pipelines are most needed.
\end{enumerate}
\end{tcolorbox}

\subsection{Comparison \& Limitations}\label{sec:comparison}

The previous sections examine traditional, learning-based, and hybrid approaches individually. This section compares them and discusses their shared limitations. It closes with an in-depth look at the oracle problem that cuts across all detection approaches. Table~\ref{tab:detection-comparison} synthesizes the trade-offs across all approaches.

Static and dynamic analysis form the two foundational signals, with complementary strengths. Static analysis can scan millions of lines in seconds. It achieves near-perfect recall for syntactic changes in statically-typed ecosystems [S6, S64]. However, it over-approximates, reporting changes that never cause actual failures. It cannot detect behavioral changes. Dynamic analysis achieves higher precision for behavioral issues [S75]. It depends on test quality and coverage, which is inadequate for the majority of transitive dependencies [S71]. Learning-based and hybrid approaches build on these foundations. Learning-based approaches can discover patterns humans miss but require large training datasets and struggle with interpretability. Hybrid approaches substantially reduce candidate callsites by combining static and dynamic signals [S19]. They inherit the limitations of their constituent techniques.

The main remaining gap is behavioral breaking changes that preserve syntactic interfaces. Static tools cannot detect that a method now throws different exceptions, returns different values, or exhibits altered side effects without executing code. Dynamic tools face the oracle problem. It is difficult to determine whether a test failure represents a breaking change or a test bug. This limitation is rooted in Rice's theorem and the undecidability of semantic equivalence. Syntactic breaking changes are now reliably detectable. Behavioral changes remain largely undetectable without comprehensive test suites and explicit oracles. This gap matters in practice because behavioral changes are common across ecosystems and routinely occur in non-major releases [S1, S12].

\paragraph{The Oracle Problem.}
The most common limitation across all detection approaches is the oracle problem. The problem is determining whether observed behavioral differences are breaking changes or intended improvements. As discussed in Section~\ref{sec:rq2}, behavioral breaks split between test errors (exceptions that automated test generators can detect) and test failures (assertion mismatches that produce incorrect results without crashing) [S12]. The latter requires explicit oracles. The challenge is amplified in domains where behavioral changes are pervasive, such as machine learning libraries where default argument changes silently alter outputs [S26]. Without ground-truth labels, automated tools cannot determine whether the new output is correct or broken. Contract-based random testing tools like QuickCheck and Randoop can verify compliance with user-defined contracts. They cannot generate assertion oracles without explicit properties. Addressing the oracle problem requires formal specifications or comprehensive test oracles that most projects lack.

\begin{table}[h]
\caption{Comparison of detection approach trade-offs.}
\label{tab:detection-comparison}
\scriptsize
\renewcommand{\arraystretch}{1.3}
\begin{tabularx}{\textwidth}{|l|X|X|X|X|}
\hline
\textbf{Approach} & \textbf{Strengths} & \textbf{Limitations} & \textbf{Best For }& \textbf{Worst For} \\ \hline
Static Analysis & High scalability; Fast (seconds); 95-98\% recall for syntactic changes & Over-approximation; Blind to behavioral changes; High false positives & Syntactic breaks in statically-typed ecosystems (Java, C\#) & Behavioral changes; Dynamic ecosystems (npm, Python) \\ \hline
Dynamic Analysis & High precision for behavioral issues; 96\% recall on seeded breaks & Test-dependent; Computationally expensive; Under-approximation & Behavioral breaks with good test coverage & Projects with poor tests (42\% direct, 79\% transitive deps uncovered [S71]) \\ \hline
Learning-based & Discovers patterns humans miss; Cross-language mapping & Requires large training data; Interpretability issues; Cold start problem & API evolution patterns; Refactoring detection & New projects without history \\ \hline
Hybrid & Reduces candidate callsites by 89\%; Context-aware; Comprehensive coverage & Inherits constituent limitations; Higher complexity & Client-specific impact analysis; Production deployments & Oracle problem; Silent behavioral changes \\ \hline
\end{tabularx}
\end{table}

\section{RQ4: Communication, Prevention, and Recovery Strategies for Breaking Changes}
\label{sec:rq4}

In this section, we examine the strategies that exist to communicate, prevent, and recover from breaking changes in software ecosystems. Following the thematic analysis procedure in Section~\ref{sec:methodology}, we extract and line-by-line code the prevention (D9) and mitigation (D10) segments of the 97 primary studies, yielding 66 distinct strategies drawn from 43 of the 97 papers.

Each strategy describes an action by a particular actor (a maintainer deprecates a method, a client pins a version, a tool runs a reachability check), so the strategies aggregate naturally around three groups: library maintainers, client developers, and automated tooling. We use these groups as the organizing axis for the remainder of this section. First, we examine \textbf{Practices for Library Maintainers}, drawn from 28 of the 43 papers. Second, we analyze \textbf{Strategies for Client Developers}, drawn from 25 of the 43 papers. Third, we survey \textbf{Automated Tool Support}, drawn from 33 of the 43 papers. Because most papers describe strategies for more than one role, these subsection counts overlap rather than partition the 43 papers.

\subsection{Practices for Library Maintainers}

Library maintainers both introduce breaking changes and apply the practices that limit their propagation. As Section~\ref{sec:rq2} shows, the dominant reasons for breaking changes are engineering decisions such as API simplification and design improvement, so the management challenge is not to avoid breaking changes entirely but to communicate them effectively, design APIs that minimize unnecessary breakage, and enforce versioning contracts that allow informed upgrade decisions. The maintainer-side strategies, drawn from 28 of the 43 management-strategy papers, span three themes covered in the following subsubsections: communication and deprecation lifecycle, API design for stability, and versioning and compatibility contracts. Table~\ref{tab:strategy-maintainer} summarizes the strategies under each theme.

\subsubsection{Communication and Deprecation Lifecycle}

\begin{figure}[htbp]
\centering
\includegraphics[width=0.95\linewidth]{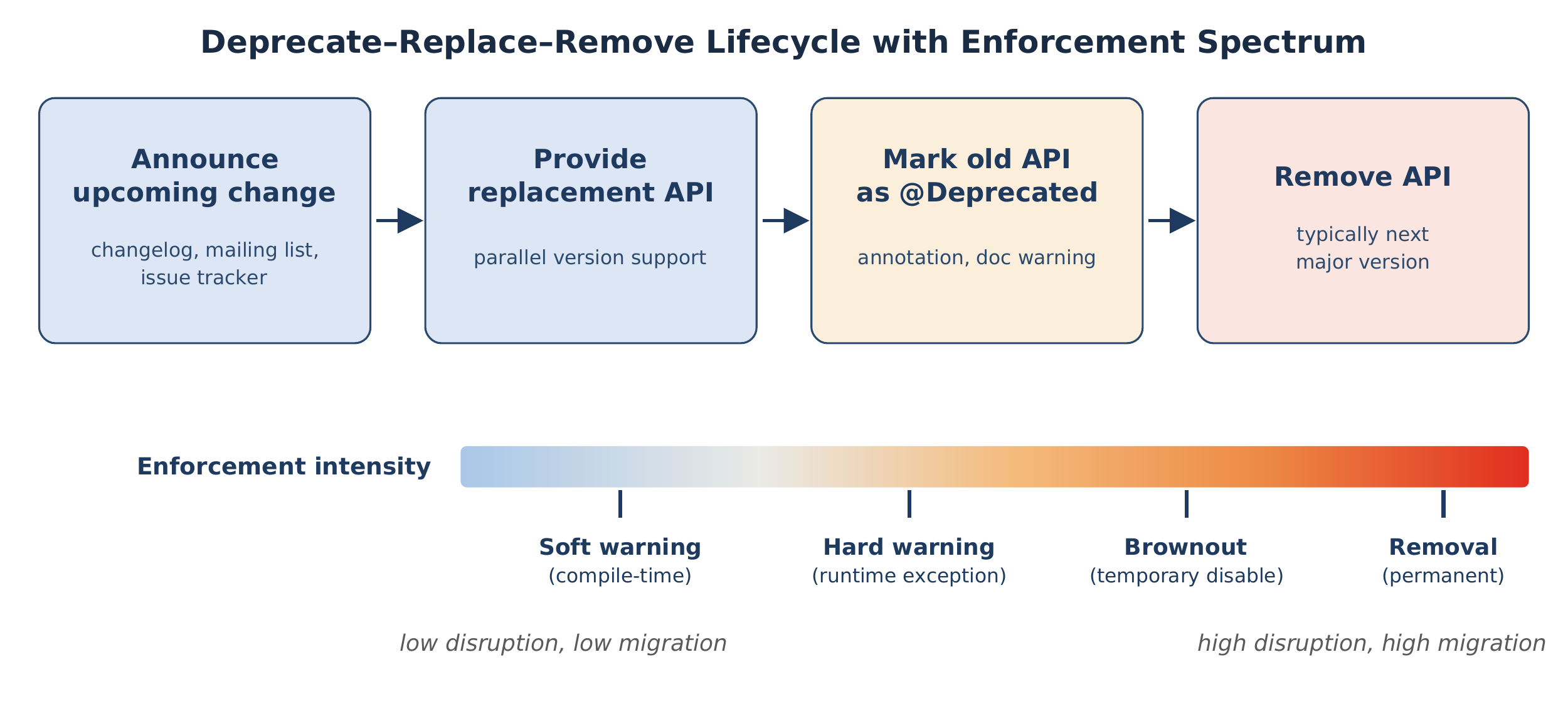}
\caption{The deprecate-replace-remove lifecycle for communicating breaking changes, and the spectrum of enforcement intensities (soft warnings, hard warnings, brownouts, removal) that maintainers can apply at the deprecation stage.}
\label{fig:deprecation-lifecycle}
\end{figure}

Within this theme, the literature describes three families of strategies, drawn from 12 of the 28 maintainer-strategy papers. The first is the \textit{deprecate-replace-remove} lifecycle. The second is proactive notification through changelogs and annotations. The third is active enforcement mechanisms such as blackout tests. Figure~\ref{fig:deprecation-lifecycle} summarizes these as a phased lifecycle. The figure also shows the enforcement-intensity spectrum that maintainers can apply at the deprecation stage. We discuss each in turn.

The standard strategy for communicating breaking changes is the \textit{deprecate-replace-remove} lifecycle, which announces the upcoming change, provides a replacement API, marks the old API as deprecated, and eventually removes it. This phased approach is the most widely discussed in the literature and is reported across Java, npm, Python, and Android ecosystems [S36, S31, S3, S9, S52, S62]. However, this lifecycle is often not followed in practice. Section~\ref{sec:rq2} reports that deprecation tags are sparsely applied in Maven and that public methods are routinely deleted without prior deprecation [S31]. Hora et al.\ find that explicit deprecation warnings nonetheless trigger faster and more frequent client reactions than silent removals [S48].

Beyond deprecation tags, proactive notification through changelogs, issue reports, mailing lists, and \texttt{@Deprecated} annotations represents the broadest category of communication strategies [S3, S9, S28, S36, S43, S47, S52, S73]. The effectiveness of these channels varies across ecosystems. Tightly governed ecosystems such as Eclipse rely on centralized release planning, whereas decentralized ecosystems such as npm treat changelogs as the primary, and often sole, communication mechanism [S43]. The volume of notifications can exceed developers' capacity to process them in fast-moving ecosystems [S3, S43].

Some ecosystems supplement passive signaling with active enforcement. Blackout tests (brownouts) temporarily disable older API versions in short time frames to force clients to notice deprecations, and are particularly effective for Web APIs where the provider controls the runtime [S40, S52]. Ko\c{c}i et al.\ describe these mechanisms as part of an enforcement-intensity spectrum that ranges from soft warnings through hard warnings to complete removal [S40].

\subsubsection{API Design for Stability}

A second class of strategies, drawn from 11 of the 28 maintainer-strategy papers, aims to prevent unnecessary breaking changes through careful API design. The surveyed studies describe four design strategies, each covered in the paragraphs below: strict visibility control and hiding internals, explicit unstable API naming, additive evolution and modularization, and parallel version support.

\vspace{-0.4\baselineskip}
\paragraph{Strict Visibility Control and Hiding Internals.}
The most fundamental design strategy minimizes the public API surface by explicitly hiding internal implementation details, covered in five primary studies [S11, S31, S36, S66, S73]. The recommended mechanism is framework-level enforcement (e.g., OSGi, Java Modules) rather than convention [S31, S73]. The limitation is twofold: standard Java (pre-9) lacks sufficient visibility modifiers, and clients still rely on internal interfaces despite explicit ``discouraged access'' warnings [S66, S11].

\vspace{-0.4\baselineskip}
\paragraph{Explicit Unstable API Naming.}
A lighter-weight alternative is naming conventions such as \texttt{.internal} or \texttt{.experimental} that signal instability and warn clients away from non-public APIs [S36, S66]. Because these conventions are social contracts rather than technical enforcement, clients ignore them when no public alternative exists or when the internal API offers superior functionality [S66].

\vspace{-0.4\baselineskip}
\paragraph{Additive Evolution and Modularization.}
Additive evolution avoids breaking changes by introducing new functionality through optional parameters or new endpoints rather than modifying existing signatures [S40]. Modularization (package splitting) decomposes monolithic libraries into smaller modules so clients import only what they need, reported across Maven, npm, CRAN, and Linux ecosystems in five primary studies [S3, S11, S28, S43, S46]. The long-term cost is API bloat, which can later require consolidation that itself introduces breaking changes.

\vspace{-0.4\baselineskip}
\paragraph{Parallel Version Support.}
A complementary strategy keeps deprecated APIs available alongside their replacements, letting clients migrate at their own pace [S3, S31, S43, S62]. The cost is maintenance burden and API bloat, since developers rarely delete deprecated methods [S31].

\subsubsection{Versioning and Compatibility Contracts}

Semantic Versioning (SemVer) is the central social contract governing breaking changes across nearly every major ecosystem, but Section~\ref{sec:rq2} shows that manual adherence has proven ineffective. The strategies in this category, contributed by 9 of the 28 maintainer-strategy papers, address the gap between SemVer's theoretical promise and its empirical failure. We describe three strategies in the paragraphs below: automated versioning enforcement, code contracts and formal specifications, and ecosystem-level coordination.

\vspace{-0.4\baselineskip}
\paragraph{Automated Versioning Enforcement.}
The prevailing strategy embeds versioning checks in automated tooling rather than leaving them to developer discipline. Tools such as Elm's \texttt{elm bump}, Rust's \texttt{rust-semverver}, and Java's RevAPI and Clirr automatically calculate the next version number from detected binary breaking changes [S56, S31, S52, S73]. This eliminates the most common source of SemVer violations, but such tools reliably detect only syntactic breaks, not behavioral ones.

\vspace{-0.4\baselineskip}
\paragraph{Code Contracts and Formal Specifications.}
To address signature-based tools' blindness to behavioral changes, Lam et al.\ propose formal specifications (e.g., JML) or lightweight type annotations (e.g., \texttt{@NonNull}, \texttt{@Pure}) to make API behavioral contracts machine-checkable [S73]. The barrier is the cost of retrofitting contracts to legacy code, which the authors call the ``Closet Contract Conjecture'' [S73]. Declarative versioning with provenance extends this by publishing machine-readable metadata on the taxonomy of changes behind a version number [S73].

\vspace{-0.4\baselineskip}
\paragraph{Ecosystem-Level Coordination.}
Beyond individual library practices, ecosystem-level mechanisms enforce compatibility through coordinated release cycles (e.g., Eclipse's simultaneous release [S43]), snapshot consistency policies (e.g., CRAN [S3, S28, S43]), and explicit conflict declaration in package metadata (\texttt{Conflicts}, \texttt{Replaces}, \texttt{Breaks} fields in Linux distributions [S46]). These strategies trade off stability against agility: coordinated releases and gatekeeping ensure compatibility but slow feature delivery, whereas decentralized ecosystems achieve rapid iteration at the cost of frequent breakage.

{\footnotesize
\begin{landscape}
\begin{longtable}{|p{2.7cm}|p{1.6cm}|p{4.2cm}|p{3.2cm}|p{3.2cm}|p{2.1cm}|}
\caption{Summary of strategies for library maintainers to communicate, prevent, and recover from breaking changes.}
\label{tab:strategy-maintainer}\\
\hline
\textbf{Strategy} & \textbf{Target} & \textbf{Description \& Mechanism} & \textbf{Rationale} & \textbf{Risks / Limitations} & \textbf{References} \\ \hline
\endfirsthead
\hline
\multicolumn{6}{c}{\tablename\ \thetable{} -- continued from previous page} \\
\textbf{Strategy} & \textbf{Target} & \textbf{Description \& Mechanism} & \textbf{Rationale} & \textbf{Risks / Limitations} & \textbf{References} \\ \hline
\endhead
\hline
\endfoot
\hline
\endlastfoot

Semantic Versioning (SemVer) & Library & Signal the nature of changes through structured version numbers (Major.Minor.Patch). & Allows downstream users to automate minor/patch updates while avoiding breakages; provides a contract for stability. & Not all developers adhere strictly (35.7\% of minor releases have breaking changes [S31]); developers struggle to spot subtle incompatible changes. & [S56,S3,S9,S15,\newline S28,S31,S32,S35,\newline S38,S43,S52,S73,\newline S76,S78] \\ \hline

Proactive Notification \& Documentation & Library & Document changes in changelogs, issue reports, mailing lists, or direct contact. Includes \texttt{@Deprecated} annotations and co-usage recommendations. & Supports client recovery; provides transparency so developers can coordinate changes. & Information overload causes developers to ignore channels; documentation is often outdated. & [S3,S9,S28,\newline S36,S38,S40,S43,\newline S45,S47,S52,S54,\newline S65,S73] \\ \hline

Maintain Old Interfaces / Parallel Version Support & Library & Keep old APIs available (marked as deprecated) or support multiple major versions simultaneously. & Delays rework cost for clients; decouples provider releases from client upgrades. & Incurs technical debt; developers rarely delete deprecated methods, leading to API bloat. & [S3,S31,S36,\newline S40,S43,S46,S52,\newline S62] \\ \hline

Strict Visibility Control / Hiding Internals & Library & Use framework capabilities (e.g., OSGi, Java Modules) to explicitly hide internal packages from public access. & Prevents reliance on unstable internal code; reduces the ``surface area'' of breaking changes. & Requires specific framework support; standard Java (pre-9) lacks sufficient visibility modifiers. & [S31,S36,S73] \\ \hline

Explicit Unstable API Naming & Library & Use naming conventions (e.g., \texttt{.internal}, \texttt{.experimental}) to indicate an API is not intended for public use. & Clear human-readable warning signal; easier to implement than strict modules. & Clients frequently ignore conventions and rely on internal code anyway. & [S36,S66] \\ \hline

Additive Evolution (Parameter Expansion) & Library & Introduce new functionality by adding optional parameters or new endpoints rather than modifying existing signatures. & Avoids breaking changes entirely (75\% of Web API changes are non-breaking additions). & Can lead to API bloat and complex parameter combinations. & [S40] \\ \hline

Modularization (Package Splitting) & Library & Split the library into smaller, separate modules (e.g., core, extensions). & Clients import only what they need, reducing unnecessary transitive risks. & Requires significant architectural effort; complex metadata updates for package mapping. & [S11,S3,S28,S43,\newline S46] \\ \hline

Mark as Optional & Library & Add optional flags to dependencies in the library configuration. & Prevents the dependency from being transitively pulled into client projects by default. & Forces clients to manually declare the dependency if they need that feature. & [S11] \\ \hline

Coordinated Release Cycles & Ecosystem & Coordinate release schedules across the ecosystem (e.g., Eclipse Simultaneous Release). & Predictable planning horizon; ensures mutual compatibility at specific points in time. & High coordination overhead; slows feature releases for individual packages. & [S43] \\ \hline

Explicit Conflict Declaration & Library & Use package manager metadata fields (\texttt{Conflicts}, \texttt{Replaces}, \texttt{Breaks}) to prevent co-installation of incompatible packages. & Proactively prevents installation failures by blocking known incompatible combinations. & Requires manual maintenance; declarations are often coarse-grained; risk of spurious conflicts. & [S46] \\ \hline

Virtual Packages / Service Indirection & Ecosystem & Define a ``virtual package'' representing a capability, allowing multiple concrete packages to provide it interchangeably. & Decouples dependencies from specific implementations; avoids conflicts. & Lacks formal definition linking virtual packages to system resources; relies on maintainer consensus. & [S46] \\ \hline

File Diversions (Resource Relocation) & Library & Use system mechanisms (e.g., \texttt{dpkg-divert}) to force conflicting files to different locations. & Resolves file path conflicts at the installation level. & Increases installation script complexity; requires specific package manager support. & [S46] \\ \hline

Snapshot Consistency / Gatekeeping & Platform & Enforce rules where a new version must not break existing packages (e.g., CRAN policies). & Ensures ``latest'' versions are always mutually compatible. & Places high coordination burden on upstream developers; slows feature releases. & [S3,S28,S43] \\ \hline

Blackout Tests (Brownouts) & Library & Temporarily disable older API versions in short time frames to force clients to notice deprecation. & Reveals hidden dependencies on outdated versions; signals urgency of migration. & Disruptive to client operations; can decrease consumer satisfaction. & [S40,S52] \\ \hline

Usage-Driven Evolution (Feedback Loops) & Library & Collect and analyze usage data from the ecosystem to identify high-impact APIs and guide deprecation decisions. & Minimizes ecosystem disruption by preserving most-used features; enables data-driven debt removal. & Data availability/privacy issues; survival bias may not represent enterprise users. & [S52,S73] \\ \hline

Script-Based Migration Injection & Library & Provide mechanisms for developers to embed custom scripts into migration guides for complex changes. & Enables resolution of complex logic changes that automated tools cannot infer. & Requires manual developer effort; scripts may lack context if the root model changes. & [S25] \\ \hline

API Promotion (Internal to Public) & Library & Systematically identify stable internal interfaces via clone detection and officially promote them to public API. & Mitigates compatibility failures for clients forced to use internal code. & Promotion occurs slowly; requires effort to distinguish truly stable vs.\ coincidentally stable interfaces. & [S66] \\ \hline

Code Contracts (Formal \& Lightweight) & Library & Use formal specifications (e.g., JML) or lightweight annotations (\texttt{@NonNull}, \texttt{@Pure}) to explicitly define API behavioral contracts. & Transforms implicit expectations into machine-checkable rules; enables behavioral break detection. & High effort to retrofit legacy code; requires sophisticated checkers. & [S73] \\ \hline

Declarative Versioning with Provenance & Library & Publish machine-readable metadata detailing the reasoning and taxonomy of changes behind a version number. & Establishes trust in automated versioning; allows filtering upgrades by specific change types. & Requires ecosystem-wide agreement on change taxonomies; adds release overhead. & [S73] \\ \hline

Automated Versioning Enforcement & Library & Use plugins/tools (e.g., \texttt{elm bump}, Semantic Version Calculators) to automatically calculate version numbers based on detected breaking changes. & Removes human error in versioning; ensures version number accurately reflects compatibility. & Relies on accuracy of diffing tools which may miss semantic changes. & [S56,S31,S52,S73] \\
\hline
\end{longtable}
\end{landscape}
}

\begin{tcolorbox}[colback=yellow!10!white, colframe=yellow!75!black, title=\textbf{Key Findings: Practices for Library Maintainers}]
\begin{enumerate}
    \item \textbf{Maintainer strategies fall along three themes}: Across the surveyed studies, library-maintainer strategies cover communication and deprecation lifecycle, API design for stability, and versioning and compatibility contracts (see Table~\ref{tab:strategy-maintainer}).
    \item \textbf{The deprecate-replace-remove pipeline is widely advocated but rarely followed end-to-end}: Across multiple Maven, npm, and Pharo studies, deprecation tags are sparsely applied, public methods are routinely deleted without prior deprecation, methods that are deprecated are rarely actually removed in subsequent versions, and clients tend to ignore deprecation warnings even when present.
    \item \textbf{Manual SemVer compliance fails systematically, driving a shift to automated enforcement}: Across multiple ecosystem studies, the most common SemVer violations trace to maintainer discretion rather than malice, which has led the field toward tools that compute version increments and detect contract violations from analysis output rather than rely on developer judgment.
\end{enumerate}
\end{tcolorbox}

\subsection{Strategies for Client Developers}

In the previous subsection, we examine practices for library maintainers. In this subsection, we examine strategies for client developers, who must absorb breaking changes from dependencies they may not have explicitly declared. The client-side strategies, drawn from 25 of the 43 management-strategy papers, span three themes covered in the following subsubsections: dependency configuration and version management, update decision-making, and adaptation and isolation techniques. Table~\ref{tab:strategy-client} summarizes the strategies under each theme.

\subsubsection{Dependency Configuration and Version Management}

\begin{figure}[htbp]
\centering
\includegraphics[width=0.85\linewidth]{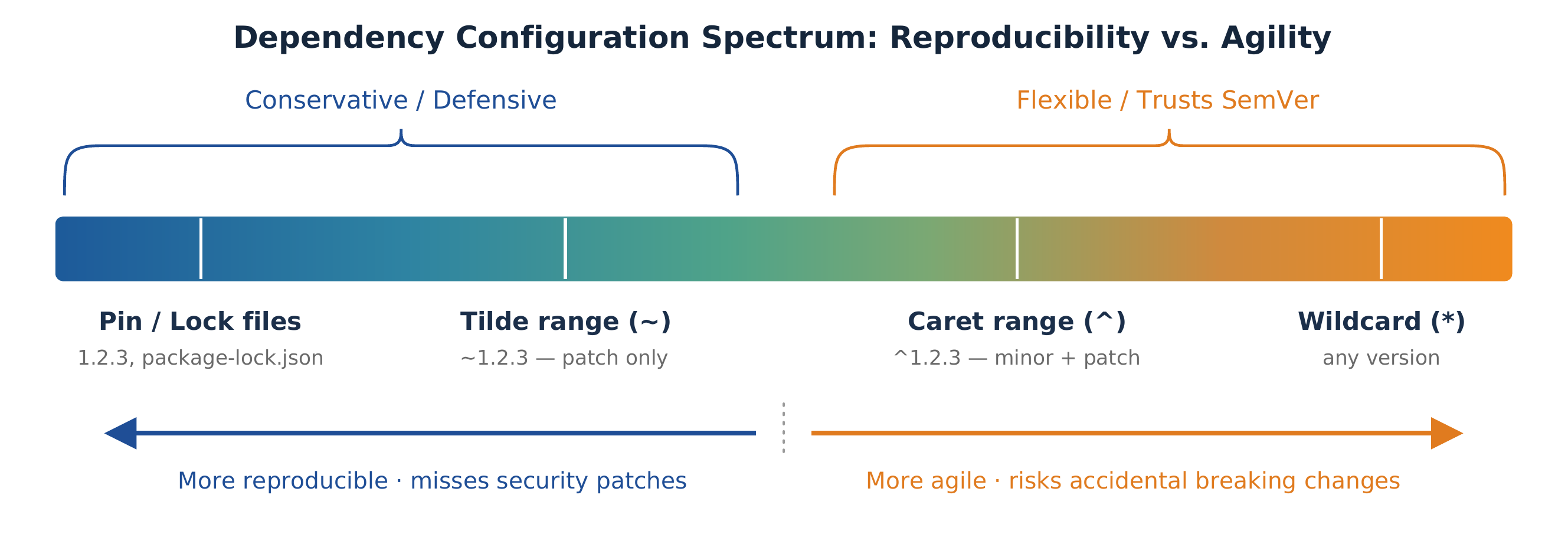}
\caption{Spectrum of client-side dependency configuration strategies. Strict pinning maximizes reproducibility but loses access to upstream patches and bug fixes; flexible ranges enable agility but rely on the upstream's adherence to semantic versioning.}
\label{fig:dep-config-spectrum}
\end{figure}

The most immediate lever available to client developers is how they declare dependency versions. The empirical evidence, drawn from 11 of the 25 client-strategy papers, reveals a spectrum of strategies with different risk profiles. Figure~\ref{fig:dep-config-spectrum} illustrates this along an axis of reproducibility versus agility. We cover three families in the paragraphs below: version pinning and lock files, version ranges and range narrowing, and dependency exclusion and shadowing.

\paragraph{Version Pinning and Lock Files.}
Fixing version numbers and using lock files (\texttt{package-lock.json}, \texttt{yarn.lock}) are the conservative end of the spectrum, prioritizing reproducibility over automatic updates [S56, S11, S3, S9]. The trade-off is that pinned projects miss automatic security patches and accumulate technical lag. Dietrich et al.\ observe that developers typically adopt pinning after experiencing a break, treating it as a defensive measure rather than a proactive strategy [S56].

\paragraph{Version Ranges and Range Narrowing.}
At the opposite end, flexible versioning uses range operators (\texttt{\^{}}, \texttt{\~{}}, \texttt{*}) to enable automatic deployment of bug fixes and security patches [S56, S73]. This requires trust in upstream SemVer adherence, which Section~\ref{sec:rq2} shows to be empirically unwarranted. Range narrowing replaces broad caret (\texttt{\^{}}) ranges with narrower tilde (\texttt{\~{}}) ranges and is documented as a common reactive strategy in npm after experiencing a breaking update [S53].

\paragraph{Dependency Exclusion and Shadowing.}
For transitive dependency conflicts, clients escalate from blocking a dependency to isolating it entirely [S11, S3, S38, S77, S46]. Excluding a transitive dependency manually blocks it from the build but risks runtime crashes if the direct dependency relies on the excluded code [S11]. Shadowing declares the transitive dependency as a direct dependency to pin its version [S11, S3, S38, S77]. Shading (vendoring) takes isolation further by copying and renaming the dependency's bytecode into the project, completely preventing version conflicts at the cost of artifact size and debugging complexity [S11, S3, S28, S43, S46].

\subsubsection{Update Decision-Making}

Clients must also decide \textit{when} and \textit{whether} to upgrade. The literature on this question, drawn from 11 of the 25 client-strategy papers, reveals that the dominant strategy is deliberate avoidance. We describe two families of strategies in the paragraphs below: selective upgrading and reactive maintenance, and incremental and holistic updates.

\paragraph{Selective Upgrading and Reactive Maintenance.}
The dominant client philosophy is to skip most available releases, upgrading only when substantial benefits appear or a build actually breaks [S47, S92, S28, S43, S56, S62]. Bavota et al.\ find that Apache developers intentionally skip the majority of dependency releases, and that adopted releases contain significantly more bug fixes than ignored ones [S47]. The extreme is reactive maintenance (``wait for breakage''), documented by Bogart et al.\ as the dominant philosophy in CRAN and Node.js [S28]. The trade-off is technical debt accumulation: the longer clients defer upgrades, the more breaking changes accumulate between versions and the more costly each eventual upgrade becomes.

\paragraph{Incremental and Holistic Updates.}
Five primary studies present strategies to mitigate the risks of upgrading [S35, S38, S76, S77, S78]. Incremental (minimal-delta) updates upgrade to the closest safe version rather than the latest, minimizing the size of each change [S35, S78]. Holistic multi-library updates use graph algorithms (min-$(s,t)$-cut, topological sorting) to identify simultaneous updates that keep the entire dependency subgraph compatible [S38, S76, S77, S78]. Per-tool figures are reported in Table~\ref{tab:detection-hybrid}.

\subsubsection{Adaptation and Isolation Techniques}

When upgrading is unavoidable and breakage occurs, clients employ adaptation strategies described across 8 of the 25 client-strategy papers, each covered in the paragraphs below: abstraction layers and the tolerant reader pattern, runtime version checking, and library migration.

\paragraph{Abstraction Layers and the Tolerant Reader Pattern.}
Wrapping a dependency in a local abstraction layer (facade or wrapper) isolates the impact of a breaking change to a single shim class. Six primary studies cover this strategy across npm, Web API, Maven, CRAN, and Python ecosystems [S3, S25, S31, S43, S45, S52]. The tolerant reader pattern complements this for message-based APIs by extracting only necessary data and ignoring unknown fields, allowing additive API evolution without breakage [S25]. Both strategies are effective only for structural changes; if the underlying logic changes, no wrapper can preserve the original semantics.

\paragraph{Runtime Version Checking.}
In ecosystems where multiple API versions coexist at runtime (e.g., Android), clients use conditional logic to check the runtime environment version before invoking specific methods [S62]. The strategy enables a single application to support multiple API levels but increases cyclomatic complexity and creates dead code paths that must be tested across all supported versions.

\paragraph{Library Migration.}
When a library is unmaintained, has unpatched vulnerabilities, or has diverged from client needs, complete library migration replaces it with a functionally equivalent alternative [S54]. He et al.\ find that the primary reasons are end-of-life status, security concerns, and superior alternatives, but that migrations carry high cost and risk for identifying suitable targets and validating semantic equivalence [S54].

{\scriptsize
\begin{longtable}{|p{2.5cm}|p{3.5cm}|p{2.5cm}|p{2.5cm}|p{1.5cm}|}
\caption{Summary of strategies for client developers to communicate, prevent, and recover from breaking changes.}
\label{tab:strategy-client}\\
\hline
\textbf{Strategy} & \textbf{Description \& Mechanism} & \textbf{Rationale} & \textbf{Risks / Limitations} & \textbf{References} \\ \hline
\endfirsthead
\hline
\multicolumn{5}{c}{\tablename\ \thetable{} -- continued from previous page} \\
\textbf{Strategy} & \textbf{Description \& Mechanism} & \textbf{Rationale} & \textbf{Risks / Limitations} & \textbf{References} \\ \hline
\endhead
\hline
\endfoot
\hline
\endlastfoot

Exclude Transitive Dependency & Manually configure the build file (e.g., \texttt{pom.xml}) to explicitly block a specific transitive dependency. & Prevents version conflicts and removes unwanted/vulnerable dependencies (most common action). & High risk: if the direct dependency relies on the excluded code, runtime crashes result. & [S11] \\ \hline

Define as Direct Dependency (Shadowing) & Explicitly declare the transitive dependency as a direct dependency to pin a specific version. & Forces Maven's ``shortest path'' rule to shadow other versions; ensures version stability. & Can lead to dependency hell by bloating the project list and increasing maintenance burden. & [S11,S3,S38,S77] \\ \hline

Transitive-Aware Compatibility Upgrading & Upgrade transitive dependencies by analyzing their compatibility against the client's usage independently of the direct parent. & Mitigates technical lag deep in the dependency tree; prevents ``invisible'' breaking changes. & Computationally expensive; requires analyzing the entire dependency tree. & [S76,S77,S78] \\ \hline

Library Migration / Replacement & Completely replace a library with a functionally equivalent alternative. & Necessary when the source library is unmaintained or has unpatched vulnerabilities. & High cost and effort; risk of semantic mismatches; performance improvements not guaranteed. & [S54] \\ \hline

Use SBOM / Lock Files & Use lock files (\texttt{package-lock.json}) to fix exact versions of all transitive dependencies. & Ensures version consistency and awareness of all provider versions from the last successful build. & Can cause conflicts when multiple dependencies require different versions. & [S11,S3,S9] \\ \hline

Fix Version Numbers (Steady Range) & Use explicit version numbers instead of ranges (``pin'' dependencies). & Prevents resolution to incompatible ``latest'' version; ensures reproducible builds. & Projects miss automatic security patches and bug fixes; increases risk of dependency hell. & [S56,S11,S3,S9,\newline S31,S32,S53] \\ \hline

Flexible Versioning (Version Ranges) & Declare dependencies using range operators (\texttt{\^{}}, \texttt{\~{}}, \texttt{*}) to allow best-version resolution. & Enables agility; ensures automatic deployment of bug fixes and security patches. & Builds become less deterministic; range updates can introduce accidental breaking changes. & [S56,S73] \\ \hline

Range Narrowing (Restrictive Ranges) & Replace broad range operators with narrower ones (e.g., caret to tilde). & Reduces surface area of potential breaking changes by accepting only patch updates. & Limits access to new features; patch updates can still contain breaking changes. & [S53] \\ \hline

Automated Dependency Pruning & Automatically filter out upgrade candidates that introduce new, unused transitive dependencies. & Reduces project bloat and attack surface; maintains a leaner dependency graph. & Requires accurate metadata; might skip versions with desirable features. & [S76] \\ \hline

Dependency Tree Logging (Historical Auditing) & Maintain daily logs of the dependency tree or commit dependency states to VCS. & Enables tracing the exact implicit update that caused a breakage; reduces diagnosis latency. & Requires automated routine setup; logs can become verbose. & [S53] \\ \hline

Provider-Focused Regression Testing & Intensify testing on functionalities that specifically rely on providers, including corner cases and cross-version testing. & Safeguards against flexible versioning where implicit updates might introduce bugs. & Increases test suite complexity and maintenance cost. & [S53,S71,S73] \\ \hline

Abstraction Layers (Facade/Wrapper) & Wrap a dependency in a local abstraction layer (e.g., \texttt{rxjs-compat}) to isolate breaking changes. & Isolates impact to a single shim class; enables stability by translating new versions to old interfaces. & Requires additional effort; ineffective if underlying logic changes fundamentally. & [S3,S25,S31,S43,\newline S45,S52] \\ \hline

Tolerant Reader Pattern & Design clients to extract only necessary data and ignore unknown fields or extensions. & Allows fields to be added/removed without breaking the client (``liberal'' receiving). & Syntactically non-breaking removals may still have significant semantic implications. & [S25] \\ \hline

Reactive Maintenance (Wait for Breakage) & Deliberately ignore upstream updates, taking action only when a build actually breaks. & Avoids information overload; assumes breaks are infrequent. & Increases risk of emergency fixes; creates technical debt as dependencies drift. & [S56,S28,S32,S43,\newline S62] \\ \hline

Selective Upgrading & Intentionally skip specific library releases, upgrading only when substantial changes are present (69\% of releases ignored). & Avoids risk of breaking changes in minor updates; focuses effort on high-value upgrades. & May result in technical debt; requires manual evaluation of release notes. & [S47] \\ \hline

Wait for Bug Fixes & Prioritize upgrading when the new release contains a high number of bug fixes (mean 32 in upgraded vs.\ 15 in ignored releases). & Maximizes system stability by targeting proven improvements. & Delays access to new features; assumes bug fixes warrant the risk. & [S47,S62] \\ \hline

Avoid Deleted APIs (Structural Check) & Actively avoid upgrading to versions that contain deleted public methods. & Directly prevents compile-time errors and immediate breakage. & Can lead to being stuck on old versions indefinitely. & [S47] \\ \hline

Downgrade Dependency & Temporarily revert to a previous stable version of a provider. & Provides an immediate fix to restore build stability after a break. & May miss security patches or new features; introduces technical lag. & [S9,S53] \\ \hline

Runtime Version Checking (Conditional Execution) & Use conditional logic (e.g., \texttt{Build.VERSION.SDK\_INT}) to check the runtime API version before invocation. & Allows a single application to support multiple API versions simultaneously. & Increases code complexity; requires testing across all supported versions. & [S62] \\ \hline

Metric-Based Library Selection & Utilize historical stability metrics, retention rates, and migration trends for evidence-based decisions. & Avoids adopting unstable or frequently abandoned libraries; uses ``wisdom of the crowd.'' & Relies on past data which may not predict future trends; limits choice to popular libraries. & [S54,S55] \\ \hline

Crowd-Sourced Constraint Recommendation & Analyze dependency constraints used by other dependents to determine safe constraint levels. & Provides empirical confidence score; bypasses unreliable SemVer claims by following community consensus. & Relies on the crowd being correct; high latency waiting for enough adoption data. & [S32] \\ \hline

Holistic Multi-Library Update & Use graph algorithms (min-$(s,t)$-cut, topological sorting) to identify simultaneous updates for interconnected libraries. & Minimizes total breaking changes; ensures ``blossom compatibility.'' & Computationally expensive; relies on complete ecosystem database. & [S38,S76,S77,S78] \\ \hline

Incremental / Minimal-Delta Updates & Upgrade to the closest safe version rather than the latest. & Minimizes change size and likelihood of multiple simultaneous breaks. & May miss security patches in the latest version; requires iterative steps. & [S35,S78] \\ \hline

Shade (Shadow) Classes / Vendoring & Copy and rename the dependency's bytecode or source code directly into the project (static linking). & Completely isolates the dependency; prevents version conflicts. & Increases artifact size; makes debugging harder; requires manual maintenance. & [S11,S3,S28,S43,\newline S46] \\ \hline

Change Dependency Scope & Change scope to \texttt{test} or \texttt{provided} to prevent the dependency from leaking into the runtime classpath. & Prevents classpath pollution; limits exposure to test or build time. & Cannot be used if code is needed at runtime. & [S11] \\ \hline

Interactive Static Analysis Filtering & Use tools that prompt developers with context-specific questions to resolve static analysis ambiguities. & Overcomes imprecision of static analysis in dynamic languages; filters false positives. & Relies on developer knowledge; potential for user fatigue. & [S45] \\ \hline

Cross-Language API Mapping & Use learning-based models (e.g., GANs) to map APIs between programming languages for migration. & Enables adaptation across ecosystem boundaries; helps in platform shifts. & High complexity; limited accuracy for behavioral differences; requires large training corpora. & [S52] \\
\hline
\end{longtable}
}

\begin{tcolorbox}[colback=yellow!10!white, colframe=yellow!75!black, title=\textbf{Key Findings: Strategies for Client Developers}]
\begin{enumerate}
    \item \textbf{Client strategies fall along three themes}: Across the surveyed studies, client-developer strategies cover dependency configuration and version management, update decision-making, and adaptation and isolation techniques (see Table~\ref{tab:strategy-client}).
    \item \textbf{The dominant client strategy is to not upgrade at all}: Across multiple Maven, npm, and CRAN studies, clients deliberately skip most available releases and adopt the ``wait for breakage'' philosophy, treating upstream evolution as a risk to be avoided rather than as an opportunity to absorb improvements.
    \item \textbf{Defensive configuration is adopted reactively after a break, not proactively}: Across multiple ecosystem studies, version pinning, range narrowing, and lock-file adoption all spread among developers as recovery responses to specific breakage incidents, rather than as upfront design choices made before any breakage is observed.
\end{enumerate}
\end{tcolorbox}

\subsection{Automated Tool Support}

The preceding subsections have shown that manual strategies such as SemVer adherence, changelog reading, and proactive upgrading are often insufficient at ecosystem scale. Automated tools address this gap by embedding compatibility analysis, version management, and code adaptation into programmatic workflows. The evolution of tooling mirrors the three generations of detection approaches identified in Section~\ref{sec:rq3}, progressing from static analysis tools, through dependency bots, to AI-driven approaches that automatically repair client code. These automated tools, drawn from 33 of the 43 management-strategy papers, span four themes covered in the following subsubsections: dependency resolution and optimization, automated code repair, refactoring replay and migration tools, and analysis and verification. Table~\ref{tab:strategy-tooling} summarizes the strategies under each theme.

\subsubsection{Dependency Resolution and Optimization}

The first category of automated tool support, covered by 4 of the 33 tool-support papers, operates at the version-resolution level without modifying client code. The two approaches differ in scope: dependency bots monitor upstream releases and open update pull requests one library at a time, while multi-objective dependency optimization models updating as a graph-level problem to find globally compatible version sets.

\paragraph{Dependency Bots.}
Dependency bots monitor upstream releases and automatically create pull requests to update dependencies, accelerating updates relative to manual processes [S9]. As discussed in Section~\ref{sec:rq3}, the tests that bot-generated PRs trigger cover only a minority of direct and especially transitive dependency calls, so a ``green build'' does not guarantee compatibility [S71]. A further limitation is that bots typically update one library at a time and ignore the graph-level interactions that holistic strategies address. When bot-generated updates do introduce breakage, downgrades to a stable release are the most frequent recovery action [S9].

\paragraph{Multi-Objective Dependency Optimization.}
Two primary studies model dependency updating as a graph-level optimization problem [S77, S78]. Jaime et al.\ aggregate diverse metrics (freshness, popularity, CVE count) and minimize the cost of change under user-defined weights [S77]. Zhang et al.\ combine dependency graph partitioning with SMT solvers and backtracking to find globally compatible version sets [S78]. These approaches are limited by the accuracy of the underlying cost metrics and by scalability as the dependency graph grows.

\subsubsection{Automated Code Repair}

\begin{figure}[htbp]
\centering
\includegraphics[width=0.7\textwidth]{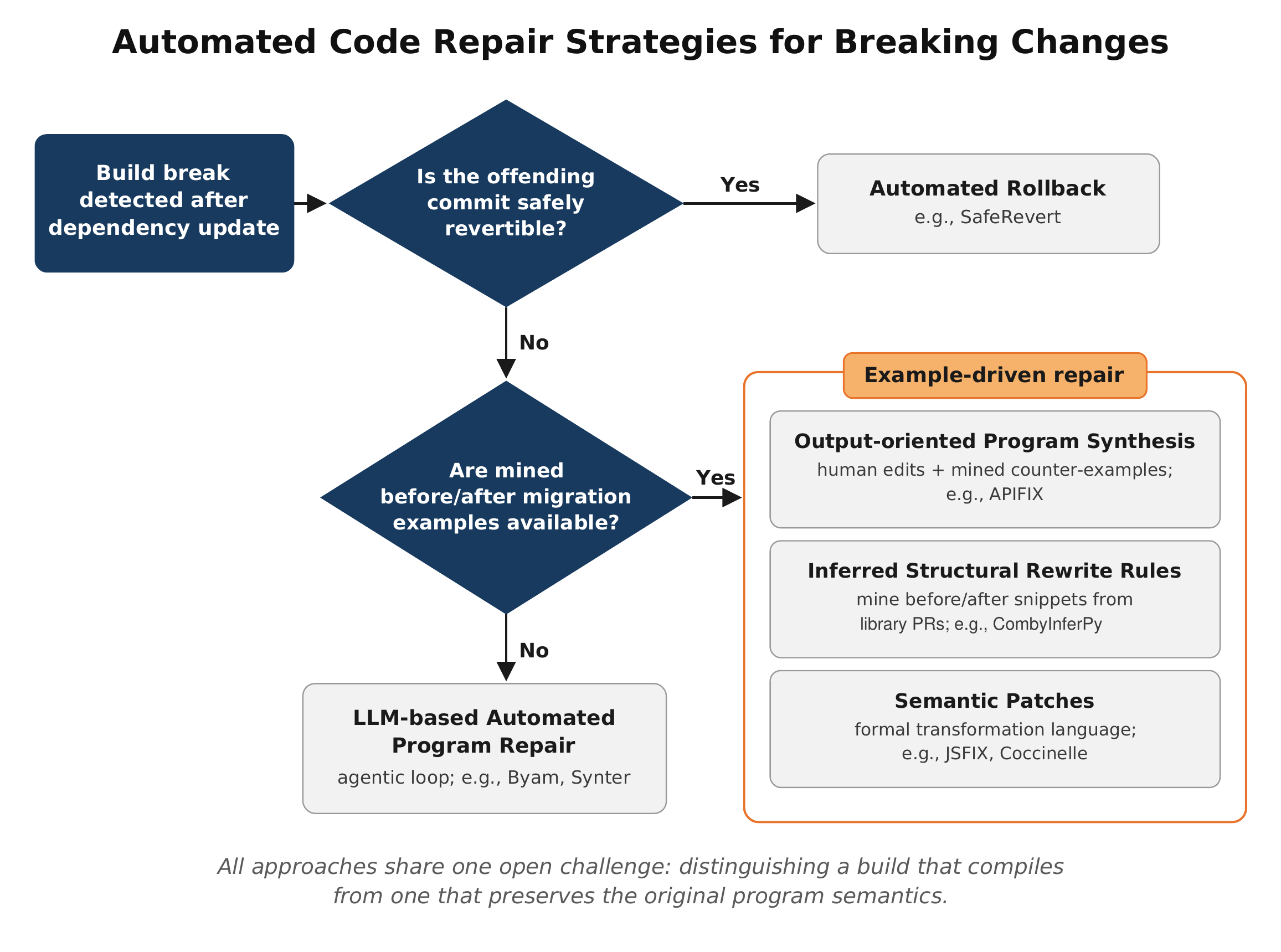}
\caption{Decision flow over the five families of automated code repair strategies for breaking changes. Branches escalate from reverting the offending commit, to example-driven synthesis when before/after examples are available, to LLM-based repair when no such examples exist.}
\label{fig:repair-escalation}
\end{figure}

Dependency bots cannot fix the code breakages that version updates produce. The 8 of the 33 tool-support papers that address this gap describe an escalation of options, from reverting the offending change to applying automatically synthesized patches. Figure~\ref{fig:repair-escalation} arranges these options as a decision flow keyed on whether a safe revert is possible and whether mined migration examples are available. We describe five families in the paragraphs below: automated rollback, LLM-based automated program repair, output-oriented program synthesis, inferred structural rewrite rules, and semantic patches.

\paragraph{Automated Rollback.}
The simplest response to a detected breakage is to revert the offending change. Henderson et al.'s SafeRevert at Google automatically reverts commits that cause test failures in CI and substantially improves recall on revertible-commit identification compared to the baseline [S84]. Rollback restores build stability but does not fix the underlying breaking change.

\paragraph{LLM-Based Automated Program Repair.}
Three primary studies apply Large Language Models to client code adaptation [S4, S94, S86]. Fruntke and Krinke evaluate LLM-based repair on the BUMP benchmark and find that an agentic feedback loop substantially reduces cost and time for the best-performing model [S4]. Reyes et al.'s Byam and Liu et al.'s Synter target related but distinct settings: Byam fixes broken production code on the BUMP dataset [S94], while Synter augments LLMs with static analysis to repair client test cases instead [S86]. Per-tool figures are reported in Table~\ref{tab:strategy-tooling}. The common gap across these studies is between compilation-level fixes and semantic correctness, which requires a reliable test oracle.

\paragraph{Output-Oriented Program Synthesis.}
APIFIX synthesizes transformation rules from human edit examples paired with ``additional outputs'' mined from new-API usages, which act as counter-examples to prevent overfitting [S21]. The approach achieves high precision and recall on C\# API migration tasks but requires useful additional outputs and lacks context analysis.

\paragraph{Inferred Structural Rewrite Rules.}
Mitchell's CombyInferPy mines ``before/after'' snippets from a library's GitHub pull requests to infer structural search-and-replace templates for client code adaptation [S22]. The approach grounds transformations in the library's own evolution history, avoiding the need for large training datasets, but inferred rules may require generalization to match diverse client usage patterns.

\paragraph{Semantic Patches.}
Semantic patches use a formal language to write transformation rules that are written once and applied across all affected clients [S45, S65]. Tools like JSFIX (JavaScript) and Coccinelle (C/Linux) are highly precise for common structural changes, although a small fraction of patterns cannot be expressed in the template language. JSFIX further integrates interactive static analysis filtering to resolve ambiguities in dynamic languages [S45].

\subsubsection{Refactoring Replay and Migration Tools}

A distinct class of tools, drawn from 5 of the 33 tool-support papers, bridges the gap between detection and resolution by reusing migration information that the library itself already provides. Rather than synthesizing patches from scratch, these tools capture the maintainer's refactoring operations or mine before-and-after migration examples and replay them on client code. The paragraph below covers graph-based adaptation and client adapter generation, which together represent the dominant approaches in this category.

\paragraph{Graph-Based Adaptation and Client Adapter Generation.}
The five studies in this category adopt graph-based and template-based approaches that learn or encode migration rules and replay them on client code [S25, S65, S85, S87, S96]. LIBSYNC mines API usage adaptation patterns from already-migrated client programs, with high precision but a cold-start dependency on training data [S65]. Three follow-on tools reduce this dependency: Meditor infers migration edits from before/after examples [S87], LibCatch encodes compiler-driven migration operators triggered by specific compiler error types [S85], and VSCode Migrate offers IDE-integrated semi-automatic suggestions for projects with low test coverage [S96]. For Web APIs, Schmiedmayer et al.\ generate machine-readable migration guides by diffing OpenAPI specifications, although semantic changes still require manual intervention [S25]. Per-tool figures are reported in Table~\ref{tab:detection-learning}.

\subsubsection{Analysis and Verification}

The final category of automated tools, drawn from 10 of the 33 tool-support papers, supports management decisions through analysis rather than direct repair. We describe three families in the paragraphs below: reachability and confidence scoring, predictive stability analysis, and social and community-based strategies.

\paragraph{Reachability and Confidence Scoring.}
Static change impact analysis combines AST differencing with call-graph and Class Dependency Graph (CDG) analysis to determine whether modified library functions are actually invoked by the client [S71, S76, S77, S78]. Hejderup and Gousios find that reachability analysis is substantially more effective than tests alone at detecting direct and transitive faults [S71]. Confidence scoring extends this signal by augmenting bot-generated PRs with a score based on test coverage of changed functions, making the reliability of a ``green build'' more transparent [S71]. Per-tool figures are reported in Table~\ref{tab:detection-combined}.

\paragraph{Predictive Stability Analysis.}
A second family predicts which interfaces are likely to remain stable in future releases. Businge et al.\ use code metrics (method length, parameter count) and clone history to predict the future stability of internal interfaces [S66] and develop a separate compatibility prediction approach for Eclipse plug-ins based on historical evolution data [S89]. Both approaches are probabilistic with moderate precision.

\paragraph{Social and Community-Based Strategies.}
Social strategies complement tool-based analysis by leveraging community signals [S32, S47, S28, S43]. Discussion through mailing lists and issue trackers allows communities to collectively assess upgrade safety [S47]. Crowd-sourced constraint recommendation analyzes the dependency constraints used by other dependents to bypass unreliable SemVer claims [S32]. Reputation-based selection that relies on personal trust in ``classic'' or ``core'' package authors is effective when technical tools fail but does not scale to large ecosystems [S28, S43].

{\footnotesize
\begin{longtable}{|p{2.2cm}|p{1.0cm}|p{3.0cm}|p{2.3cm}|p{2.3cm}|p{1.4cm}|}
\caption{Summary of automated tool support for communicating, preventing, and recovering from breaking changes.}
\label{tab:strategy-tooling}\\
\hline
\textbf{Strategy} & \textbf{Target} & \textbf{Description \& Mechanism} & \textbf{Rationale} & \textbf{Risks / Limitations} & \textbf{References} \\ \hline
\endfirsthead
\hline
\multicolumn{6}{c}{\tablename\ \thetable{} -- continued from previous page} \\
\textbf{Strategy} & \textbf{Target} & \textbf{Description \& Mechanism} & \textbf{Rationale} & \textbf{Risks / Limitations} & \textbf{References} \\ \hline
\endhead
\hline
\endfoot
\hline
\endlastfoot

Update Direct Dependency & Both & Upgrade the direct parent dependency to a newer version. & Often brings a newer, compatible version of the transitive dependency automatically. & The new version might introduce its own breaking changes. & [S11,S3,S9,S38,\newline S76] \\ \hline

Automated Dependency Bots & Tooling & Use tools like Dependabot, Renovate, Greenkeeper to monitor and auto-update dependencies. & Increases update speed by 1.6x; triggers tests automatically; provides automated PRs. & May trigger build failures; lacks ability to fix code breakages; focuses on single libraries. & [S9,S4,S32,S35,\newline S38,S43,S53,S71,\newline S73,S76,S77,S78] \\ \hline

Multi-Objective Dependency Optimization & Tooling & Model dependency updating as a mathematical optimization problem (LP, SMT Solver) aggregating diverse metrics. & Enables customized balance between conflicting goals; finds global optimum rather than local greedy choice. & Computationally expensive; depends on accuracy of cost/quality metrics. & [S77,S78] \\ \hline

Static Analysis \& Linting & Both & Use linters, API tools (Revapi, japicmp, \texttt{rust-semverver}), or LSP to catch errors before release. & Prevents semantically wrong code; enforces compatibility policies via deep API surface analysis. & Cannot verify dynamic issues; often limited to shallow syntactic checking. & [S56,S9,S4,S35,\newline S43,S46,S52,S54,\newline S71,S73,S76,S77,\newline S78] \\ \hline

Static Change Impact Analysis (Reachability) & Tooling & Use AST differencing combined with call-graph and CDG reachability to determine if modified functions are invoked by client. & More effective than tests (74\% vs.\ 47\% detection rate); reduces false negatives. & Prone to false positives from over-approximation; misses reflection and dynamic class loading. & [S71,S76,\newline S77,S78] \\ \hline

Confidence Scoring / Reliability Metrics & Tooling & Augment automated PRs with a confidence score based on test coverage of changed dependency functions. & Increases transparency; helps developers assess the reliability of ``green build'' signals. & Requires sophisticated tooling integration; metrics may be misleading with shallow tests. & [S71] \\ \hline

LLM-Based Automated Program Repair & Tooling & Use LLMs to automatically generate code patches adapting client code to new dependency versions. & Automates manual adaptation effort; handles API usage updates. & Susceptible to hallucinations; performance is model-dependent. & [S4,S94,S86] \\ \hline

Iterative Feedback \& Refinement (Agentic) & Tooling & Employ an agentic loop that feeds compiler/test errors back to the LLM to refine patches. & Reduces cost by 57\% and time by 67\% for best model; enables self-debugging. & Higher token cost for weaker models; requires reliable test suite; results are model-dependent. & [S4] \\ \hline

Output-Oriented Program Synthesis & Tooling & Synthesize transformation rules using human edit examples and ``additional outputs'' (mined new-API usages). & Reduces overfitting via counter-examples; achieves 98.7\% precision and 91.5\% recall. & Requires useful additional outputs; lacks context analysis. & [S21] \\ \hline

Automated Client Adapter Generation & Tooling & Compare API versions and generate a machine-readable migration guide driving creation of a self-updating client library. & Automates migration of syntactic changes; provides default/fallback values. & Semantic changes still require manual intervention; relies on accurate API specifications. & [S25] \\ \hline

Refactoring Replay (Recording) & Tooling & Library developers record refactoring operations (renames, moves) as ``edit scripts'' that clients can automatically apply. & Produces correct, unambiguous traces for supported refactorings; reduces client effort to zero. & Requires upstream adoption; only supports a subset of refactoring types; cannot capture logic/semantic changes. & [S52,S65] \\ \hline

Inferred Structural Rewrite Rules & Tooling & Mine ``before/after'' code snippets from the library's internal PRs to infer structural search-and-replace templates. & Infers rules from library source code; avoids need for large training datasets. & Inferred rules may be redundant or require generalization to match diverse usages. & [S22] \\ \hline

Semantic Patches (Template-Based Transformation) & Tooling & Use a formal language (e.g., JSFIX, Coccinelle) to write detection patterns and code templates for automated client transformation. & Amortizes repair effort (write once, fix many); highly precise for common structural changes. & Cannot express complex behavioral changes ($\sim$2\% of patterns); requires patch collection maintenance. & [S45,S65] \\ \hline

Graph-Based Adaptation Recommendation & Tooling & Use graph mining (e.g., LIBSYNC) to learn API usage adaptation patterns from migrated code. & Handles complex adaptations involving multiple objects and control structures; 100\% precision. & Requires cold-start set of already migrated programs; depends on graph alignment thresholds. & [S65] \\ \hline

Dependency Graph Partitioning & Tooling & Divide the dependency graph into smaller independent subgraphs (vertical/horizontal) to solve the update problem incrementally. & Reduces exponential complexity of global optimization; makes SMT solving scalable. & Potential to miss global optimal solutions across partition boundaries. & [S78] \\ \hline

Backtracking Dependency Resolution & Tooling & Implement mechanisms to revert version decisions when a dead end is encountered during graph traversal. & Avoids local optima; ensures constraints are satisfied globally. & Increases computational time; requires state management of visited nodes. & [S78] \\ \hline

Social Negotiation / Reputation-Based Selection & Both & Rely on personal trust in ``classic'' or ``core'' package authors rather than technical metrics. & Effective when tools fail or standards are inconsistent; builds socio-technical congruence. & Not scalable for large ecosystems; relies heavily on personal networks. & [S28,S43] \\ \hline

Discussion \& Consensus & Ecosystem & Engage in mailing list or issue tracker discussions to collectively decide on upgrade safety. & Leverages community knowledge to identify risks; facilitates shared decision-making. & Time-consuming; relies on active community participation. & [S47] \\ \hline

Predictive Interface Stability Analysis & Tooling & Use code metrics and clone history to predict future stability of specific internal interfaces. & Helps estimate promotion effort and risk of depending on non-APIs (Precision > 56\%). & Probabilistic; interface age is found to be an insignificant predictor. & [S66] \\
\hline
\end{longtable}
}

\begin{tcolorbox}[colback=yellow!10!white, colframe=yellow!75!black, title=\textbf{Key Findings: Automated Tool Support}]
\begin{enumerate}
    \item \textbf{Automated tools fall along four themes}: Across the surveyed studies, tooling support covers dependency resolution and optimization, automated code repair, refactoring replay and migration, and analysis and verification (see Table~\ref{tab:strategy-tooling}).
    \item \textbf{A passing dependency-bot build does not imply compatibility}: Across multiple ecosystem studies, bots accelerate updates but rely on test suites that fail to exercise most direct and transitive dependency calls, so a ``green build'' routinely co-exists with undetected breakage.
    \item \textbf{Automated repair tools achieve syntactic success but stall at semantic correctness}: Across the surveyed LLM-based, synthesis-based, and rule-based repair tools, every approach can fix a substantial share of compilation errors, yet none reliably distinguishes a build that compiles from one that preserves the original program semantics, since no automated test oracle is available.
\end{enumerate}
\end{tcolorbox}

\section{Challenges and Opportunities}
\label{sec:challenges}

In this section, we synthesize the open challenges and research opportunities that emerge from cross-cutting themes across the four research questions. The remainder of this section is organized into two subsections. First, we discuss three challenges (Section~\ref{sec:challenges-list}): Behavioral Breaking Change Detection at Scale, the Systematic Failure of Semantic Versioning as a Trust Mechanism, and Transitive Dependency Propagation and Information Asymmetry. Second, we discuss three opportunities (Section~\ref{sec:opportunities}): LLM-Augmented Behavioral Contract Inference and Automated Repair, Ecosystem-Level Dependency Graph Intelligence, and Domain-Specific Breaking Change Management for ML and Data Science.

\subsection{Challenges}
\label{sec:challenges-list}

In the previous section, we summarize 66 strategies for communicating, preventing, and recovering from breaking changes. In this subsection, we synthesize three challenges that remain unresolved across the four research questions. \textbf{Behavioral Breaking Change Detection at Scale} addresses the persistent gap between detection capabilities for syntactic and behavioral breaks. \textbf{Systematic Failure of Semantic Versioning as a Trust Mechanism} concerns the empirical evidence that version numbers are unreliable signals of compatibility. \textbf{Transitive Dependency Propagation and Information Asymmetry} concerns breakage that arrives from libraries the downstream consumer never declared. It also concerns the diagnostic information gap that downstream consumers face when breakage occurs. We discuss each in the following subsubsections.

\subsubsection{Behavioral Breaking Change Detection at Scale}
\label{sec:challenge-detection}

As discussed in RQ3, static analysis tools for syntactic breaks in statically-typed ecosystems have reached high maturity. Sigtest achieves a 98.7\% detection rate, and Maracas achieves 96.3\% precision and 98.5\% recall [S64, S6]. However, behavioral breaking changes, which account for 68.1\% of npm breaks [S1], remain difficult to detect automatically. The root cause is theoretical. Determining whether two program versions are semantically equivalent is undecidable (Rice's theorem). Static tools cannot detect logic drift, implicit calculation errors, or exception behavior changes without executing code. Dynamic tools can achieve higher precision on specific cases. For example, GILESI detects 96\% of seeded behavioral breaks [S75]. However, dynamic tools depend entirely on test quality. Only 58\% of direct dependencies and 21\% of transitive dependencies have any test coverage [S71].

This detection gap is compounded by the \textit{oracle problem}. Given an observed behavioral difference between two versions, it is difficult to determine algorithmically whether the difference represents a breaking change or an intentional improvement. Jayasuriya et al.\ find that 59.35\% of behavioral breaks manifest as test errors (exceptions) that automated tools can detect. The remaining 40.65\% are test failures (assertion mismatches) that produce incorrect results without crashing [S12]. The problem is more pronounced in ML libraries. Montandon et al.\ find that 75\% of Scikit-Learn default parameter changes produce behavioral effects that change the function's behavior. The accuracy differences reach up to 77 percentage points on benchmark datasets, without triggering any exception [S26]. Without formal or semi-formal specifications, no algorithm can distinguish a bug fix from a regression. The ``Closet Contract Conjecture'' [S73] observes that most real-world code lacks such specifications. Contract-based random testing tools (QuickCheck, Randoop) can verify compliance with user-defined contracts. They cannot help when no explicit properties exist [S12]. As a result, the dominant category of breaking changes remains partially undetectable by current tools.

\subsubsection{Systematic Failure of Semantic Versioning as a Trust Mechanism}
\label{sec:challenge-semver}

Semantic versioning (SemVer) is the central social contract for communicating breaking changes across ecosystems. However, the empirical evidence shows systematic and persistent violations. In Maven, 35.7\% of minor releases and 23.8\% of patch releases contain at least one breaking change [S31]. In npm, approximately 95\% (61 out of 64) of manifesting breaking changes appear in non-major releases (minor, patch, or pre-release) rather than major ones [S9]. In both ecosystems, non-major releases routinely contain breaking changes that the version number does not advertise.

SemVer violations have downstream consequences beyond individual breakage events. When clients cannot rely on version numbers, the rational response is to delay upgrades. Bavota et al.\ find that 69\% of new Apache releases are ignored by direct dependents [S47]. Hora et al.\ find that the median reaction time to API changes is 34 days. API changes related to method improvements take longer (median 47 days) than those related to method replacements (median 20 days) [S48]. These data suggest that upgrade reluctance is a response to unreliable version signals.

The root cause is partly cognitive. RQ2 finds that the dominant reasons for breaking changes are internal maintenance (API simplification at 29\% and maintainability improvement at 24\% in Java [S8]) and API design improvement (61.8\% in npm [S1]). Maintainers cannot reliably identify all breaking changes in a release. Dietrich et al.\ show that experienced Java developers correctly predict binary compatibility only approximately 60\% of the time. The rate drops to 27\% for source/binary asymmetry cases such as return type specialization [S57]. Automated SemVer tools (elm bump, rust-semverver, RevAPI) can detect syntactic breaks reliably but miss behavioral changes, which may produce a false sense of compliance. The rate of SemVer violations has decreased over time (from 67.7\% in 2005 to 16.0\% in 2018 [S6]). However, the residual rate remains high enough to affect upgrade decisions across ecosystems.

\subsubsection{Transitive Dependency Propagation and Information Asymmetry}
\label{sec:challenge-transitive}

Even clients who carefully manage their direct dependencies are exposed to breaking changes through transitive propagation. In npm, 57.8\% of manifesting breaking changes originate from indirect providers [S9]. In Maven, transitive dependency changes are the single most common source of client-impacting source code breaking changes, accounting for 20.36\% of all cases [S11]. The structural difficulty is that 61.24\% of clients call functionality from transitive dependencies without declaring them [S11]. A client may carefully pin direct dependencies, monitor changelogs, and follow SemVer conventions. It can still be broken by a patch release from a library two hops away in the dependency graph. Only 21\% of transitive dependencies have any test coverage [S71]. UPPDATERA's static reachability analysis detects 64\% of transitive faults (compared to 35\% by tests alone) [S71]. Over one-third of transitive breaks remain undetected.

This propagation problem is amplified by a structural information asymmetry between maintainers and consumers. Maintainers know what changes and why. Consumers often cannot determine either. Only 5.4\% of Maven artifacts use deprecation tags, and 33.03\% of public methods are deleted without prior deprecation [S31]. When breakage occurs, 46.51\% of precondition-violation exceptions lack meaningful error messages. Nearly all such exceptions appear deep in the call stack, far from the dependency boundary where the actual change occurs [S12]. As noted in the previous subsubsection, even experienced developers correctly predict binary compatibility only approximately 60\% of the time [S57]. This information asymmetry compounds the previous two challenges. Behavioral breaks that are difficult to detect (Section~\ref{sec:challenge-detection}) are also difficult to diagnose. SemVer violations (Section~\ref{sec:challenge-semver}) that are undocumented leave no trace for clients to act on. Transitive breaks from undeclared dependencies are invisible to the client until they manifest.

\subsection{Opportunities}
\label{sec:opportunities}

In this subsection, we discuss three research opportunities that emerge from the gaps identified above. \textbf{LLM-Augmented Behavioral Contract Inference and Automated Repair} addresses the oracle problem from the first challenge. It uses large language models to infer lightweight specifications and to connect detection output to repair input. \textbf{Ecosystem-Level Dependency Graph Intelligence} addresses the transitive propagation problem from the third challenge. It treats the dependency graph itself as a source of compatibility evidence. \textbf{Domain-Specific Breaking Change Management for ML and Data Science} addresses the silent behavioral drift that current general-purpose tools cannot detect. We discuss each in the following subsubsections.

\subsubsection{LLM-Augmented Behavioral Contract Inference and Automated Repair}
\label{sec:opp-llm}

Section~\ref{sec:challenge-detection} establishes that behavioral breaking changes are difficult to detect without specifications, and the oracle problem cannot be solved by better tools alone. The missing element is formal or semi-formal behavioral contracts. Large language models (LLMs) offer a potential path to bridge this gap. They serve not as direct detectors of breaking changes, but as contract inference engines whose output can be verified by existing tools. The argument rests on three capabilities required by the breaking change scenario. Each capability has quantitative evidence from a related task. The first capability is translating informal natural language (docstrings, commit messages, release notes) into formally checkable assertions. Endres et al.\ study this problem under the name nl2postcond. They find that LLM-generated postconditions distinguish 70 buggy methods, covering 64 real-world historical bugs in Defects4J, when used as runtime assertions~\cite{endres2024nl2postcond}. The second capability is recovering candidate oracles from code structure when documentation is sparse. Sparse documentation is the prevailing condition in real-world code [S73]. Dinella et al.'s TOGA achieves 96\% accuracy on a held-out assertion-inference test set. It also detects 57 bugs in Defects4J, of which 30 are not found by any other automated testing method evaluated~\cite{dinella2022toga}. TOGA uses a specialized transformer that predates current general-purpose LLMs. Still, its result establishes that learned models can recover usable oracles from code structure alone. The third capability is editing across multiple functions, classes, and files in an unfamiliar codebase given only a textual description of the change. Jimenez et al.\ introduce SWE-bench, a benchmark of 2,294 real GitHub issues from 12 Python projects. The reference solutions edit on average 1.7 files, 3.0 functions, and 32.8 lines per task~\cite{jimenez2024swebench}. On SWE-bench Verified, GPT-4o resolves 33.2\% of instances~\cite{openai2024swebenchverified}. SWE-bench Verified is a 500-instance subset of SWE-bench that OpenAI released in 2024, after human annotators removed underspecified problems and unfair unit tests. Repository-level adaptation given a textual description is the same operation that a client performs when absorbing a breaking change. Contracts inferred from documentation and method context can then serve as lightweight specifications for cross-version verification by existing contract checking tools (QuickCheck, Randoop, GILESI). When dynamic tools detect a behavioral difference between versions, LLMs can classify it as a breaking change, a bug fix, or an intentional improvement. The classification leverages code intent recovered from documentation and commit messages. This partially addresses the oracle problem.

Several findings from the surveyed literature support this direction. LLM-based repair has been explored by multiple groups, with results that show meaningful code comprehension. Examples include Fruntke and Krinke [S4], Reyes et al.\ with Byam [S94], and Liu et al.\ with Synter for test repair [S86]. SENSOR's context-driven approach shows that mining client source code provides rich signal for understanding expected behavior [S58]. Test evolution analysis can identify 90.2\% of behavioral changes from library-internal test updates [S42]. Code contracts (JML, \texttt{@NonNull}, \texttt{@Pure}) are already supported by existing verification infrastructure [S73]. The bottleneck is contract generation, not verification. Beyond contract inference, connecting detection output directly to repair input would reduce the manual effort required to absorb breaking changes. Current detection tools identify what breaks (RQ3), and current repair tools can sometimes fix it (RQ4). However, the two capabilities exist in isolation. A detection-informed repair system would combine three inputs: breaking change type classification (from RQ1), structured detection output (which API changes and how), and examples of how other clients adapt to the same change. Such a system would transform repair from open-ended code generation into constrained adaptation. APIFIX's output-oriented program synthesis achieves 98.7\% precision and 91.5\% recall when given appropriate examples [S21]. LIBSYNC's graph-based adaptation achieves 100\% precision and 91\% recall [S65]. The BUMP benchmark provides 571 reproducible breaking dependency updates for standardized evaluation [S69].

\subsubsection{Ecosystem-Level Dependency Graph Intelligence}
\label{sec:opp-graph}

Section~\ref{sec:challenge-transitive} establishes that transitive dependency propagation is the largest single source of client-impacting breaks. However, current strategies operate at the individual library or client level. The dependency graph itself is an underutilized data source for predicting, preventing, and diagnosing transitive breakage. Before releasing a new version, ecosystem-level tools could analyze the complete transitive closure of a library's dependents. The analysis would predict which clients will be affected and through which paths. UPCY's graph-based minimal-cut approach reduces incompatibilities in 41.1\% of cases [S38], demonstrating feasibility at moderate scale. Lu et al.\ show that holistic multi-library updates using topological sorting and graph algorithms minimize total incompatibility compared to naive single-library updates [S76]. Extending these approaches to ecosystem-level services would enable clients to compute optimal update paths that simultaneously satisfy all compatibility constraints.

Combining dependency graph analysis with breaking change detection (from RQ3 tools) could enable real-time propagation tracking. The tracking would show how a breaking change at depth 3 in the graph reaches a client at depth 0, including the specific API call chains that transmit the incompatibility. Additionally, analyzing the dependency constraints used by other dependents can provide empirical evidence of upgrade safety. Decan and Mens show that observing other packages' dependency constraints provides useful compatibility information [S32]. At ecosystem scale, the cumulative record of successful upgrades by other clients constitutes empirical evidence about compatibility that complements the SemVer declaration. Hejderup and Gousios's confidence scoring for bot-generated pull requests adds test coverage metrics [S71]. This represents an initial step toward graph-aware upgrade recommendations. Automated SemVer enforcement can complement graph intelligence. The approach would shift from optional compliance checking to mandatory CI/CD gate enforcement. Version validation would be embedded into build pipelines as a blocking check. ROSEAU achieves F1 = 0.99 and runs two orders of magnitude faster than compilation-based tools [S29]. This shows that enforcement need not be a performance bottleneck. Elm and Rust's built-in version tools have already achieved widespread adoption [S12]. This indicates that developers accept automated version computation when it is integrated into their workflow.

\subsubsection{Domain-Specific Breaking Change Management for ML and Data Science}
\label{sec:opp-ml}

The surveyed literature reveals that ML and data science libraries exhibit a distinct breaking change profile that general-purpose tools do not address. Default parameter changes in Scikit-Learn can shift model accuracy by up to 77 percentage points without triggering any exception [S26]. 35\% of Scikit-Learn clients and 21\% of Pandas clients are affected by such silent behavioral drift [S26]. No existing detection tool specifically targets this domain. The oracle problem (Section~\ref{sec:challenge-detection}) is most acute here because ``correct'' model behavior lacks a fixed specification.

However, the ML domain has unique properties that enable specialized solutions. In general-purpose software, behavioral correctness is often underspecified. ML pipelines have natural oracles: model performance metrics (accuracy, loss, F1 score). A breaking change detector for ML libraries could automatically run standard benchmark suites before and after a library upgrade. It would then flag statistically significant performance regressions as potential behavioral breaks. This bypasses the general oracle problem by leveraging domain-specific evaluation criteria. Default parameter tracking is another well-defined opportunity. A specialized tool could extract all default parameter values from each library version, compare them across versions, cross-reference with client code to identify affected call sites that rely on defaults, and generate explicit warnings with old and new default values. Montandon et al.\ demonstrate that default parameter changes are the primary mechanism of silent behavioral drift in data science libraries [S26]. This is a bounded, well-defined problem with high practical impact. ML reproducibility is a first-class requirement for regulatory compliance, scientific reproducibility, and model governance. A domain-specific dependency management tool could enforce complete environment specifications (library versions, default parameters, random seeds, hardware configuration) as machine-readable artifacts. The ML community has already built experiment tracking infrastructure (MLflow, Weights \& Biases, Neptune). Integrating breaking change detection into these platforms would flag when a library upgrade coincides with unexplained metric drift. This approach leverages existing developer workflows rather than requiring new tool adoption. AexPy provides a foundation for Python-specific detection with 86.9\% recall and 93.5\% precision [S23]. Extending it with ML-specific heuristics is architecturally feasible. The BUMP benchmark focuses on Java [S69]. Creating an equivalent benchmark for Python ML libraries would catalyze research in this area.

\section{Threats to Validity}
\label{sec:threats}

We discuss the threats to the validity of this systematic literature review along four common types: construct validity, internal validity, external validity, and conclusion validity.

\textit{Construct validity} concerns whether our search and selection process collects the right body of evidence. The main constructs of this survey are ``breaking change'' and the related concepts of backward incompatibility and API incompatibility. A narrow or imprecise search string could omit relevant studies. To mitigate this threat, we build the search string from 14 terminological variants that span different research communities. The variants include ``Breaking Change'', ``Backward Incompatible Change'', ``API Incompatibility'', ``Breaking Update'', and ``Breaking Dependency'' (Section~\ref{sec:methodology}). We then apply the search string to the title, abstract, and keyword fields of three major databases (IEEE Xplore, ACM Digital Library, and Springer) that cover the main software engineering venues. Keyword search alone may miss relevant studies. To reduce this risk, we perform both backward and forward snowballing on the 50 seed papers, following Wohlin's guidelines~\cite{Wohlin2014-snowballing}. The snowballing phase adds 47 papers, which suggests that database search alone would under-cover the topic. We also relax the 2010 cutoff during snowballing. This allows us to recover influential earlier studies that are heavily cited by papers within our search window.

\textit{Internal validity} refers to factors that may influence the analysis of the extracted data. The screening, data extraction, and coding stages all involve subjective judgment that could introduce bias. To mitigate this threat, we define the inclusion and exclusion criteria (I1--I2 and E1--E5) before screening. We then apply them consistently to all candidate papers assessed against the inclusion and exclusion criteria, both the 55 papers that proceed to full-text reading from database search and the additional candidates surfaced through snowballing. In the first-stage screening, the first author uses a large language model (Claude Opus 4.7) to assess each candidate against the exclusion criteria. The model may misclassify papers, for example by excluding a relevant study. To mitigate this threat, the first author manually reviews every assessment and makes all final decisions. The model is not used in the full-text stage, and the snowballing phase further recovers relevant studies that the first-stage screening may miss. We also define a fixed data extraction schema of ten items (D1--D10). All 97 selected studies are recorded against this schema before further analysis (Table~\ref{tab:data_items}). The first two authors carry out the extraction and coding iteratively. Disagreements are resolved through discussion until consensus is reached. For taxonomy construction (RQ1), we follow the open card sorting method~\cite{Spencer2009-cardsorting} in three rounds. This method makes the dimensions and subcategories traceable to specific terms in the original papers. For reasons, impacts, detection methods, and communication, prevention, and recovery strategies (RQ2--RQ4), we follow the thematic analysis guidelines of Cruzes and Dyba~\cite{cruzes2011recommended}. The analysis uses three explicit phases: initial coding, theme generation, and synthesis. To further reduce extraction bias, we cross-check every quantitative claim and tool attribution against the source PDF before including it in the survey.

\textit{External validity} refers to the degree to which the findings can be generalized. Our review selects 97 primary studies in total. 50 come from database search across IEEE Xplore, ACM Digital Library, and Springer, restricted to papers published between 2010 and 2026. The other 47 are added through backward and forward snowballing. Studies indexed only in other databases or published before 2010 without being cited within our search window may be missed. We choose 2010 as the start year because modern package ecosystems such as npm and Maven Central reach widespread adoption around that time. Most breaking change research has appeared since then. The snowballing phase partly compensates for this restriction by recovering pre-2010 work that is heavily cited within our search window. Our findings cover five software ecosystems (Maven/Java, npm/JavaScript, Python, Web APIs, and Linux distributions). Other ecosystems such as Go, Rust, and .NET are underrepresented in the selected studies, so the quantitative results may not generalize to them. The empirical numbers we report are also unevenly distributed across ecosystems. Most quantitative findings come from the Maven/Java and npm/JavaScript literature. Findings for Python, Web APIs, and Linux distributions rely on a smaller number of studies. We acknowledge these limits in scope when reporting cross-ecosystem comparisons.

\textit{Conclusion validity} refers to whether the study would yield similar results if other researchers replicate it. The taxonomy construction and the synthesis of reasons, impacts, detection techniques, and communication, prevention, and recovery strategies involve qualitative judgment. Other reviewers might group the same evidence differently. To mitigate this threat, we make the data collection and analysis process explicit in Section~\ref{sec:methodology}, so that each step can be replicated. We also list all 97 primary studies in Appendix~\ref{appendix:primary-studies}. We use [S\#\#] citations throughout the survey, so that every claim can be traced back to a specific source paper. For each subcategory in the taxonomy, we report the number of supporting studies. We rely on descriptive statistics and direct quotes from the cited papers rather than on inferential claims. This allows readers to verify our interpretations against the original studies.

\section{Conclusion}
\label{sec:conclusion}

Modern software ecosystems make extensive use of reusable libraries. Breaking changes propagate through dependency graphs and remain a persistent source of failure for downstream consumers. In this systematic literature review, we survey 97 primary studies on breaking changes. The survey covers five software ecosystems: Maven/Java, npm/JavaScript, Python, Web APIs, and Linux distributions.

Our analysis produces four main results. For RQ1, we propose a four-dimensional taxonomy of breaking changes along Nature, Detectability, Scope, and Visibility. The taxonomy integrates the heterogeneous categorizations used by individual studies. For RQ2, we identify five reason categories and five impact dimensions. Across the Java and npm studies that quantify maintainer-stated reasons, maintenance and design improvements account for a larger share of breaking changes than new feature work. For RQ3, we survey 43 detection approaches. Static analysis tools have reached high accuracy on syntactic breaks in the Java/Maven ecosystem, while behavioral break detection remains an open problem because of the undecidability of semantic equivalence. For RQ4, we synthesize 66 strategies for communicating, preventing, and recovering from breaking changes. We organize them by the role of the actor that applies them: library maintainers, client developers, and automated tooling.

The synthesis also reveals three open challenges and three research opportunities. The challenges are behavioral break detection at scale, the systematic failure of semantic versioning as a trust mechanism, and transitive dependency propagation combined with information asymmetry. The opportunities are LLM-augmented behavioral contract inference and automated repair, ecosystem-level dependency graph intelligence, and domain-specific breaking change management for ML and data science. These directions point to future research in three areas. The first is closing the gap between syntactic and behavioral detection. The second is treating the dependency graph itself as a source of compatibility evidence. The third is building specialized tooling for application domains where silent behavioral drift produces the largest measurable impact. We believe that this work will help researchers and practitioners reason about breaking changes more systematically across ecosystems.

\appendix
\section{Selected Studies}
\label{appendix:primary-studies}

\noindent[S1] Kong, D., Liu, J., Bao, L., and Lo, D. Toward Better Comprehension of Breaking Changes in the NPM Ecosystem. \textit{ACM Transactions on Software Engineering and Methodology}, 34(4):111, 2025.

\noindent[S2] Jayasuriya, D., Terragni, V., Dietrich, J., Ou, S., and Blincoe, K. Understanding Breaking Changes in the Wild. In \textit{Proceedings of the 32nd ACM SIGSOFT International Symposium on Software Testing and Analysis (ISSTA)}, pp.~1433--1444, 2023.

\noindent[S3] Bogart, C., Kästner, C., Herbsleb, J., and Thung, F. When and How to Make Breaking Changes: Policies and Practices in 18 Open Source Software Ecosystems. \textit{ACM Transactions on Software Engineering and Methodology}, 30(4):42, 2021.

\noindent[S4] Fruntke, L., and Krinke, J. Automatically Fixing Dependency Breaking Changes. \textit{Proceedings of the ACM on Software Engineering}, 2(FSE):96, 2025.

\noindent[S5] Ochoa Venegas, L.M. Break the Code? Breaking Changes and Their Impact on Software Evolution. \textit{PhD thesis, Eindhoven University of Technology}, 2023.

\noindent[S6] Ochoa, L., Degueule, T., Falleri, J.-R., and Vinju, J. Breaking Bad? Semantic Versioning and Impact of Breaking Changes in Maven Central. \textit{Empirical Software Engineering}, 27(3):61, 2022.

\noindent[S7] Ochoa, L., Degueule, T., and Falleri, J.-R. BreakBot: Analyzing the Impact of Breaking Changes to Assist Library Evolution. In \textit{Proceedings of the 44th International Conference on Software Engineering: New Ideas and Emerging Results (ICSE-NIER)}, pp.~26--30, 2022.

\noindent[S8] Brito, A., Valente, M.T., Xavier, L., and Hora, A. You Broke My Code: Understanding the Motivations for Breaking Changes in APIs. \textit{Empirical Software Engineering}, 25(2):1458--1492, 2020.

\noindent[S9] Venturini, D., Cogo, F.R., Polato, I., Gerosa, M.A., and Wiese, I.S. I Depended on You and You Broke Me: An Empirical Study of Manifesting Breaking Changes in Client Packages. \textit{ACM Transactions on Software Engineering and Methodology}, 32(4):94, 2023.

\noindent[S10] Jayasuriya, D. Breaking Changes and their Effect on Client Projects. \textit{PhD thesis, University of Auckland}, 2025.

\noindent[S11] Jayasuriya, D., Ou, S., Hegde, S., Terragni, V., Dietrich, J., and Blincoe, K. An Extended Study of Syntactic Breaking Changes in the Wild. \textit{Empirical Software Engineering}, 30(2):42, 2025.

\noindent[S12] Jayasuriya, D., Terragni, V., Dietrich, J., and Blincoe, K. Understanding the Impact of APIs Behavioral Breaking Changes on Client Applications. \textit{Proceedings of the ACM on Software Engineering}, 1(FSE):56, 2024.

\noindent[S13] Xavier, L., Brito, A., Hora, A., and Valente, M.T. Historical and Impact Analysis of API Breaking Changes: A Large-Scale Study. In \textit{Proceedings of the 24th IEEE International Conference on Software Analysis, Evolution and Reengineering (SANER)}, pp.~138--147, 2017.

\noindent[S14] Brito, A., Xavier, L., Hora, A., and Valente, M.T. APIDiff: Detecting API Breaking Changes. In \textit{Proceedings of the 25th IEEE International Conference on Software Analysis, Evolution and Reengineering (SANER)}, pp.~507--511, 2018.

\noindent[S15] Raemaekers, S., van Deursen, A., and Visser, J. Semantic Versioning and Impact of Breaking Changes in the Maven Repository. \textit{Journal of Systems and Software}, 129:140--158, 2017.

\noindent[S16] Keshani, M., Vos, S., and Proksch, S. On the Relation of Method Popularity to Breaking Changes in the Maven Ecosystem. \textit{Journal of Systems and Software}, 203:111738, 2023.

\noindent[S17] Møller, A., and Torp, M.T. Model-Based Testing of Breaking Changes in Node.js Libraries. In \textit{Proceedings of the ACM Joint Meeting on European Software Engineering Conference and Symposium on the Foundations of Software Engineering (ESEC/FSE)}, pp.~409--419, 2019.

\noindent[S18] Reyes, F., Baudry, B., and Monperrus, M. Breaking-Good: Explaining Breaking Dependency Updates with Build Analysis. In \textit{Proceedings of the IEEE International Conference on Source Code Analysis and Manipulation (SCAM)}, pp.~36--46, 2024.

\noindent[S19] Sharma, V., and Lam, P. Detecting Exception-Related Behavioural Breaking Changes with UnCheckGuard. In \textit{Proceedings of the IEEE International Conference on Source Code Analysis and Manipulation (SCAM)}, pp.~1--12, 2025.

\noindent[S20] Møller, A., Nielsen, B.B., and Torp, M.T. Detecting Locations in JavaScript Programs Affected by Breaking Library Changes. \textit{Proceedings of the ACM on Programming Languages}, 4(OOPSLA):187, 2020.

\noindent[S21] Gao, X., Radhakrishna, A., Soares, G., Shariffdeen, R., Gulwani, S., and Roychoudhury, A. APIfix: Output-Oriented Program Synthesis for Combating Breaking Changes in Libraries. \textit{Proceedings of the ACM on Programming Languages}, 5(OOPSLA), Article 161, 2021.

\noindent[S22] Mitchell, H. Automatically Fixing Breaking Changes of Data Science Libraries. In \textit{Proceedings of the 37th IEEE/ACM International Conference on Automated Software Engineering (ASE)}, Article 193, pp.~1--3, 2022.

\noindent[S23] Du, X., and Ma, J. AexPy: Detecting API Breaking Changes in Python Packages. In \textit{Proceedings of the 33rd IEEE International Symposium on Software Reliability Engineering (ISSRE)}, pp.~470--481, 2022.

\noindent[S24] Kong, D., Liu, J., Ni, C., Lo, D., and Bao, L. More Effective JavaScript Breaking Change Detection via Dynamic Object Relation Graph. In \textit{Proceedings of the ACM on Software Engineering}, 2(ISSTA), Article ISSTA103, 2025.

\noindent[S25] Schmiedmayer, P., Bauer, A., and Bruegge, B. Reducing the Impact of Breaking Changes to Web Service Clients During Web API Evolution. In \textit{Proceedings of the 10th IEEE/ACM International Conference on Mobile Software Engineering and Systems (MOBILESoft)}, pp.~1--11, 2023.

\noindent[S26] Montandon, J.E., Silva, L.L., Politowski, C., Prates, D., Bonifácio, A., and El-Boussaidi, G. Unboxing Default Argument Breaking Changes in 1~+~2 Data Science Libraries. \textit{Journal of Systems and Software}, 229:112460, 2025.

\noindent[S27] Mezzetti, G., Møller, A., and Torp, M.T. Type Regression Testing to Detect Breaking Changes in Node.js Libraries. In \textit{Proceedings of the 32nd European Conference on Object-Oriented Programming (ECOOP)}, Article 7, pp.~7:1--7:24, 2018.

\noindent[S28] Bogart, C., Kästner, C., and Herbsleb, J. When It Breaks, It Breaks: How Ecosystem Developers Reason about the Stability of Dependencies. In \textit{Proceedings of the 30th IEEE/ACM International Conference on Automated Software Engineering Workshops (ASEW) -- New Ideas Track}, pp.~86--89, 2015.

\noindent[S29] Latappy, C., Degueule, T., Falleri, J.-R., Robbes, R., and Ochoa, L. ROSEAU: Fast, Accurate, Source-Based API Breaking Change Analysis in Java. In \textit{Proceedings of the 41st IEEE International Conference on Software Maintenance and Evolution (ICSME)}, pp.~517--528, 2025.

\noindent[S30] Lercher, A., Bauer, C., Macho, C., and Pinzger, M. AutoGuard: Reporting Breaking Changes of REST APIs from Java Spring Boot Source Code. In \textit{Proceedings of the IEEE International Conference on Software Analysis, Evolution and Reengineering (SANER)}, pp.~814--818, 2025.

\noindent[S31] Raemaekers, S., van Deursen, A., and Visser, J. Semantic Versioning versus Breaking Changes: A Study of the Maven Repository. In \textit{Proceedings of the 14th IEEE International Working Conference on Source Code Analysis and Manipulation (SCAM)}, pp.~215--224, 2014.

\noindent[S32] Decan, A., and Mens, T. What Do Package Dependencies Tell Us About Semantic Versioning? \textit{IEEE Transactions on Software Engineering}, 47(6):1226--1240, 2021.

\noindent[S33] Mujahid, S., Abdalkareem, R., Shihab, E., and McIntosh, S. Using Others' Tests to Identify Breaking Updates. In \textit{Proceedings of the 17th International Conference on Mining Software Repositories (MSR)}, pp.~466--476, 2020.

\noindent[S34] Dietrich, J., Jezek, K., and Brada, P. Broken Promises: An Empirical Study into Evolution Problems in Java Programs Caused by Library Upgrades. In \textit{Proceedings of the 2014 Software Evolution Week -- IEEE Conference on Software Maintenance, Reengineering, and Reverse Engineering (CSMR-WCRE)}, pp.~64--73, 2014.

\noindent[S35] Foo, D., Chua, H., Yeo, J., Ang, M.Y., and Sharma, A. Efficient Static Checking of Library Updates. In \textit{Proceedings of the ACM Joint Meeting on European Software Engineering Conference and Symposium on the Foundations of Software Engineering (ESEC/FSE)}, pp.~791--796, 2018.

\noindent[S36] Brito, A., Xavier, L., Hora, A., and Valente, M.T. Why and How Java Developers Break APIs. In \textit{Proceedings of the 25th IEEE International Conference on Software Analysis, Evolution and Reengineering (SANER)}, pp.~255--265, 2018.

\noindent[S37] Jezek, K., Dietrich, J., and Brada, P. How Java APIs Break -- An Empirical Study. \textit{Information and Software Technology}, 65:129--146, 2015.

\noindent[S38] Dann, A., Hermann, B., and Bodden, E. UpCy: Safely Updating Outdated Dependencies. In \textit{Proceedings of the 45th International Conference on Software Engineering (ICSE)}, pp.~233--244, 2023.

\noindent[S39] Chen, L., Hassan, F., Wang, X., and Zhang, L. Taming Behavioral Backward Incompatibilities via Cross-Project Testing and Analysis. In \textit{Proceedings of the 42nd International Conference on Software Engineering (ICSE)}, pp.~112--124, 2020.

\noindent[S40] Ko\c{c}i, R., Franch, X., Jovanovic, P., and Abell\'{o}, A. Classification of Changes in API Evolution. In \textit{Proceedings of the 23rd IEEE International Enterprise Distributed Object Computing Conference (EDOC)}, pp.~243--249, 2019.

\noindent[S41] Zhang, L., Liu, C., Xu, Z., Chen, S., Fan, L., Chen, B., and Liu, Y. Has My Release Disobeyed Semantic Versioning? Static Detection Based on Semantic Differencing. In \textit{Proceedings of the 37th IEEE/ACM International Conference on Automated Software Engineering (ASE)}, Article 51, pp.~51:1--51:12, 2022.

\noindent[S42] Mostafa, S., Rodriguez, R., and Wang, X. Experience Paper: A Study on Behavioral Backward Incompatibilities of Java Software Libraries. In \textit{Proceedings of the 26th ACM SIGSOFT International Symposium on Software Testing and Analysis (ISSTA)}, pp.~215--225, 2017.

\noindent[S43] Bogart, C., Kästner, C., Herbsleb, J., and Thung, F. How to Break an API: Cost Negotiation and Community Values in Three Software Ecosystems. In \textit{Proceedings of the 24th ACM SIGSOFT International Symposium on Foundations of Software Engineering (FSE)}, pp.~109--120, 2016.

\noindent[S44] Wang, J., Li, L., Liu, K., and Cai, H. Exploring How Deprecated Python Library APIs Are (Not) Handled. In \textit{Proceedings of the 28th ACM Joint Meeting on European Software Engineering Conference and Symposium on the Foundations of Software Engineering (ESEC/FSE)}, pp.~233--244, 2020.

\noindent[S45] Nielsen, B.B., Torp, M.N., and Møller, A. Semantic Patches for Adaptation of JavaScript Programs to Evolving Libraries. In \textit{Proceedings of the 43rd International Conference on Software Engineering (ICSE)}, pp.~74--85, 2021.

\noindent[S46] Artho, C., Suzaki, K., Di Cosmo, R., Treinen, R., and Zacchiroli, S. Why Do Software Packages Conflict? In \textit{Proceedings of the 9th IEEE Working Conference on Mining Software Repositories (MSR)}, pp.~141--150, 2012.

\noindent[S47] Bavota, G., Canfora, G., Di Penta, M., Oliveto, R., and Panichella, S. How the Apache Community Upgrades Dependencies: An Evolutionary Study. \textit{Empirical Software Engineering}, 20(5):1275--1317, 2015.

\noindent[S48] Hora, A., Robbes, R., Anquetil, N., Etien, A., Ducasse, S., and Valente, M.T. How Do Developers React to API Evolution? The Pharo Ecosystem Case. In \textit{Proceedings of the 31st IEEE International Conference on Software Maintenance and Evolution (ICSME)}, pp.~251--260, 2015.

\noindent[S49] Harrand, N., Benelallam, A., Soto-Valero, C., Bettega, F., Barais, O., and Baudry, B. API Beauty is in the Eye of the Clients: 2.2 Million Maven Dependencies Reveal the Spectrum of Client-API Usages. \textit{Journal of Systems and Software}, 184:111134, 2022.

\noindent[S50] Cossette, B.E., and Walker, R.J. Seeking the Ground Truth: A Retroactive Study on the Evolution and Migration of Software Libraries. In \textit{Proceedings of the 20th ACM SIGSOFT International Symposium on the Foundations of Software Engineering (FSE)}, Article 55, 2012.

\noindent[S51] Serbout, S., Muñoz Hurtado, D.C., and Pautasso, C. Interactively Exploring API Changes and Versioning Consistency. In \textit{Proceedings of the 2023 IEEE Working Conference on Software Visualization (VISSOFT)}, pp.~28--39, 2023.

\noindent[S52] Lamothe, M., Gu\'{e}h\'{e}neuc, Y.-G., and Shang, W. A Systematic Review of API Evolution Literature. \textit{ACM Computing Surveys}, 54(8):171, 2021.

\noindent[S53] Cogo, F.R., Oliva, G.A., and Hassan, A.E. An Empirical Study of Dependency Downgrades in the npm Ecosystem. \textit{IEEE Transactions on Software Engineering}, 47(11):2457--2470, 2021.

\noindent[S54] He, H., He, R., Gu, H., and Zhou, M. A Large-scale Empirical Study on Java Library Migrations: Prevalence, Trends, and Rationales. In \textit{Proceedings of the 29th ACM Joint Meeting on European Software Engineering Conference and Symposium on the Foundations of Software Engineering (ESEC/FSE)}, pp.~478--490, 2021.

\noindent[S55] Raemaekers, S., van Deursen, A., and Visser, J. Measuring Software Library Stability Through Historical Version Analysis. In \textit{Proceedings of the 28th IEEE International Conference on Software Maintenance (ICSM)}, pp.~378--387, 2012.

\noindent[S56] Dietrich, J., Pearce, D., Stringer, J., Tahir, A., and Blincoe, K. Dependency Versioning in the Wild. In \textit{Proceedings of the 16th International Conference on Mining Software Repositories (MSR)}, pp.~349--359, 2019.

\noindent[S57] Dietrich, J., Jezek, K., and Brada, P. What Java Developers Know About Compatibility, and Why This Matters. \textit{Empirical Software Engineering}, 21(3):1371--1396, 2016.

\noindent[S58] Wang, Y., Wu, R., Wang, C., Wen, M., Liu, Y., Cheung, S.-C., Yu, H., Xu, C., and Zhu, Z. Will Dependency Conflicts Affect My Program's Semantics? \textit{IEEE Transactions on Software Engineering}, 48(7):2295--2316, 2022.

\noindent[S59] Zhu, C., Zhang, M., Wu, X., Xu, X., and Li, Y. Client-Specific Upgrade Compatibility Checking via Knowledge-Guided Discovery. \textit{ACM Transactions on Software Engineering and Methodology}, 32(4):98, 2023.

\noindent[S60] Hou, D., and Yao, X. Exploring the Intent behind API Evolution: A Case Study. In \textit{Proceedings of the 18th Working Conference on Reverse Engineering (WCRE)}, pp.~131--140, 2011.

\noindent[S61] Linares-V\'{a}squez, M., Bavota, G., Bernal-C\'{a}rdenas, C., Di Penta, M., Oliveto, R., and Poshyvanyk, D. API Change and Fault Proneness: A Threat to the Success of Android Apps. In \textit{Proceedings of the 9th Joint Meeting on Foundations of Software Engineering (ESEC/FSE)}, pp.~477--487, 2013.

\noindent[S62] McDonnell, T., Ray, B., and Kim, M. An Empirical Study of API Stability and Adoption in the Android Ecosystem. In \textit{Proceedings of the 29th IEEE International Conference on Software Maintenance (ICSM)}, pp.~70--79, 2013.

\noindent[S63] Wu, W., Khomh, F., Adams, B., Gu\'{e}h\'{e}neuc, Y.-G., and Antoniol, G. An Exploratory Study of API Changes and Usages Based on Apache and Eclipse Ecosystems. \textit{Empirical Software Engineering}, 21(6):2366--2412, 2016.

\noindent[S64] Jezek, K., and Dietrich, J. API Evolution and Compatibility: A Data Corpus and Tool Evaluation. \textit{Journal of Object Technology}, 16(4):2:1--2:23, 2017.

\noindent[S65] Nguyen, H.A., Nguyen, T.T., Wilson Jr., G., Nguyen, A.T., Kim, M., and Nguyen, T.N. A Graph-Based Approach to API Usage Adaptation. In \textit{Proceedings of the ACM International Conference on Object-Oriented Programming Systems Languages and Applications (OOPSLA)}, pp.~302--321, 2010.

\noindent[S66] Businge, J., Kawuma, S., Openja, M., Bainomugisha, E., and Serebrenik, A. How Stable Are Eclipse Application Framework Internal Interfaces? In \textit{Proceedings of the 26th IEEE International Conference on Software Analysis, Evolution and Reengineering (SANER)}, pp.~117--127, 2019.

\noindent[S67] Wu, W., Gu\'{e}h\'{e}neuc, Y.-G., Antoniol, G., and Kim, M. AURA: A Hybrid Approach to Identify Framework Evolution. In \textit{Proceedings of the 32nd ACM/IEEE International Conference on Software Engineering (ICSE)}, pp.~325--334, 2010.

\noindent[S68] Hora, A., Etien, A., Anquetil, N., Ducasse, S., and Valente, M.T. APIEvolutionMiner: Keeping API Evolution Under Control. In \textit{Proceedings of the Software Evolution Week -- IEEE Conference on Software Maintenance, Reengineering, and Reverse Engineering (CSMR-WCRE)}, pp.~420--424, 2014.

\noindent[S69] Reyes, F., Gamage, Y., Skoglund, G., Baudry, B., and Monperrus, M. BUMP: A Benchmark of Reproducible Breaking Dependency Updates. In \textit{Proceedings of the 31st IEEE International Conference on Software Analysis, Evolution and Reengineering (SANER)}, pp.~159--170, 2024.

\noindent[S70] Xu, X., Zhu, C., and Li, Y. CompSuite: A Dataset of Java Library Upgrade Incompatibility Issues. In \textit{Proceedings of the 38th IEEE/ACM International Conference on Automated Software Engineering (ASE)}, pp.~2098--2101, 2023.

\noindent[S71] Hejderup, J., and Gousios, G. Can We Trust Tests to Automate Dependency Updates? A Case Study of Java Projects. \textit{Journal of Systems and Software}, 183:111097, 2022.

\noindent[S72] Elizalde Zapata, R., Kula, R.G., Chinthanet, B., Ishio, T., Matsumoto, K., and Ihara, A. Towards Smoother Library Migrations: A Look at Vulnerable Dependency Migrations at Function Level for npm JavaScript Packages. In \textit{Proceedings of the 34th IEEE International Conference on Software Maintenance and Evolution (ICSME)}, pp.~559--563, 2018.

\noindent[S73] Lam, P., Dietrich, J., and Pearce, D.J. Putting the Semantics into Semantic Versioning. In \textit{Proceedings of the 2020 ACM SIGPLAN International Symposium on New Ideas, New Paradigms, and Reflections on Programming and Software (Onward!)}, pp.~157--179, 2020.

\noindent[S74] Zhang, Z., Zhu, H., Wen, M., Tao, Y., Liu, Y., and Xiong, Y. How Do Python Framework APIs Evolve? An Exploratory Study. In \textit{Proceedings of the 27th IEEE International Conference on Software Analysis, Evolution and Reengineering (SANER)}, pp.~81--92, 2020.

\noindent[S75] Monce, G., Degueule, T., Falleri, J.-R., and Robbes, R. Client-Library Compatibility Testing with API Interaction Snapshots. In \textit{Proceedings of the 41st IEEE International Conference on Software Maintenance and Evolution (ICSME)}, pp.~791--796, 2025.

\noindent[S76] Lu, R., Zhang, L., Li, K., Zhang, M., and Chen, Y. Minimizing Breaking Changes and Redundancy in Mitigating Technical Lag for Java Projects. \textit{arXiv preprint arXiv:2511.06762}, 2025. To appear in \textit{Proceedings of the 47th International Conference on Software Engineering (ICSE)}, 2026.

\noindent[S77] Jaime, D., Poizat, P., El Haddad, J., and Degueule, T. Balancing the Quality and Cost of Updating Dependencies. In \textit{Proceedings of the 39th IEEE/ACM International Conference on Automated Software Engineering (ASE)}, pp.~1834--1845, 2024.

\noindent[S78] Zhang, L., Liu, C., Xu, Z., Chen, S., Fan, L., Zhao, L., Wu, J., and Liu, Y. Compatible Remediation on Vulnerabilities from Third-Party Libraries for Java Projects. In \textit{Proceedings of the 45th International Conference on Software Engineering (ICSE)}, pp.~2540--2552, 2023.

\noindent[S79] Qiu, D., Li, B., and Su, Z. An Empirical Analysis of the Co-evolution of Schema and Code in Database Applications. In \textit{Proceedings of the 9th Joint Meeting on Foundations of Software Engineering (ESEC/FSE)}, pp.~125--135, 2013.

\noindent[S80] Dagenais, B., and Robillard, M.P. Recommending Adaptive Changes for Framework Evolution. In \textit{Proceedings of the 30th International Conference on Software Engineering (ICSE)}, pp.~481--490, 2008.

\noindent[S81] Espinha, T., Zaidman, A., and Gross, H.-G. Web API Growing Pains: Loosely Coupled yet Strongly Tied. \textit{Journal of Systems and Software}, 100:27--43, 2015.

\noindent[S82] Li, W., Wu, F., Fu, C., and Zhou, F. A Large-Scale Empirical Study on Semantic Versioning in Golang Ecosystem. In \textit{Proceedings of the 38th IEEE/ACM International Conference on Automated Software Engineering (ASE)}, pp.~1604--1614, 2023.

\noindent[S83] Wang, Y., Wen, M., Liu, Y., Wang, Y., Li, Z., Wang, C., Yu, H., Cheung, S.-C., Xu, C., and Zhu, Z. Watchman: Monitoring Dependency Conflicts for Python Library Ecosystem. In \textit{Proceedings of the 42nd International Conference on Software Engineering (ICSE)}, pp.~125--135, 2020.

\noindent[S84] Henderson, T.A.D., Kondareddy, A., Azad, S., and Nickell, E. SafeRevert: When Can Breaking Changes be Automatically Reverted? In \textit{Proceedings of the IEEE Conference on Software Testing, Verification and Validation (ICST)}, pp.~395--406, 2024.

\noindent[S85] Zhong, H., and Meng, N. Compiler-Directed Migrating API Callsite of Client Code. In \textit{Proceedings of the IEEE/ACM 46th International Conference on Software Engineering (ICSE)}, pp.~2796--2807, 2024.

\noindent[S86] Liu, J., Yan, J., Xie, Y., Yan, J., and Zhang, J. Fix the Tests: Augmenting LLMs to Repair Test Cases with Static Collector and Neural Reranker. In \textit{Proceedings of the IEEE 35th International Symposium on Software Reliability Engineering (ISSRE)}, pp.~367--378, 2024.

\noindent[S87] Xu, S., Dong, Z., and Meng, N. Meditor: Inference and Application of API Migration Edits. In \textit{Proceedings of the IEEE/ACM 27th International Conference on Program Comprehension (ICPC)}, pp.~335--346, 2019.

\noindent[S88] Opdebeeck, R., Zerouali, A., Vel\'{a}zquez-Rodr\'{i}guez, C., and De Roover, C. Does Infrastructure as Code Adhere to Semantic Versioning? An Analysis of Ansible Role Evolution. In \textit{Proceedings of the IEEE 20th International Working Conference on Source Code Analysis and Manipulation (SCAM)}, pp.~238--248, 2020.

\noindent[S89] Businge, J., Serebrenik, A., and van den Brand, M. Compatibility Prediction of Eclipse Third-Party Plug-ins in New Eclipse Releases. In \textit{Proceedings of the IEEE 12th International Working Conference on Source Code Analysis and Manipulation (SCAM)}, pp.~164--173, 2012.

\noindent[S90] Pinckney, D., Cassano, F., Guha, A., and Bell, J. A Large Scale Analysis of Semantic Versioning in NPM. In \textit{Proceedings of the IEEE/ACM 20th International Conference on Mining Software Repositories (MSR)}, pp.~485--497, 2023.

\noindent[S91] Dorsch, R., Freund, M., and Harth, A. Analyzing Breaking Changes in IoT Systems: A Taxonomy and Empirical Study on System Stability and Longevity. In \textit{Proceedings of the 14th International Conference on the Internet of Things (IoT)}, pp.~194--199, 2024.

\noindent[S92] Javan Jafari, A., Costa, D.E., Shihab, E., and Abdalkareem, R. Dependency Update Strategies and Package Characteristics. \textit{ACM Transactions on Software Engineering and Methodology}, 32(6):149, 2023.

\noindent[S93] Godefroid, P., Lehmann, D., and Polishchuk, M. Differential Regression Testing for REST APIs. In \textit{Proceedings of the 29th ACM SIGSOFT International Symposium on Software Testing and Analysis (ISSTA)}, pp.~312--323, 2020.

\noindent[S94] Reyes, F., Mahmoud, M., Bono, F., Nadi, S., Baudry, B., and Monperrus, M. Byam: Fixing Breaking Dependency Updates with Large Language Models. \textit{Empirical Software Engineering}, 31(4):113, 2026.

\noindent[S95] Serbout, S., and Pautasso, C. How Many Web APIs Evolve Following Semantic Versioning? In \textit{Proceedings of the International Conference on Web Engineering (ICWE)}, pp.~344--359, 2024.

\noindent[S96] Vahlbrock, T., Guddat, M., and Vierjahn, T. VSCode Migrate: Semi-Automatic Migrations for Low Coverage Projects. In \textit{Proceedings of the IEEE International Conference on Software Maintenance and Evolution (ICSME)}, pp.~514--518, 2022.

\noindent[S97] Zaitsev, O., Ducasse, S., Anquetil, N., and Thiefaine, A. How Libraries Evolve: A Survey of Two Industrial Companies and an Open-Source Community. In \textit{Proceedings of the 29th Asia-Pacific Software Engineering Conference (APSEC)}, pp.~309--317, 2022.

\bibliographystyle{ACM-Reference-Format}
\bibliography{sample-base}
\end{document}